\documentclass[%
 reprint,
superscriptaddress,
 amsmath,amssymb,
 aps,
]{revtex4-2}

\usepackage{CJK}
\usepackage{graphicx}%
\usepackage{dcolumn}%
\usepackage{bm}%
\usepackage{textcomp}
\usepackage{xspace}

\newcommand{\munc}{{\textmu}NC\xspace}

\begin{document}
\begin{CJK*}{UTF8}{}
\CJKfamily{min}

\title{Energy spectra of light charged particles emitted following muon nuclear capture on $^\mathrm{nat}$Si}

\newcommand{\kyushusoriko}{Department of Advanced Energy Science and Engineering, Kyushu University, Kasuga, Fukuoka 816-8580, Japan}
\newcommand{\quasr}{Quantum and Spacetime Research Institute, Kyushu University, Fukuoka, Fukuoka 819-0395, Japan}
\newcommand{\kyushuphys}{Department of Physics, Kyushu University, Fukuoka, Fukuoka 819-0395, Japan}
\newcommand{\nishina}{RIKEN Nishina Center, Wako, Saitama 351-0198, Japan}
\newcommand{\kek}{High Energy Accelerator Research Organization (KEK), Tsukuba, Ibaraki 305-0801, Japan}
\newcommand{\ut}{Department of Physics, Graduate School of Science, The University of Tokyo, Bunkyo, Tokyo 113-0033, Japan.}
\newcommand{\rcnp}{Research Center for Nuclear Physics, Osaka University, Mihogaoka 10-1, Ibaraki, Osaka 567-0047, Japan}
\newcommand{\ral}{Science and Technology Facilities Council Rutherford Appleton Laboratory, Didcot, Oxfordshire OX11 0QX, United Kingdom}
\newcommand{\jaea}{Japan Atomic Energy Agency (JAEA), Tokai, Ibaraki 319-1195, Japan}

\author{Shoichiro Kawase (川瀬頌一郎)}
\email{shkawase@kyudai.jp}
\affiliation{\kyushusoriko}%
\affiliation{\quasr}%
\affiliation{\nishina}%

\author{Kentaro Kitafuji (北藤健太郎)}%
\affiliation{\kyushusoriko}%

\author{Teppei Kawata~(\mbox{川田哲平})}%
\affiliation{\kyushusoriko}%

\author{Yukinobu Watanabe (渡辺幸信)}%
\affiliation{\kyushusoriko}%
\affiliation{\quasr}%

\author{Megumi Niikura (新倉潤)}%
\affiliation{\nishina}%

\author{Teiichiro~Matsuzaki~(\mbox{松崎禎市郎})}%
\affiliation{\nishina}%

\author{Katsuhiko Ishida (石田勝彦)}%
\affiliation{\kek}%

\author{Rurie Mizuno (水野るり惠)}%
\affiliation{\ut}%

\author{Dai Tomono (友野大)}%
\affiliation{\rcnp}%

\author{Adrian D. Hillier}%
\affiliation{\ral}%

\author{Futoshi Minato (湊太志)}%
\affiliation{\kyushuphys}%

\author{Shin-ichiro Abe (安部晋一郎)}%
\affiliation{\jaea}%

\date{\today}%

\begin{abstract}
\noindent
\textbf{Background:} Charged-particle emission following muon nuclear capture (\munc) provides unique information on the de-excitation dynamics of highly excited nuclei, particularly on the interplay between preequilibrium and evaporation processes. While proton emission has been relatively well studied, experimental data on composite charged particles remain limited, especially in the low-energy region for $\alpha$ particles.

\noindent\textbf{Purpose:} This work aims to measure energy spectra for individual charged-particle species emitted following \munc on silicon and use them to constrain theoretical descriptions of preequilibrium, evaporation, and composite-particle emission.

\noindent\textbf{Method:} An experiment was performed at the RIKEN-RAL Muon Facility. Charged particles were identified using a combination of $\Delta E$-$E$ telescopes and digital pulse-shape analysis with nTD-Si detectors. The initial energy spectra were reconstructed through an unfolding procedure and compared with calculations based on the microscopic and evaporation model (MEM),
as well as the PHITS code incorporating the surface coalescence model
and the meson-exchange-current extension.

\noindent\textbf{Results:} Energy spectra of protons, deuterons, tritons, and $\alpha$ particles were successfully extracted over a broad energy range. In particular, the low-energy $\alpha$-particle spectrum was measured for
the first time. MEM more closely reproduces the proton, deuteron, and triton spectral shapes and gives a good description of the low-energy $\alpha$-particle spectrum. PHITS reproduces the overall slope of the $\alpha$-particle spectrum but exhibits particle-dependent discrepancies in the absolute yields, including an overestimation of the evaporation component for all four particle species.

\noindent\textbf{Conclusion:} The present results demonstrate clear particle-species-dependent differences in charged-particle emission following \munc. The comprehensive energy spectra obtained in this work provide valuable constraints on theoretical descriptions of
preequilibrium and evaporation processes and highlight the need for improved modeling of composite-particle emission.
\end{abstract}

\maketitle
\end{CJK*}

\section{Introduction}

When a negative muon stops in matter, it may be captured by an atomic nucleus rather than decay; this process is known as muon nuclear capture (\munc)~\cite{Measday2001}.
This process excites the nucleus to energies ranging from several to several tens of MeV, typically well above the particle emission threshold.
The highly excited nucleus subsequently de-excites through the emission of various particles, including neutrons, protons, deuterons, and $\alpha$ particles, and finally, gamma rays.
In this work, \munc is used as a general term for nuclear capture of a negative muon, encompassing ordinary muon capture (OMC) and radiative muon capture (RMC).
OMC proceeds without emission of a prompt radiative photon, whereas RMC includes such photon emission during the capture process.
Since no prompt-photon tag was used, the measured charged-particle spectra were not experimentally separated into OMC and RMC components.

Particle emission following \munc proceeds via multiple reaction mechanisms that occur on different time scales --- namely, direct reactions, preequilibrium processes, and statistical evaporation. Each mechanism contributes differently to the energy spectra of the emitted particles: direct reactions contribute to the high-energy component, typically above a couple of tens of MeV, with an exponential fall-off, the evaporation process dominates the low-energy region below a few MeV, and the preequilibrium process contributes to the intermediate energy range~\cite{Schroder1974}.
Neutron emission from light nuclei can also contain important nonstatistical components, including giant-resonance contributions observed for $^{12}$C and $^{16}$O~\cite{Plett1971} and the selective population of specific one-neutron-emission channels in $^{40}$Ca~\cite{Measday2006}.

Compared with neutron emission, charged-particle emission is additionally affected by the Coulomb barrier.
It is also sensitive to the manner in which the particles are formed and emitted, thereby providing complementary information on the particle-emission mechanism.
In particular, the emission of composite charged particles, such as deuterons, tritons,
and $\alpha$ particles, raises fundamental questions about their formation mechanism.
These particles may be emitted as preformed clusters reflecting intrinsic correlations
inside the nucleus, or they may be formed dynamically through the coalescence of
independently emitted nucleons during the emission process.
Disentangling these scenarios requires detailed information on both the shapes and
absolute normalizations of charged-particle energy spectra, making
charged-particle emission a uniquely sensitive probe of
correlation effects in nuclear decay.
In the case of \munc, where the nucleus is driven to high excitation
energy with relatively low angular momentum, such questions regarding the formation
mechanism of composite charged particles can be addressed under well-defined experimental conditions,
as \munc does not involve incident nucleon or heavy-ion beams
and thus avoids backgrounds originating from beam-induced nuclear reactions.

However, indirect approaches based on the production probabilities of residual nuclei,
as measured by activation measurements~\cite{Wyttenbach1978, Niikura2024, Yamaguchi2025a, Mizuno2025b}
and prompt gamma-ray detection~\cite{Backenstoss1971, Johnson1996, Gorringe1999, Stocki2002, Measday2006, Measday2007a, Measday2007, Zinatulina2019},
cannot distinguish between different particle emission processes leading to the same reaction channels;
for example, the emission of one neutron and one proton as separate particles
($1n1p$ emission) cannot be distinguished from deuteron emission.
Moreover, such approaches do not resolve how the available decay strength is distributed
among the various charged-particle channels. Similarly, discussions based solely on
integrated branching ratios, while informative, do not provide access to the underlying
emission dynamics encoded in the energy spectra. A direct experimental study of
charged-particle emission is therefore essential to fully exploit the well-defined reaction condition provided by \munc,
to disentangle the underlying emission mechanisms, and to access the detailed information encoded in the energy spectra.

In the present work, charged-particle emission following \munc on silicon was investigated.
The motivation for focusing on silicon in the present study is threefold.
Firstly, charged-particle emission after \munc is suppressed by the Coulomb barrier and becomes progressively less probable with increasing atomic number~\cite{Heusser1972,Wyttenbach1978,Heisinger2002,Niikura2024,Yamaguchi2025a}.
As a result, light to medium-mass nuclei, up to around the Ca region,
provide favorable conditions under which charged-particle emission
can be observed with sufficient yield to enable systematic discussion of the emission mechanisms.
Secondly, although silicon has been investigated in several previous experiments~\cite{Sobottka1968,Budyashov1971,Edmonds2022,Manabe2023},
the available experimental data remain limited.
In particular, existing data sets are fragmented in energy coverage and lack information in the low-energy part of the $\alpha$ spectrum.
Thirdly, $\alpha$ particles are expected to play a central role in the statistical evaporation process.
Since $\alpha$-particle emission occurs universally above a certain excitation-energy threshold, the shape of the $\alpha$-particle energy spectrum provides essential constraints on the evaporation component and on the formation mechanisms of composite charged particles.
In this respect, $^{28}$Si, the dominant isotope in natural silicon, is a self-conjugate nucleus ($N=Z$), for which $\alpha$-cluster correlations are expected to be relatively favored compared to non-self-conjugate systems.
Although the present work does not aim to probe nuclear structure directly, this property makes Si suitable 
for investigating $\alpha$ and other composite charged-particle emissions following \munc.

Beyond its relevance to nuclear reaction dynamics, the study of low-energy charged-particle emission following muon capture in silicon is also important from an applied perspective.
In semiconductor devices, cosmic-ray muons have been identified as a potential source
of radiation-induced failures, and low-energy charged particles produced via \munc can
contribute to soft error rates~\cite{Sierawski2014, Trippe2016, Manabe2018, Liao2019, Autran2024, Fabero2026}.

This paper is organized as follows.  
Section II describes the experimental setup, including the muon beam conditions, target configuration, and the detector system optimized for low-energy charged-particle measurements.
Section III outlines the data reduction procedures, including event selection, energy calibration, and particle identification methods.  
The experimental results are presented in Section IV, with a focus on the energy spectra of emitted charged particles.  
In Section V, the results are discussed in the context of nuclear reaction mechanisms and compared with existing data and theoretical model calculations.  
Finally, Section VI summarizes the conclusions of this work and provides perspectives for future studies.

\section{Experiment}

The experiment was performed at the RIKEN-RAL Muon Facility located at the Rutherford Appleton Laboratory (RAL), the UK~\cite{Matsuzaki2001}.
A schematic overview of the RIKEN-RAL beamline, from the pion-production target to the experimental ports, is shown in Fig.~3 of Ref.~\cite{Hillier2018}.
At this facility, 800\,MeV protons accelerated by the ISIS synchrotron were directed onto a 10-mm-thick graphite target installed at the Target Station 1 (TS1) to produce negative pions.  
Negative pions were extracted using quadrupole magnets and bent into a superconducting solenoid, where they decayed in flight to negative muons.
The resulting negative-muon beam was transported to Port 4 through a series of quadrupole and dipole magnets and a DC separator.
The negative-particle transport setting excludes positrons from the beam, while the DC separator suppresses prompt electrons.
Hereafter, the term ``muon'' refers to a negative muon.

The ISIS synchrotron operates at a repetition rate of 50\,Hz with a double-bunch structure.
Of these pulses, four out of every five were delivered to the TS1 target, resulting in an effective repetition rate of 40 pulses per second at TS1.

\begin{figure*}[t]
    \centering
    \includegraphics[width=\textwidth]{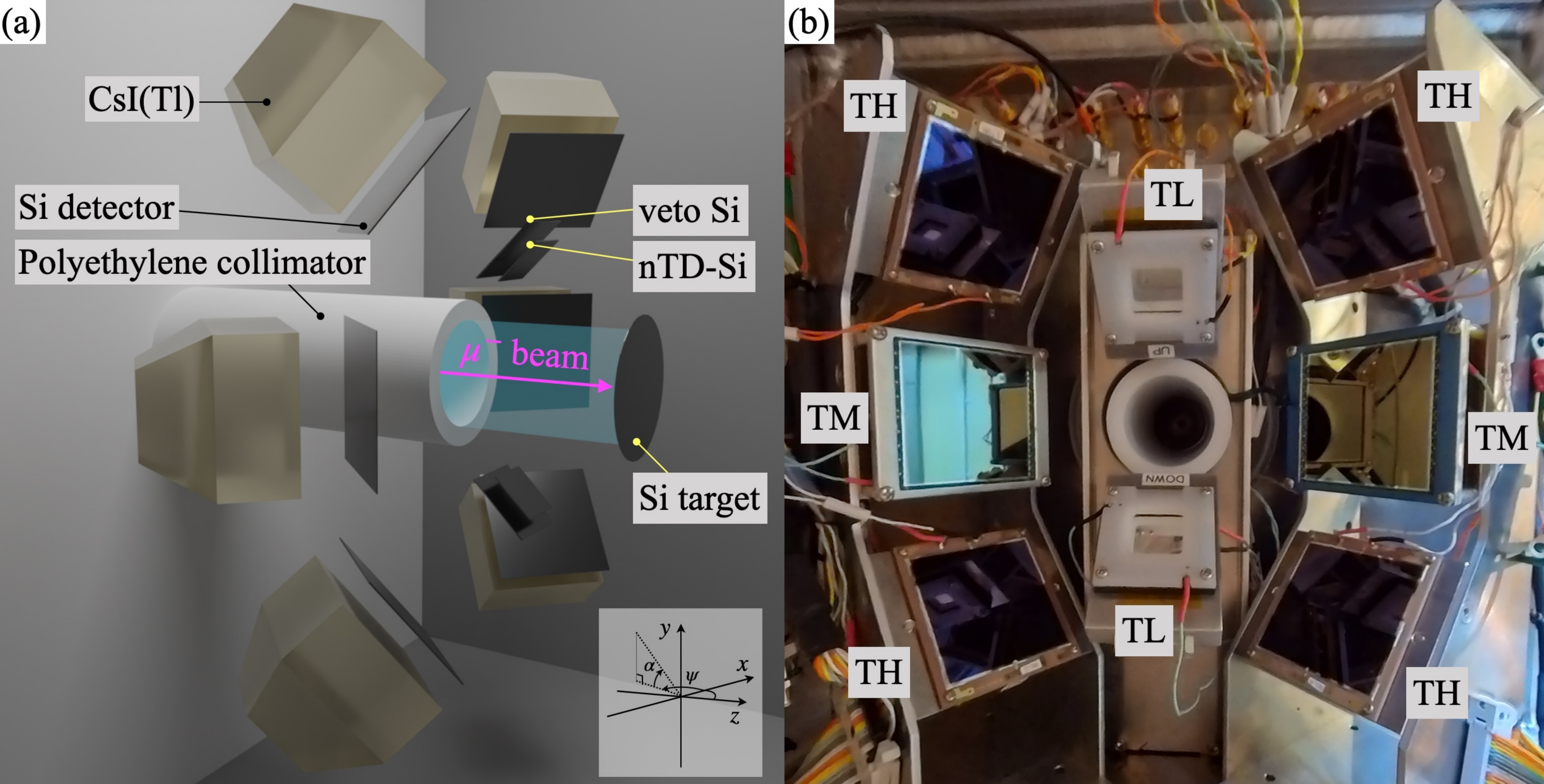}
  \caption{
    Experimental setup around the Si target.
    (a)~Schematic view of the detector arrangement inside the vacuum chamber. The muon beam enters the illustrated setup from the left along the $+z$ direction. Only the sensitive region of each detector is illustrated for simplicity.
    (b)~Photograph of the actual setup inside the vacuum chamber, viewed from downstream.
  }
  \label{fig:setup}
\end{figure*}

The reaction target and detectors were installed inside an aluminum vacuum chamber placed at Port~4 of the RIKEN-RAL facility.
The vacuum chamber was not directly connected to the beam duct; instead, the muon beam passed through two polyimide vacuum windows of 50 and 75\,{\textmu}m thickness, separated by a 5\,cm air gap, before entering the chamber.
The chamber was evacuated to a pressure below 10\,Pa during the measurements.
A schematic diagram and photograph of the experimental setup around the target region are shown in Fig.~\ref{fig:setup}.

To restrict the beam size, a tubular collimator with an inner diameter of 40\,mm
was installed in the vacuum chamber upstream of the target.
Both the target ladder and the collimator were made of polyethylene, and polyethylene sheets were also attached to the inner walls of the vacuum chamber.
These components were used to suppress background events from \munc outside the target region because the \munc probability in carbon ($\sim 6\%$) is substantially lower than that in Si ($\sim 66\%$).

Two natural-silicon wafers, both with a diameter of 50.8\,mm but different thicknesses, were used as reaction targets.
The two targets were used in separate measurements, with each target positioned 282\,mm downstream of the 50-{\textmu}m-thick polyimide window at the Port~4 beam outlet.
A thin target with a thickness of 27\,{\textmu}m was employed to measure low-energy spectra, while a thick target of 200\,{\textmu}m was used to improve statistics in the high-energy region, where the particle emission probability is relatively low. 
To maximize the number of muons stopping within each target,
the nominal muon beam momenta were set to 18.0\,MeV/$c$ for the thin target and 21.5\,MeV/$c$ for the thick target.
Approximate beam momenta were first estimated using G4beamline simulations of muon transport and stopping~\cite{Roberts2007}.
During beam tuning, the momentum was scanned around each estimated value while the numbers of muons reaching the downstream plastic scintillator were measured with and without the corresponding Si target.
The final momentum was selected so as to maximize the difference between the target-out and target-in muon yields, which provides an estimate of the number of muons stopping in the target.

To detect charged particles over a wide range of kinetic energies, measure their deposited energies,
and identify particle species, three detector-telescope configurations were employed,
as summarized in Table~\ref{table:detector_spec}.
Hereafter, they are referred to as TL, TM, and TH, where L, M, and H indicate the low-, medium-, and high-energy regions primarily covered by each configuration, respectively.

The TM and TH telescopes employed conventional $\Delta E$-$E$ configurations,
each consisting of a front silicon (Si) detector and a rear CsI(Tl) scintillator,
and were used for energy measurement and particle identification in the high-energy region.
Two Si photodiodes with different thicknesses were employed as the $\Delta E$ detector: one was a Hamamatsu Photonics 50x50x325T-PD 0899 with a thickness of 325\,{\textmu}m, and the other was a Hamamatsu Photonics S14536-500 with a thickness of 500\,{\textmu}m.  
Using detectors with different thicknesses allowed complementary coverage in energy and helped eliminate insensitive gaps in the energy spectra.  
The scintillation light from the CsI(Tl) was read out using a PIN photodiode (Hamamatsu Photonics S3584-08).  

Each of the two TL telescopes had an identical two-layer configuration consisting
of a front neutron-transmutation-doped (nTD) Si detector
(Micron Semiconductor MSX04-500 Type NTD) and a rear Si detector
(Hamamatsu Photonics S14537-500) used as a veto for penetrating particles.
Both Si detectors had a thickness of 500\,{\textmu}m.
The nTD-Si detectors were mounted with their ohmic sides facing the target,
a configuration essential for pulse-shape discrimination.
In this configuration, signals from charged particles that stop in the nTD-Si detector
exhibit particle-species-dependent pulse shapes even at the same deposited energy~\cite{Kawase2024}.
In the digital pulse-shape analysis (DPSA), two observables were extracted from each digitized preamplifier waveform:
the deposited energy $E$, derived from the total collected charge after trapezoidal filtering,
and the maximum current $I_{\mathrm{max}}$, obtained from the time derivative of the smoothed waveform.
Particle species were identified from their distinct loci in the $E$--$I_{\mathrm{max}}$ correlation.

\begin{table*}[htbp]
  \centering
  \caption{Specifications of the detector components constituting each telescope type and their applied bias voltages. All telescopes of a given type had the same detector configuration. ``Front'' and ``rear'' denote the detector layers nearer to and farther from the target, respectively.}\label{table:detector_spec}
  \begin{tabular}{ccccc}
    \hline\hline
    Telescope type (layer) & Detector & Entrance-face size & Thickness & Bias Voltage \\
    \hline
    TL (front) & nTD-Si detector MSX04-500 (Micron) & $20 \times 20$\,mm$^2$ & 500\,{\textmu}m & 260\,V \\
    TL (rear) & Si detector S14537-500 (Hamamatsu) & $28 \times 28$\,mm$^2$ & 500\,{\textmu}m & 100\,V \\
    TM (front) & Si detector 50x50x325T-PD 0899 (Hamamatsu) & $50 \times 50$\,mm$^2$ & 325\,{\textmu}m & 70\,V \\
    TH (front) & Si detector S14536-500 (Hamamatsu) & $48 \times 48$\,mm$^2$ & 500\,{\textmu}m & 150\,V \\
    TM, TH (rear) & CsI(Tl) scintillator\footnote{The crystal consists of a $50 \times 50 \times 25$\,mm$^3$ rectangular section followed by a 22-mm-thick truncated-pyramid section tapering from a $50 \times 50$\,mm$^2$ to a $28 \times 28$\,mm$^2$ cross section.} & $50 \times 50$\,mm$^2$ & 47\,mm & 100\,V \\
    \hline\hline
  \end{tabular}
\end{table*}

\begin{table}[htbp]
  \centering
  \caption{Number of telescopes and their geometric configurations.
  The detector positions are described using two angular coordinates: $\psi$ and $\alpha$. Here, $\psi$ is the azimuthal angle measured anticlockwise in the $zx$-plane from the $z$-axis, and $\alpha$ is the elevation angle from the $zx$-plane.
  The distance corresponds to that between the center of the target
  and the center of the surface of each detector.}\label{table:telescope_spec}
  \begin{tabular}{ccccc}
    \hline\hline
    Telescope & Number & ($\psi$, $\alpha$) & Distance & $\Delta\Omega$\\
              &        &                    & (mm)     & (msr) \\
    \hline
    TH & 4 & $\pm130^\circ$, $\pm40^\circ$ & 113.2 & 144.8\\
    TM & 2 & $\pm130^\circ$, $\phantom{\pm 0}0^\circ$ & 100.7 & 143.6 \\
    TL & 2 & $\phantom{\pm}180^\circ$, $\pm 45^\circ$ & \phantom{0}64.8 & \phantom{0}75.6 \\
    \hline\hline
  \end{tabular}
\end{table}

To optimize particle identification capability, a low bias voltage is preferred for the nTD-Si detectors, as long as sufficient charge collection efficiency and energy resolution are maintained.
In this experiment, a bias voltage of 260\,V was applied to the nTD-Si detectors.
In addition, a 3-mm-thick polyethylene collimator with an
$18 \times 18$\,mm$^2$ square aperture was placed in front of each nTD-Si detector.
This collimator was employed primarily to limit the effective sensitive area of the nTD-Si detector
and to prevent the deterioration of particle identification performance near its edges~\cite{Kawase2024}.

The geometric configuration of the detector telescopes is summarized in Table~\ref{table:telescope_spec}.
The detector positions are described using two angular coordinates, $\psi$ and $\alpha$.
A right-handed Cartesian coordinate system is adopted, with the $z$ axis defined along the muon beam direction and the $y$ axis along the vertical direction.
Here, $\psi$ is defined as the angle in the $zx$-plane measured from the $z$ axis, and $\alpha$ is the elevation angle measured from the $zx$-plane.
These angles are used solely to specify the detector geometry in this experiment and differ from the conventional spherical coordinate definitions.
To reduce background events caused by muons scattered in the target, all detector telescopes were placed upstream of the target.

To measure the number of incoming muons and those penetrating the target per pulse, a plastic scintillator with dimensions of $50 \times 70$\,mm$^2$ and a thickness of 0.5\,mm was placed 1\,cm downstream of the target.
The scintillation light was read out from one side using a photomultiplier tube (Hamamatsu Photonics H11934-100).
This scintillator was used only during beam tuning and was removed during the physics measurement runs, as it would otherwise contribute to background events.

Analog signals from Si detectors and CsI(Tl) scintillators were amplified using charge-sensitive preamplifiers (Fuji-diamond International Co., Ltd. 0380-16).  
The amplified signals were then digitized with 14-bit resolution at a sampling rate of 500\,MHz using waveform digitizers (CAEN SpA V1730SB), and transferred to a PC.
The waveform record length was set to 20\,{\textmu}s, which is sufficiently longer than the lifetime of the negative muon.

In the V1730SB~\cite{compass}, waveform readout is performed on a channel-pair basis, and the associated buffer capacity is insufficient to record 20\,{\textmu}s waveforms from all 16 channels simultaneously.
To circumvent this limitation, two digitizer modules were employed,
and detector signals were connected to every other input channel,
\textit{i.e.}, only even-numbered channels were used.
This configuration ensured that each channel pair processed at most a single detector signal, allowing full-length waveform acquisition for all channels without loss.
Waveform data were recorded using the CAEN Multi-Parameter Spectroscopy Software (CoMPASS)
and transferred to a PC for offline analysis.

To synchronize data acquisition with the pulse structure of the negative muon beam, a 50-Hz timing signal synchronized with the primary proton beam was fed into the V1730SB digitizer as an external trigger.
Upon each trigger, waveforms from all detector channels were simultaneously recorded.
Thus, all detector telescopes were synchronized by the common external trigger, irrespective of detector type or manufacturer.
The timing of the external trigger was adjusted so that the recorded waveform included at least 1\,{\textmu}s of data prior to the muons' arrival, ensuring adequate coverage of the baseline and signal rise.
Waveform acquisition was thus performed at fixed beam timing, regardless of whether a charged particle was detected in the telescopes or not, and even during beam cycles in which no protons were delivered to TS1.
Since the digitizer continuously recorded waveforms upon each trigger without requiring per-event decisions, the data acquisition was effectively free from dead time.

\section{Analysis}

This section describes the data analysis procedures used to extract the energy spectra of particles emitted following \munc.  
The primary objective of the analysis is to obtain the initial energy spectrum of charged particles emitted following \munc, denoted as $dN_{\mathrm{init}}/dE$.

Experimentally, however, what is directly measured is a distorted spectrum, $dN_{\mathrm{meas}}/dE$, due to energy loss and straggling in the target material.  
The measured spectrum was calculated from the experimental yields using the following formula:
\begin{equation}
\frac{dN_{\mathrm{meas}}}{dE}(E) = \frac{4\pi}{\Delta\Omega} \cdot \frac{Y(E)}{N_{\mathrm{capture}} \cdot \Delta E_\mathrm{bin}},
\end{equation}
where $Y(E)$ is the number of counts in each energy bin of width $\Delta E_\mathrm{bin}$, $\Delta\Omega$ is the solid angle covered by the detector, and $N_{\mathrm{capture}}$ is the total number of muon capture events.  
The factor $4\pi/\Delta\Omega$ accounts for the extrapolation of the measured yield to the full solid angle, assuming isotropic emission.

To correct for the distortion caused by energy loss in the target, the initial spectrum $dN_{\mathrm{init}}/dE$ was obtained by applying an unfolding analysis to the measured spectrum $dN_{\mathrm{meas}}/dE$.

In the following, we describe the procedures for determining $N_{\mathrm{capture}}$, identifying charged particles, reconstructing their energies, performing unfolding analysis, and evaluating systematic uncertainties.

\subsection{Number of stopped muons}\label{sec:muon_number}

To determine the charged-particle energy spectrum per muon capture,
the observed spectra were normalized to the number of muon captures
in the silicon target, $N_{\mathrm{capture}}$.
The number of muon captures was obtained by multiplying the number of stopped muons,
$N_{\mathrm{stop}}$, by the muon capture probability in natural silicon,
65.9(1)\%, which was calculated from the muon capture rate reported in Ref.~\cite{Suzuki1987}.
The number of stopped muons was inferred using a plastic scintillator placed just downstream of the target.
The difference between the numbers of muons reaching the scintillator in the target-out and target-in configurations, after normalizing each to the corresponding average primary-proton beam current incident on the pion production target, was attributed to muons stopped in the Si target.

The number of stopped muons $N_{\mathrm{stop}}$ was calculated using the following formula:
\begin{equation}
  N_{\mathrm{stop}} = f \cdot \left( \frac{n_{\mathrm{out}}}{\bar{I}_{\mathrm{out}}} - \frac{n_{\mathrm{in}}}{\bar{I}_{\mathrm{in}}} \right) Q_{\mathrm{meas}}
  \label{eq:nstop}
\end{equation}
where $f$ is the frequency of proton beam pulses, i.e. 40 pulses per second,
and $n$ denotes the number of muons stopped in the plastic scintillator per pulse.
The subscripts ``in'' and ``out'' correspond to configurations with and without the Si target, respectively.
$\bar{I}$ is the average proton beam current incident on the pion production target, measured during each configuration.
$Q_{\mathrm{meas}}$ is the total proton beam charge accumulated during the spectrum measurement.
The values of $\bar{I}$ and $Q_{\mathrm{meas}}$ used in this work are summarized in Table~\ref{tab:nstop}.

\begin{table}[tb]
  \centering
  \caption{Experimental parameters used to determine $N_{\mathrm{stop}}$.}
  \label{tab:nstop}
  \begin{tabular}{cccccc}
    \hline\hline
    Target thickness\rule{0pt}{2.4ex} & beam momentum & $\bar{I}_{\mathrm{out}}$ & $\bar{I}_{\mathrm{in}}$ & $Q_{\mathrm{meas}}$ \\
    ({\textmu}m) & (MeV/$c$) & (nA) & (nA) & (mC) \\
    \hline
    200 & 21.5 & 137.4 & 137.0 & 14.48 \\
    \phantom{0}27 & 18.0 & 137.8 & 137.6 & 16.43 \\
    \hline\hline
  \end{tabular}
\end{table}

The plastic scintillator used in this study had a thickness of 0.5\,mm,
which is sufficient to stop negative muons with momenta of 21.5\,MeV/$c$ and 18.0\,MeV/$c$.
The number of stopped muons per pulse, $n$, was therefore estimated by performing peak searches on the digitized waveform signals from the plastic scintillator placed downstream of the target.

\begin{figure}[tb]
  \centering
  \includegraphics[width=\linewidth]{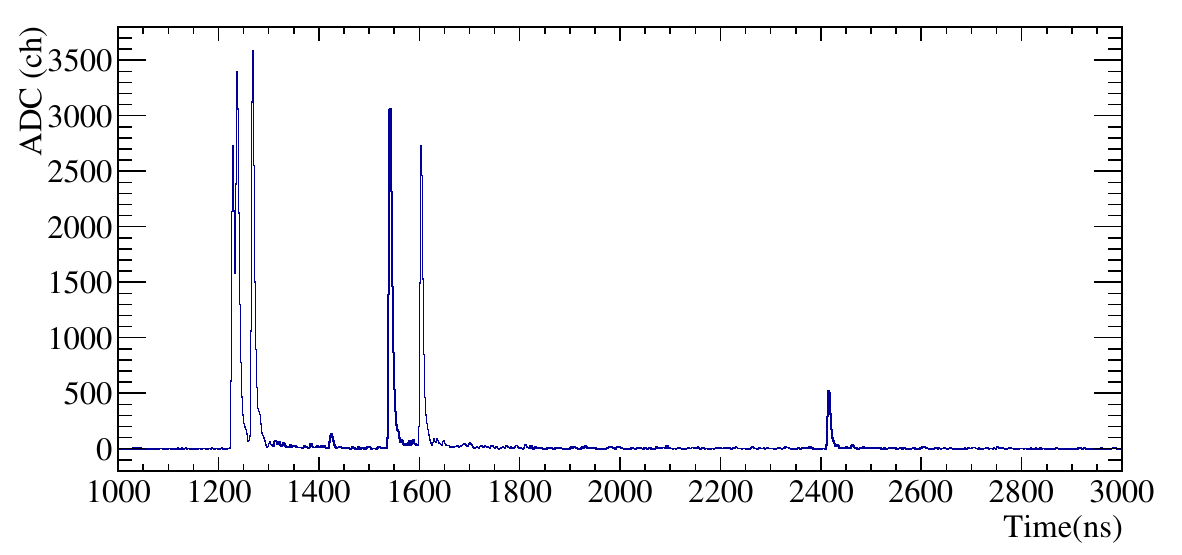}
\caption{
Digitized waveform of the plastic scintillator from the photomultiplier tube obtained for the 21.5\,MeV/$c$ muon beam run without the target.
In this waveform, three muons reach the scintillator in the first beam bunch and two muons in the second beam bunch.
Peaks observed around 1,425\,ns and 2,415\,ns are attributed to electrons originating from muon decay.
}
  \label{fig:waveform_example}
\end{figure}

Figure~\ref{fig:waveform_example} shows a typical waveform of the plastic scintillator from the PMT.
In the present analysis, a ``peak'' is defined as a local maximum in the digitized waveform exceeding a predefined threshold.
The time bin at which the waveform reaches this local maximum is referred to as the \emph{peak timing}.

To quantify the signal associated with each peak, a \emph{peak integral} was calculated by summing the waveform amplitudes over a fixed time window from $-2$ to $+3$ bins relative to the peak timing.
This integration window was chosen to capture the full pulse shape while minimizing the contribution from electronic noise and neighboring pulses.

These waveform peaks include not only negative muons but also electrons.
The electron component has two origins: electrons produced at the pion production target
by the primary proton beam and subsequently transported as contaminants in the muon beam,
and decay electrons from muons.
Electrons originating from the pion production target are strongly suppressed
by a DC separator installed in the beamline and therefore provide only a minor contribution.
However, decay electrons from muons that pass through the DC separator may still reach
the scintillator and generate spurious peaks.

The contribution from these electrons was evaluated using both the peak timing and waveform integral information.
In the following subsections, we describe the procedure used to estimate the number of stopped muons $n$ for each beam momentum and target thickness.

\subsubsection{$n_\mathrm{out}$ for the 21.5\,MeV/$c$ run}

\begin{figure}[tb]
  \centering
  \includegraphics[width=\linewidth]{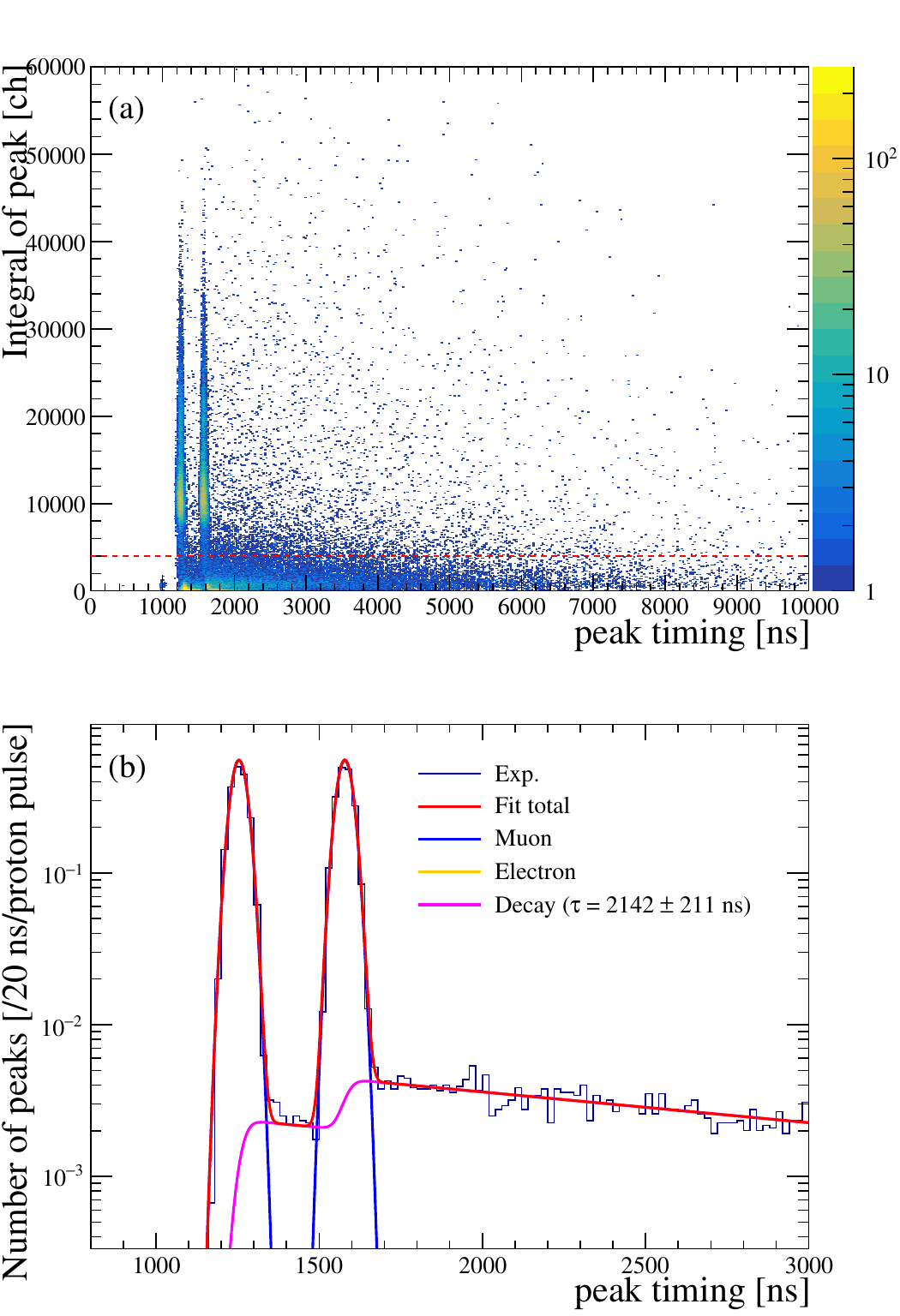}
\caption{
Beam timing structure in the 21.5\,MeV/$c$ beam run with the target removed.  
(a) Two-dimensional histogram of peak timing versus integrated waveform value obtained from the plastic scintillator. Two distinct loci corresponding to the first and second muon pulses are clearly visible.
The applied threshold on the integrated value is indicated by the red dashed line.
(b) Timing distribution of peaks with integrated values above 4,000\,ch, projected from the top panel. The data were fitted with a composite function including Gaussian components for muon and electron peaks and an exponential background for decay electrons. The decay time constant extracted from the fit is also shown in the figure.
}
  \label{fig:muon_number_200_out}
\end{figure}

Figure~\ref{fig:muon_number_200_out} (a) shows a two-dimensional histogram of the peak timing and integrated peak value obtained from the plastic scintillator waveform with the target removed. Two distinct loci with integrated values around 10,000 are observed at timing windows of 1,160--1,340\,ns and 1,500--1,680\,ns, corresponding to the arrival of negative muons in the first and second beam bunches, respectively. The double structure reflects the inherent double-bunched time structure of the primary proton beam.

In contrast, a broad distribution of peaks with low integrated values appears at later times, indicating decay electrons.
As electrons produced during muon production are largely suppressed by the DC separator in the beamline,
these later peaks are attributed primarily to decay electrons from muons that passed through the separator and subsequently stopped in the scintillator.

Figure~\ref{fig:muon_number_200_out} (b) shows the projection of the two-dimensional histogram (a) onto the timing axis, restricted to events with integrated values above 4,000\,ch. This cut reduces the contribution from decay electrons and isolates the timing structure of the muon signals.

The resulting distribution was fitted using a composite function consisting of Gaussian peaks for the muon and electron signals and a decay-convolved Gaussian component representing the contribution from decay electrons.
The explicit form of the fitting function is provided in Appendix~\ref{appendix:fit_function}.

The number of stopped muons in the scintillator was evaluated by integrating the two Gaussian components corresponding to the muon pulses. 
Before applying the pileup correction, the estimated number of stopped muons per pulse pair was $3.531 \pm 0.017$.

This indicates that events in which multiple muons arrive within a single pulse window are present, and that the resulting pileup effects are not negligible.
The pileup probability was estimated to be 15.6\% using a Monte Carlo simulation, assuming that peaks separated by more than 6\,ns can be resolved.
A correction corresponding to this pileup probability was applied, resulting in
$n_{\mathrm{out}} = 4.080 \pm 0.020$.

The statistical uncertainty was evaluated using the covariance matrix obtained from the fit. The decay time constant extracted from the exponential component was found to be consistent with the known mean lifetime of negative muons in carbon, supporting the interpretation of the tail as originating from decay electrons in the scintillator.

\subsubsection{$n_\mathrm{in}$ for the 21.5\,MeV/$c$ run with a 200-{\textmu}m target}

The determination of $n_\mathrm{in}$ for the 21.5\,MeV/$c$ run with the 200-{\textmu}m target
was performed in the same manner as in the target-removed case,
using the timing distribution of peaks identified in the plastic scintillator waveform.
However, with the 200-{\textmu}m target in place,
the energy deposition by muons in the scintillator is reduced due to energy loss in the target material. As a result, muon and electron signals cannot be clearly distinguished based on the peak integral alone.
In the present analysis, this limitation does not affect the determination of the stopped-muon yield,
since the timing distribution was analyzed by fitting both muon and electron components
above a sufficiently low threshold.

\begin{figure}[tb]
  \centering
  \includegraphics[width=\linewidth]{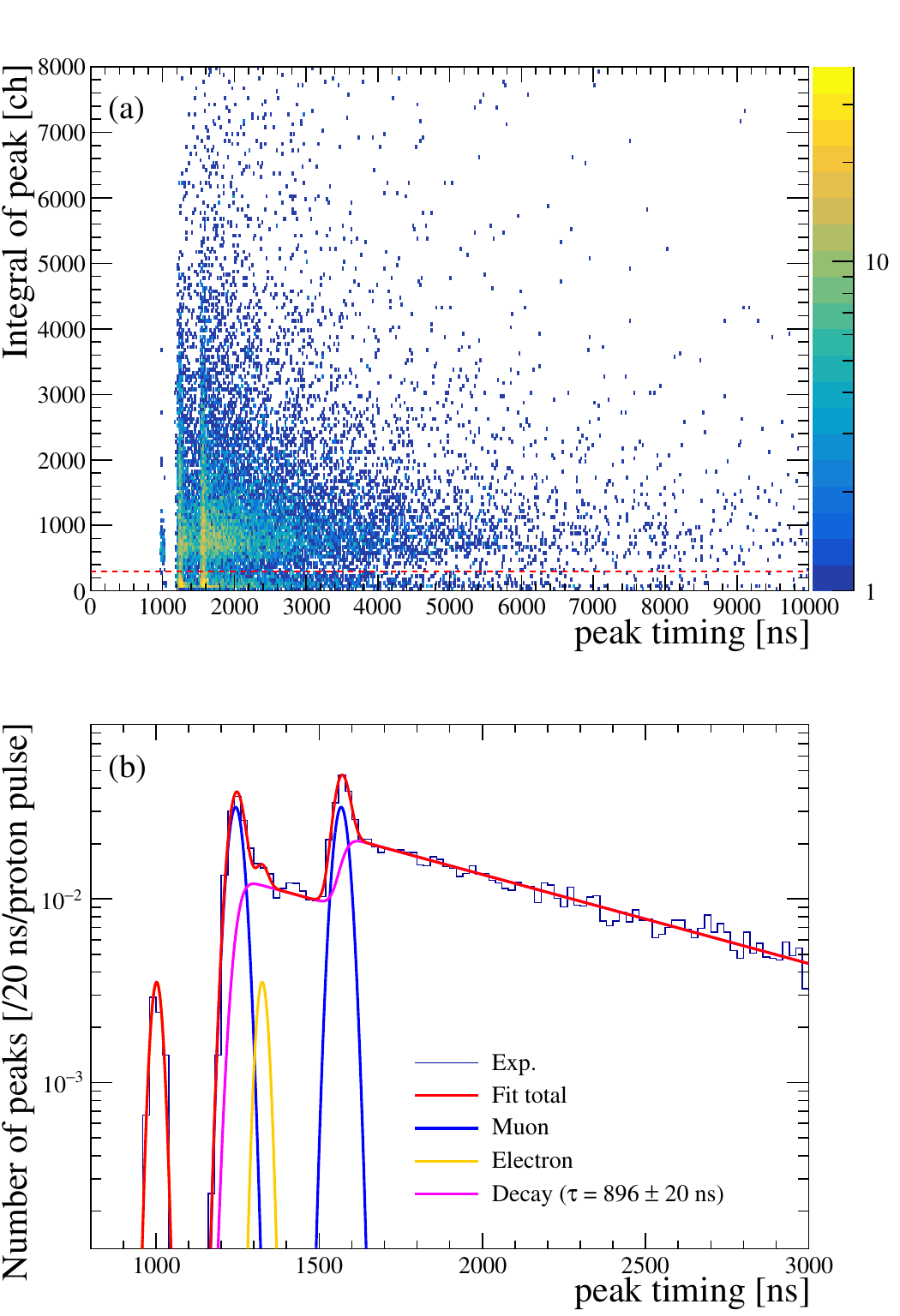}
  \caption{
  Beam timing structure for the 21.5\,MeV/$c$ run with the target in place.  
  (a) Two-dimensional histogram of peak timing versus integrated waveform value. Due to energy loss in the target, the muon loci are less distinct than in the target-removed case.  
  The applied threshold on the integrated value is indicated by the red dashed line.
  (b) Timing distribution of peaks after applying an integrated-value threshold of 300\,ch. The distribution is fitted using the model described in Appendix~\ref{appendix:fit_function}.
  }
  \label{fig:muon_number_200_in}
\end{figure}

Figure~\ref{fig:muon_number_200_in} (a) shows a two-dimensional histogram of peak timing versus integrated peak value. Unlike the target-removed case, the muon-related structures are less distinct and significantly overlap with decay-electron signals.

To extract the muon signal, a lower threshold of 300\,ch was applied to the integrated value, and the resulting peaks were projected onto the timing axis. The timing distribution is shown in Fig.~\ref{fig:muon_number_200_in} (b). The fit was performed using the same functional form as in Appendix~\ref{appendix:fit_function}.

As in the target-removed case, the number of stopped muons was evaluated by integrating the two Gaussian components corresponding to the muon pulses. The estimated number of stopped muons per pulse pair is $n_{\mathrm{in}} = 0.182 \pm 0.006$.

The statistical uncertainty was derived from the covariance matrix of the fit. 
The decay constant obtained from the exponential component is \(\tau = (896 \pm 20)\,\mathrm{ns}\). This value lies between the lifetimes of negative muons in silicon (\(\sim 756\,\mathrm{ns}\)) and in carbon (\(\sim 2{,}026\,\mathrm{ns}\))~\cite{Suzuki1987}, consistent with the interpretation that the decay component includes a mixture of muons stopped in the target and the plastic scintillator.

Because the average number of muons per bunch is significantly below one in this case, the probability of pileup is negligible.

For this beam momentum, the fitted prompt-electron-to-muon yield ratio,
$A_e/A_\mu$, was below 1\%.

\subsubsection{$n_\mathrm{out}$ for the 18.0\,MeV/$c$ run}

The determination of $n_\mathrm{out}$ for the 18.0\,MeV/$c$ run
was performed in the same manner as for the 21.5\,MeV/$c$ run,
using the timing distribution of peaks identified in the plastic scintillator waveform
without the target.

Compared with the 21.5\,MeV/$c$ target-out run, the overlap between the muon and decay-electron components in integrated peak value is greater because of the lower beam momentum. At 18.0\,MeV/$c$, muons are more strongly slowed before reaching the scintillator and therefore deposit less energy. This situation is similar to that observed in the target-in case at 21.5\,MeV/$c$.
As in that case, the stopped-muon number can be reliably determined from the timing distribution.

\begin{figure}[tb]
  \centering
  \includegraphics[width=\linewidth]{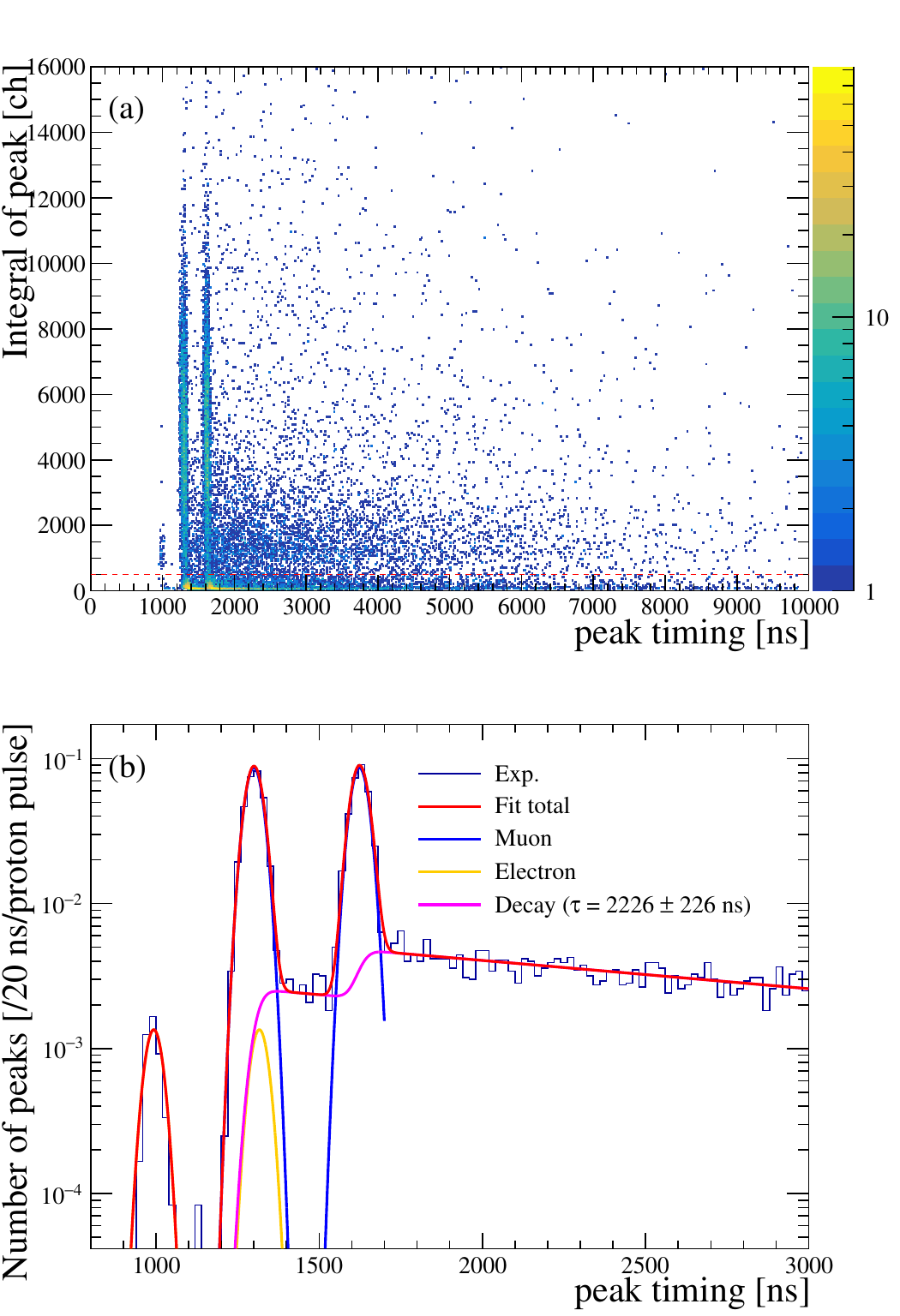}
  \caption{
  Beam timing structure for the 18.0\,MeV/$c$ run with the target removed.  
  (a) Two-dimensional histogram of peak timing versus integrated waveform value.
  The applied threshold on the integrated value is indicated by the red dashed line.
  (b) Timing distribution of peaks after applying an integrated-value threshold of 500\,ch. The distribution is fitted using the model described in Appendix~\ref{appendix:fit_function}. 
  }
  \label{fig:muon_number_25um_out}
\end{figure}

Figure~\ref{fig:muon_number_25um_out} (a) shows the two-dimensional histogram of peak timing versus integrated peak value. Two clear loci corresponding to the two beam bunches are again visible. The projection onto the timing axis with an integrated-value threshold of 500\,ch is shown in Fig.~\ref{fig:muon_number_25um_out} (b). The fit was performed using the same functional form as in Appendix~\ref{appendix:fit_function}.

As in previous cases, the number of stopped muons was extracted by integrating the two Gaussian components corresponding to the muon pulse pair. The estimated number of stopped muons per pulse pair is
$0.581 \pm 0.007$.

The decay constant obtained from the fit is consistent with the known muon lifetime in both vacuum and carbon.
For this configuration, the pileup probability was estimated to be 0.5\%, and the corresponding correction was applied,
resulting in $n_{\mathrm{out}} = 0.584 \pm 0.007$.

\subsubsection{$n_\mathrm{in}$ for the 18.0\,MeV/$c$ run with a 27-{\textmu}m target}

\begin{figure}[tb]
  \centering
  \includegraphics[width=\linewidth]{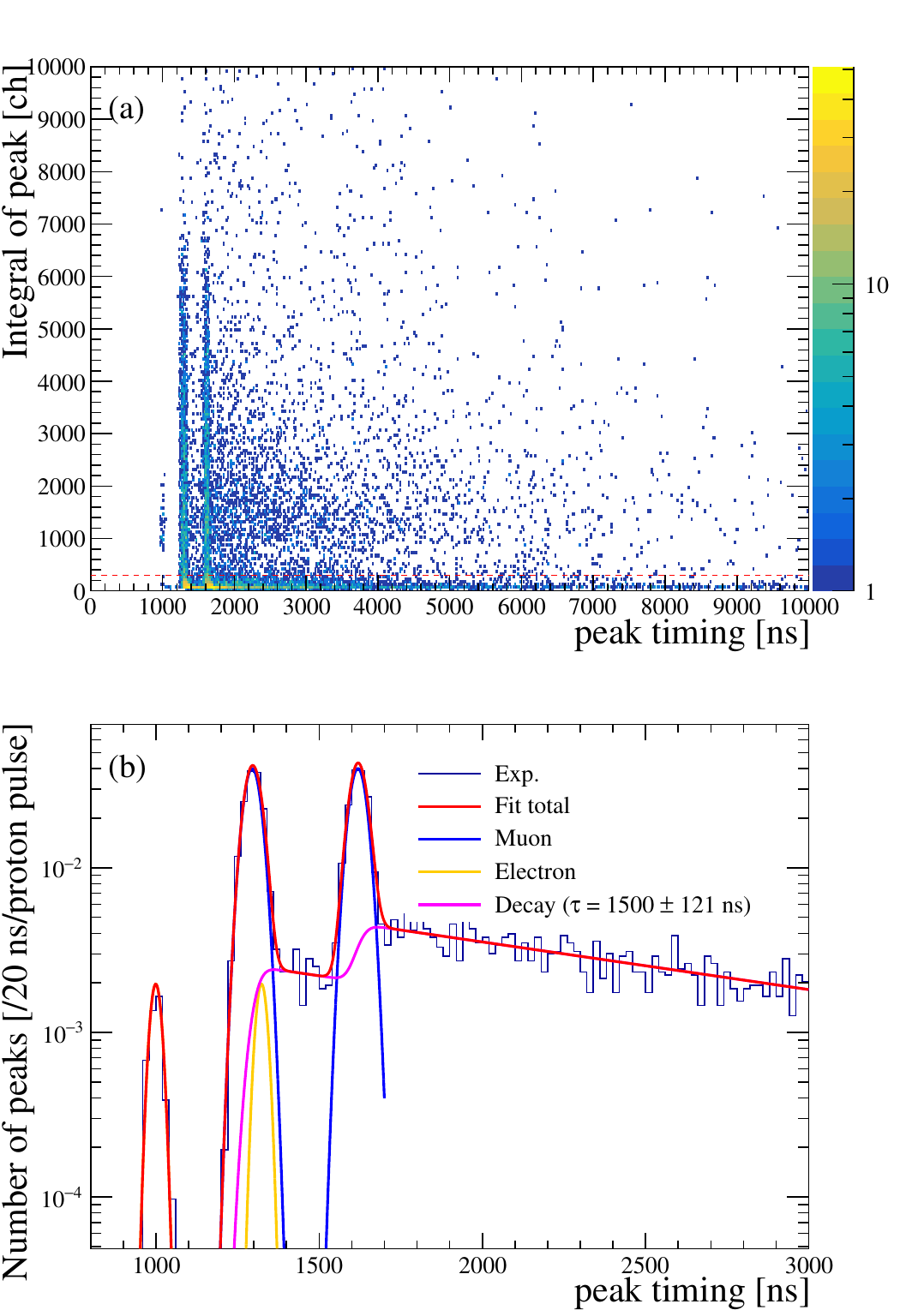}
  \caption{
  Analysis of the peak timing distribution for the 18.0\,MeV/$c$ run with the target in place.  
  (a) Two-dimensional histogram of peak timing versus integrated waveform value. Due to energy loss in the target and lower beam momentum, the muon loci are less distinct.
  The applied threshold on the integrated value is indicated by the red dashed line.
  (b) Timing distribution of peaks after applying an integrated-value threshold. The distribution is fitted using the model described in Appendix~\ref{appendix:fit_function}. 
  }
  \label{fig:muon_number_25um_in}
\end{figure}

The determination of $n_\mathrm{in}$ for the 18.0\,MeV/$c$ run with the 27-{\textmu}m target
was performed in the same manner as for the 21.5\,MeV/$c$ target-in case,
using the timing distribution of peaks identified in the plastic scintillator waveform.
Figure~\ref{fig:muon_number_25um_in} summarizes the result.

Compared to the 21.5\,MeV/$c$ run, the lower beam momentum in this setting leads to further reduction in muon energy deposition in the scintillator. As a result, the separation between muon and decay-electron signals becomes even less distinct, both in the two-dimensional histogram and in the projected timing distribution.
Nevertheless, the stopped-muon yield can be reliably determined from the timing distribution,
as in the previous cases.

Figure~\ref{fig:muon_number_25um_in} (a) shows the two-dimensional histogram of peak timing versus integrated waveform value. The muon loci are not clearly distinguishable from the background of decay electrons. As in previous cases, a threshold of 300\,ch on the integrated value was applied to suppress low-energy background, and the timing distribution of the selected peaks is shown in Fig.~\ref{fig:muon_number_25um_in} (b).

The distribution was fitted using the same functional form described in Appendix~\ref{appendix:fit_function}. The number of stopped muons was evaluated by integrating the Gaussian components corresponding to the muon pulses. The resulting number of stopped muons per pulse pair is $n_{\mathrm{in}} = 0.267 \pm 0.006$.

The decay constant extracted from the exponential component is \(\tau = (1{,}500 \pm 121)\,\mathrm{ns}\), which is again consistent with a mixture of muon decays in both silicon and carbon. Given the relatively low muon rate per pulse, the probability of pileup is negligible.

For this beam momentum, the fitted prompt-electron-to-muon yield ratio,
$A_e/A_\mu$, was at most 1\%.
At both beam momenta, electrons make a negligible contribution to nuclear reactions.

\subsubsection*{5. Evaluated numbers of stopped muons in the target}

Table~\ref{tab:muon_counts} summarizes the evaluated number of incident muons per pulse pair,
$n_{\mathrm{out}}$, the number of muons stopped in the plastic scintillator per pulse pair,
$n_{\mathrm{in}}$, and the total number of muons stopped in the target, $N_{\mathrm{stop}}$,
for each beam momentum.
The values of $N_{\mathrm{stop}}$ were calculated using Eq.~(\ref{eq:nstop}).

\begin{table}[tb]
\centering
\caption{Evaluated muon counts per pulse pair ($n$) and total stopped muon counts in the target for each beam momentum.}
\label{tab:muon_counts}
\begin{tabular}{ccccc}
\hline\hline
$p_\mathrm{beam}$ & $n_{\mathrm{out}}$ & $n_{\mathrm{in}}$ & $N_{\mathrm{stop}}$ & $N_{\mathrm{capture}}$ \\
(MeV/$c$) &  &    & ($/10^6$) & ($/10^6$)\\
\hline
21.5 & $4.080(20)$ & $0.182(6)$ & $16.43(9)$ & 10.83(6)\\
18.0 & $0.584(7)\phantom{0}$ & $0.267(6)$ & \phantom{0}$1.51(4)$ & \phantom{0}1.00(3) \\
\hline\hline
\end{tabular}
\end{table}

\subsection{Energy calibration}

The energy signals from the silicon detectors were obtained from the deposited-energy observable $E$ extracted by DPSA. For energy calibration of the silicon channels, a mixed $\alpha$-source containing $^{148}$Gd, $^{241}$Am, and $^{244}$Cm was used.

The CsI(Tl) detectors exhibit a non-linear relationship between the light output and deposited energy. Therefore, the following quadratic calibration formula was adopted:
\begin{equation}
E = a + bQ + c\,(bQ)^2
\label{eq:CsIcal}
\end{equation}
where $Q$ is the light output.

For each measured proton energy loss $\Delta E$ in the silicon detector, the corresponding light output $Q_p$ in the CsI(Tl) was obtained.
Independently, the proton energy loss $\Delta E$ in the silicon detector and the residual energy $E$ deposited in the CsI(Tl) were simulated with Geant4~\cite{Agostinelli2003, Allison2006, Allison2016}.
The parameters $a$, $b$, and $c$ were determined by matching the simulated $(\Delta E,E)$ pairs to the measured $(\Delta E,Q_p)$ pairs using Eq.~(\ref{eq:CsIcal}).
No statistically significant $\alpha$-particle population was observed among events reaching the CsI(Tl) scintillators.
The $\alpha$-particle spectrum was therefore constructed only from events that stopped in the Si detectors and did not rely on the CsI(Tl) calibration.

\begin{figure}[t]
  \centering
  \includegraphics[width=\linewidth]{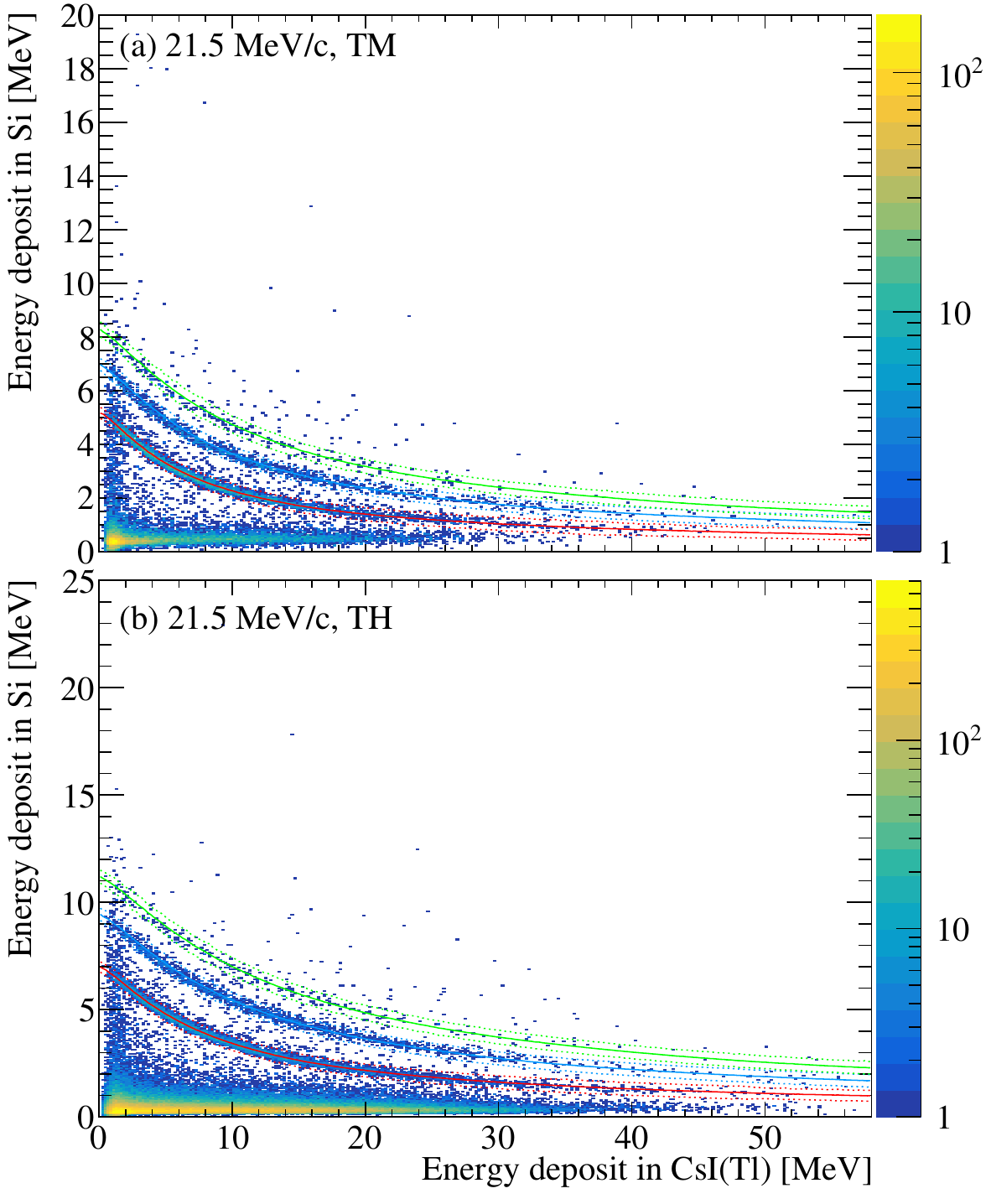}
  \caption{
  (a) $\Delta E$-$E$ plot obtained with telescope TM for a muon beam momentum of 21.5\,MeV/$c$.
  The horizontal axis represents the energy deposited in the CsI(Tl) scintillator, while the vertical axis shows the energy deposited in the Si detector. The solid lines indicate the centroid loci of the Geant4 simulations, and the dashed lines correspond to $\pm 3\sigma$ around them. Events within these bands were selected as the respective particle species: protons (red), deuterons (blue), and tritons (green). (b) Same as (a), but for telescope TH.
  }
  \label{fig:pid_dee}
\end{figure}

Figure~\ref{fig:pid_dee} shows the resulting $\Delta E$--$E$ correlation for the measured data together with the Geant4 simulation, demonstrating good agreement over the full dynamic range.

With these calibrations in place, all subsequent energy spectra of charged particles were converted from raw pulse heights into physical energy deposits.

\subsection{Particle identification}

\subsubsection{Telescope TM and TH}

For events in which particles penetrated the front Si detectors of telescopes TM and TH, particle identification was performed using the conventional $\Delta E$-$E$ method. Figure~\ref{fig:pid_dee} shows the correlation between the energy deposited in the Si detector and that in the CsI(Tl) scintillator, where clear separation of particle species is achieved up to about 50\,MeV. In the present analysis, protons, deuterons, and tritons were identified by requiring that their energy loss in the Si detector fall within $\pm 3\sigma$ of the centroid values obtained from Geant4 simulations.
A population of events attributed to signals from decay electrons or gamma rays is observed in the low-energy region, and these events were not used in the present analysis.

For particles that stopped in the Si detectors, complementary information from telescope TL,
which is described in the following subsection,
indicates that the yields of $^{3}$He, $^{6}$He, and particles with $Z>2$ are negligible compared to $\alpha$ particles.
Accordingly, events exhibiting an energy loss above 13\,MeV in TM and TH, corresponding to the penetration threshold of tritons through all Si detectors, were identified as $\alpha$ particles.

\subsubsection{Telescope TL}

\begin{figure}[h!]
  \centering
  \includegraphics[width=\linewidth]{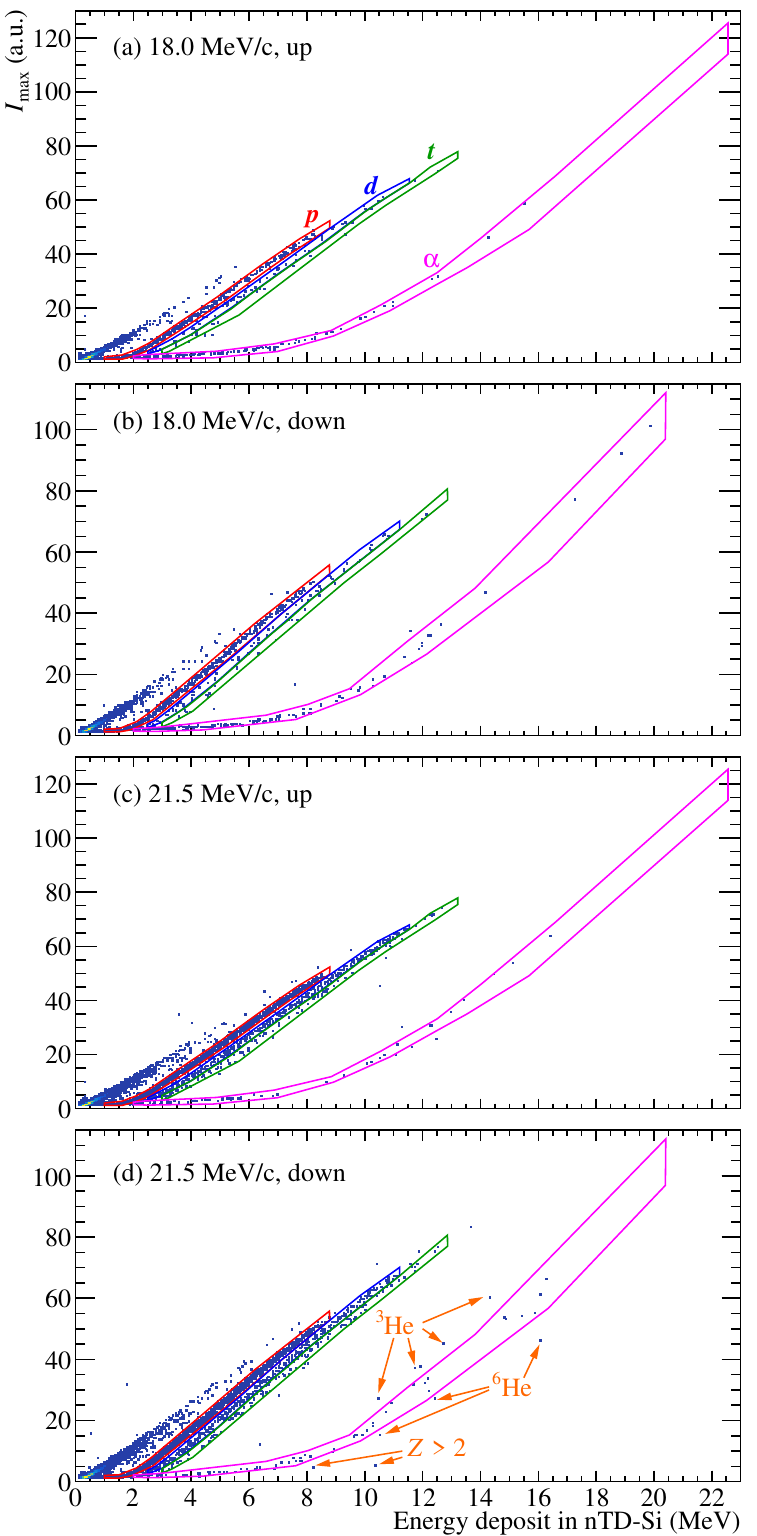}
  \caption{
  Particle identification plots obtained using digital pulse-shape analysis (DPSA) with
telescope TL. Panels (a) and (b) show the results for an incident muon momentum of 18.0\,MeV/$c$ with
the detector placed at the upper (up, $\alpha = +45^\circ$) and lower (down, $\alpha = -45^\circ$) positions relative to the target, respectively. Panels (c) and (d) show the corresponding results for 21.5\,MeV/$c$.
The correlation between the energy deposited in the nTD-Si detector and the maximum
current ($I_{\mathrm{max}}$) is shown. The loci of protons (red), deuterons (blue), tritons (green), and $\alpha$ particles
(magenta) are indicated.
Events corresponding to rare charged-particle emission channels
($^{3}$He, $^{6}$He, and $Z>2$) are indicated by orange arrows.}\label{fig:pid_psa}
\end{figure}

Particle identification in telescope TL was performed using the $E$--$I_{\mathrm{max}}$ correlation. Figure~\ref{fig:pid_psa} shows the correlation between the energy deposited in the nTD-Si detector and the maximum current ($I_{\mathrm{max}}$). Since the statistics of the 27-{\textmu}m target run were limited, graphical cuts were defined using the 200-{\textmu}m target data and subsequently applied to the 27-{\textmu}m data.
An intense component is observed in the low-energy region, which is attributed to signals induced by decay electrons or gamma rays; this component was not used in the present analysis and does not affect the results discussed below.

Below 2\,MeV, particle separation was not feasible with this method; however, given that protons dominate the energy spectrum in this region, all particles with energies below 2\,MeV were treated as protons in the analysis, as shown in Fig.~\ref{fig:pid_psa}. Similarly, in the 2--3\,MeV region, deuterons, tritons, and $\alpha$ particles could not be distinguished. In this case, non-proton particles were treated as $\alpha$ particles for the same reason.

In the 200-{\textmu}m run, a few events consistent with $^{3}$He or $^{6}$He were observed.
However, their statistics were too limited, and the effect of energy loss in the target was too large to extract their energy spectra.
In addition, two events attributable to particles with $Z>2$ were also observed, possibly corresponding to the emission of heavier clusters such as $^{8}$Be. To the best of our knowledge, such observations of $^{3}$He, $^{6}$He, and $Z>2$ cluster emission following \munc have not been reported previously, and this work therefore represents the first experimental evidence of these channels.

\subsection{Event Selection}
\subsubsection{Beam Signal}

Waveform acquisition was synchronized with the 50\,Hz timing signal of the synchrotron; however, proton beam pulses were delivered to TS1 only at 40 pulses per second. As a result, data were also recorded during periods without proton delivery to TS1. To restrict the analysis to events associated with the incoming muon beam, a coincident signal from the Cherenkov detector installed in the muon beamline was required. Only events fulfilling this condition were retained for further analysis.

\subsubsection{Signal Timing}

In the present measurement, the waveform record length was set to 20,000\,ns, with the external trigger signal fed at 2,000\,ns. Throughout this paper, the signal timing is defined relative to this trigger position.

\begin{figure}[tb]
  \centering
  \includegraphics[width=\linewidth]{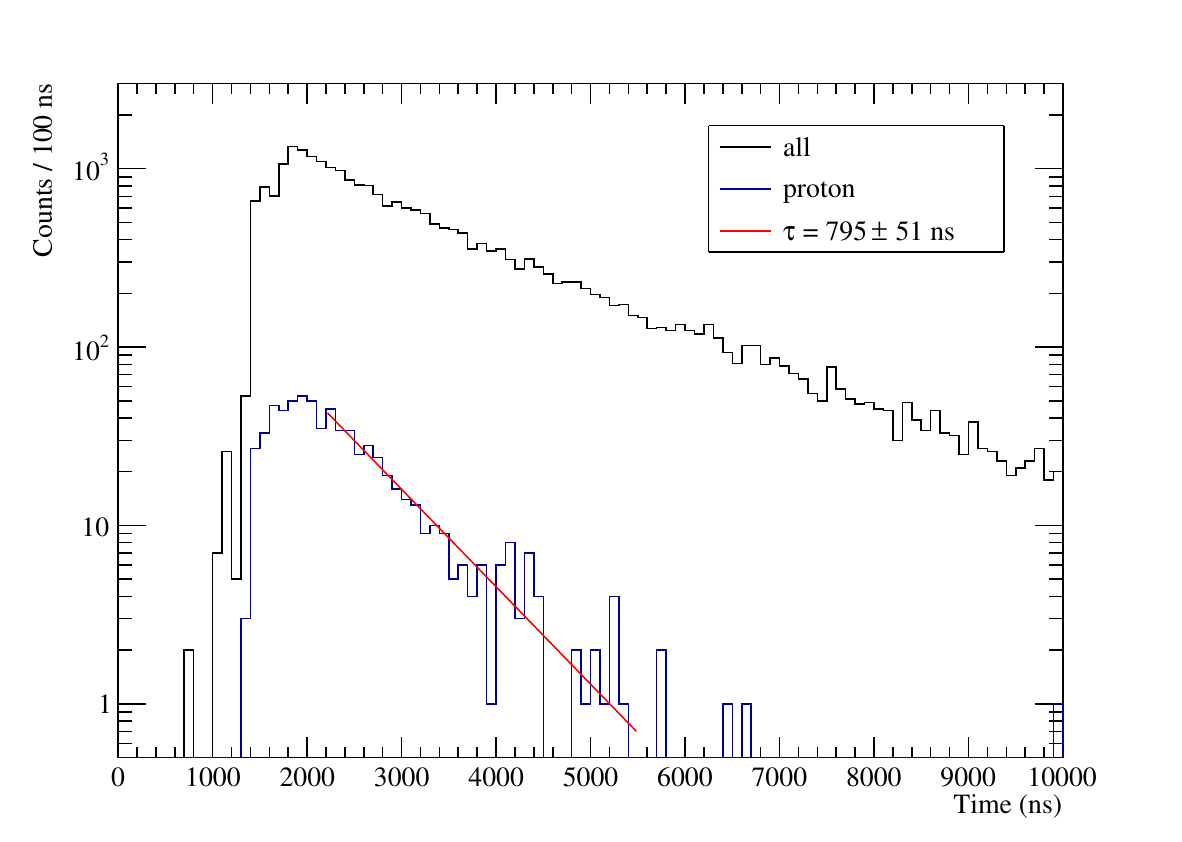}
  \caption{
  Distribution of signal rise times measured with the nTD-Si detectors in the TL telescopes for a muon beam momentum of 21.5\,MeV/$c$.}
  \label{fig:timing_TL}
\end{figure}

Figure~\ref{fig:timing_TL} shows the distribution of signal rise times measured with the nTD-Si detector in the TL telescopes, together with the distribution of events identified as protons. The overall distribution exhibits a decay constant of approximately 2\,{\textmu}s, which is attributed mainly to events in which muons stopped in surrounding polyethylene components, including the target ladder, the upstream beam collimator, and the sheets attached to the chamber walls. In contrast, the proton distribution yields a decay constant of $795 \pm 51$\,ns, consistent with the muonic atom lifetime in
silicon of 756.0(10)\,ns reported in the literature~\cite{Suzuki1987}. This agreement demonstrates that the observed protons originate from \munc in the Si target.

To retain target-originated events while reducing the contribution from muon stops in non-target materials, the present analysis was restricted to events occurring between 1,400\,ns and 6,000\,ns.
The two muon beam bunches arrive at approximately 1,450\,ns and 1,800\,ns, respectively.
The upper limit of 6,000\,ns was chosen so as to cover a time interval corresponding to approximately six lifetimes of the muonic atom in silicon.
The resulting loss of genuine target events is estimated to be at most about 0.3\%, which is negligible compared to the uncertainty in the number of stopped muons.

\begin{figure}[tb]
  \centering
  \includegraphics[width=\linewidth]{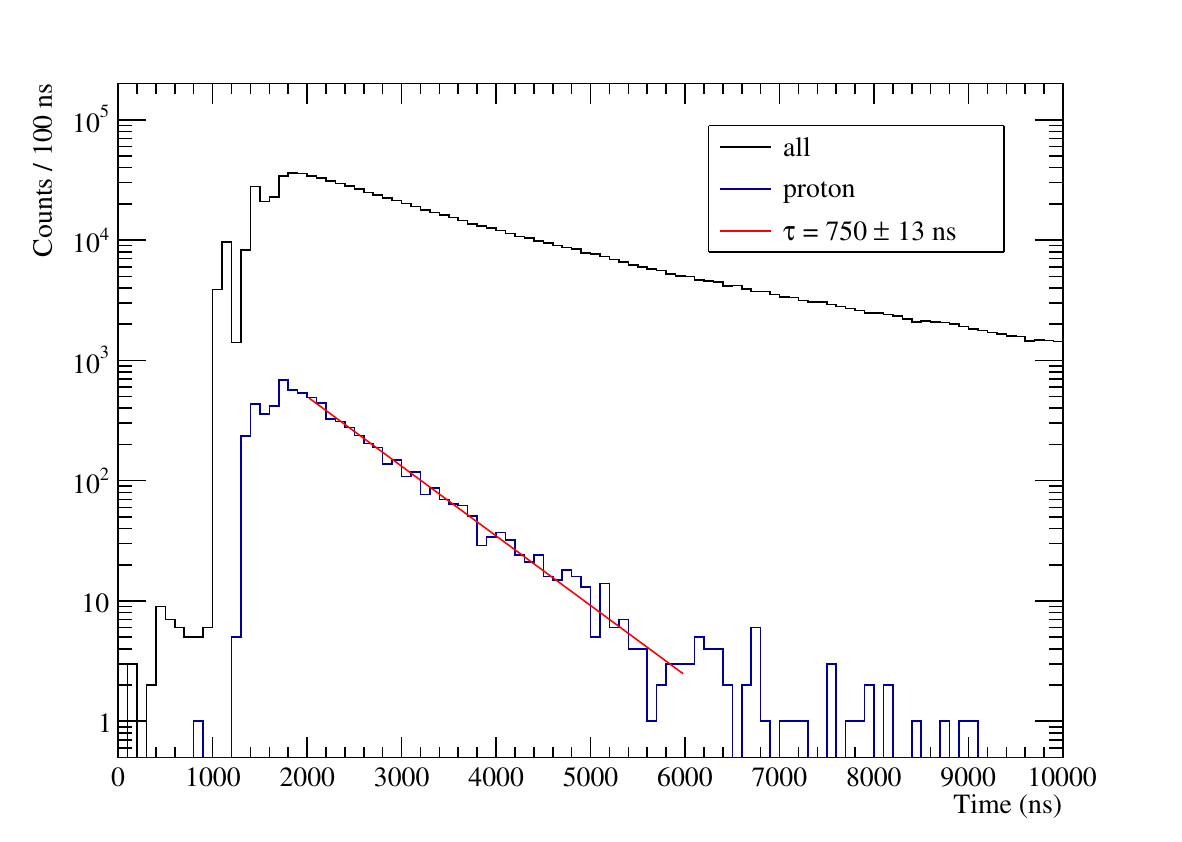}
  \caption{
  Distribution of signal rise times measured with the Si detectors in the TH telescopes for a muon beam momentum of 21.5\,MeV/$c$.}
  \label{fig:timing_TH}
\end{figure}

The timing distribution measured with the Si detectors in the TH telescopes, together with the distribution of events identified as protons, is shown in Fig.~\ref{fig:timing_TH}. The overall component includes contributions from muons stopped in surrounding materials, while the proton component yields a decay constant of $750 \pm 13$\,ns. This value is consistent with the muonic atom lifetime in silicon reported in the literature~\cite{Suzuki1987}, confirming that the protons observed in the TH telescopes also originate from \munc in the Si target. For the analysis, events occurring between 1,200\,ns and 6,000\,ns were selected.

Because of their higher velocity, prompt electrons arrive earlier than muons at the same momentum, as seen in Figs.~\ref{fig:muon_number_200_in}--\ref{fig:muon_number_25um_in}, and the first electron peak lies outside the selected time windows.
In the particle-identification-selected distributions shown in Figs.~\ref{fig:timing_TL} and \ref{fig:timing_TH}, no statistically significant events are observed at the prompt-electron peak timings, confirming that electron-induced events remaining within the time windows are rejected by the particle identification analysis.

\subsubsection{Time Difference between Si and CsI(Tl) signals}

For events in which particles penetrated a Si detector of telescopes TM and TH, an additional selection based on timing information was applied to ensure true coincidences between the Si and CsI(Tl) detectors. The leading time of each waveform was defined as the point where the signal exceeded a threshold set at $7\sigma$ above the baseline fluctuation. The difference between the leading times of the Si and CsI(Tl) signals was then used as the selection variable.

\begin{figure}[tb]
  \centering
  \includegraphics[width=\linewidth]{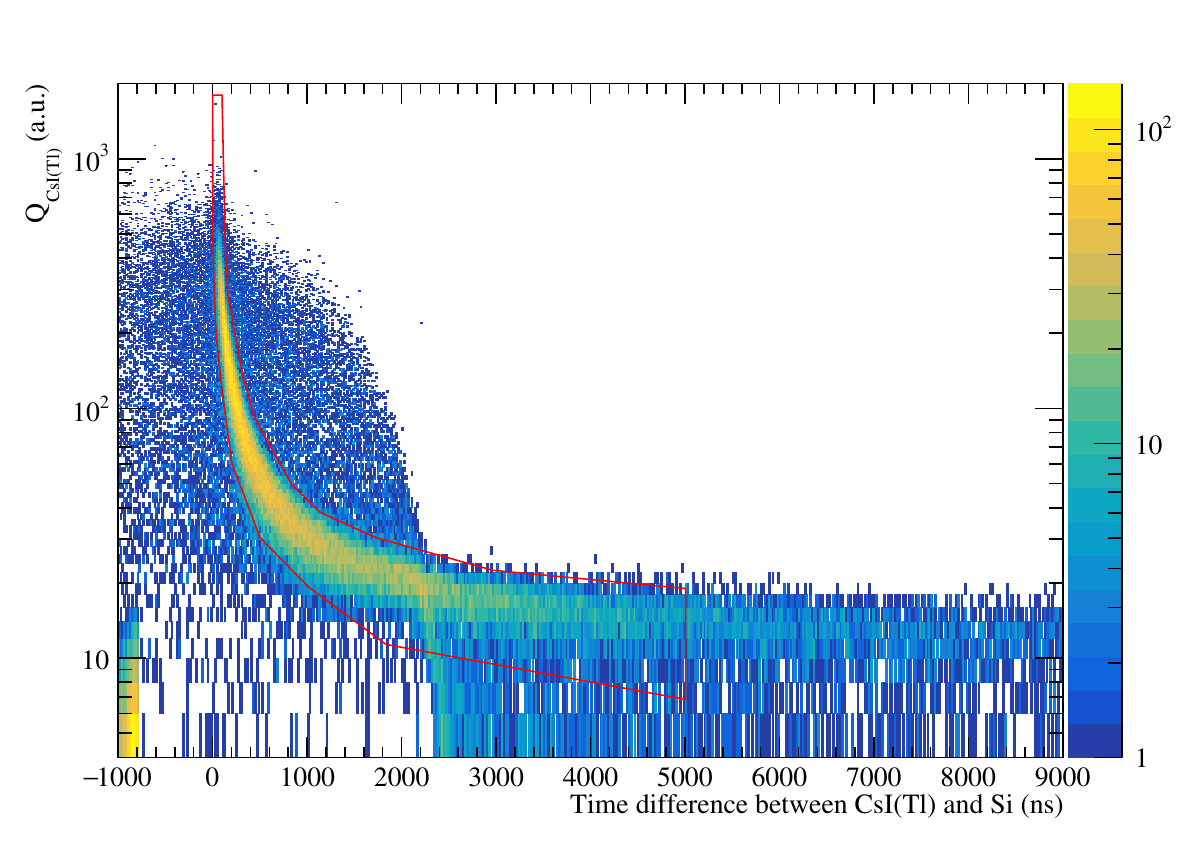}
  \caption{
  Correlation plot of the time difference and the integrated charge of the CsI(Tl) signal. The horizontal axis represents the time difference between the rise times of the CsI(Tl) and Si detector signals, while the vertical axis shows the integrated charge ($Q$) of the CsI(Tl) signal. The graphical selection region applied in the analysis is indicated by the red solid line.}
  \label{fig:tdiff}
\end{figure}

Figure~\ref{fig:tdiff} shows the correlation between this time difference and the integrated charge of the CsI(Tl) signal. A clear dependence is observed: events with lower CsI(Tl) charge exhibit larger time differences, reflecting the effect of time walk. The graphical cut adopted in the analysis is indicated by the red solid line in the figure.

\subsection{Measured energy spectra}

\begin{figure*}[htb]
  \centering
  \includegraphics[width=\textwidth]{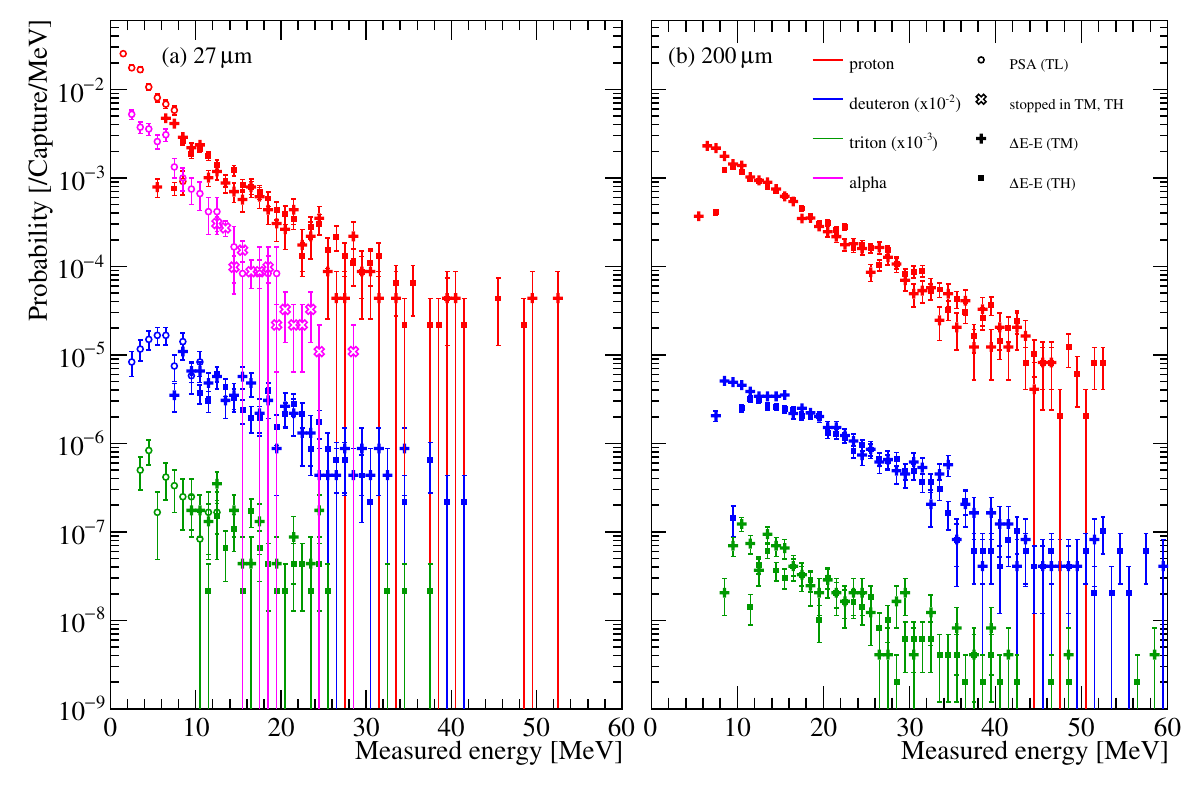}
  \caption{
  Measured energy spectra of charged particles emitted following \munc in Si before unfolding analysis with (a) the 27-{\textmu}m target and (b) the 200-{\textmu}m target.}
  \label{fig:measured_spectra}
\end{figure*}

Based on the particle identification methods and event selection procedures described above, the measured energy spectra for each particle species were obtained, as shown in Fig.~\ref{fig:measured_spectra}.

Target-removed blank runs were performed at 21.5 and 18.0\,MeV/$c$ using the same analysis selections as for the target runs. No statistically significant charged-particle emission was observed.

In the overlapping energy regions covered by different telescopes, the spectra are generally consistent with each other. However, in the overlapping region between TL and TM, the spectrum measured with TM is slightly suppressed compared to that with TL. This suppression is attributed to the relatively thick insensitive layer (20\,{\textmu}m according to the datasheet) of the Hamamatsu Si detectors used in TM and TH. The influence of such insensitive layers, as well as energy loss in the target, is corrected for in the unfolding analysis described in the following section.

\subsection{Unfolding}

Energy losses in the target material and in the inactive layers of the detectors distort the measured energy spectra of emitted particles. In order to extract the initial energy spectra at the emission point, an unfolding analysis was performed.

The detector response functions were constructed using Monte Carlo simulations
based on Geant4~\cite{Agostinelli2003, Allison2006, Allison2016}, which take into
account the full experimental geometry as well as energy loss and straggling
effects in the target and inactive layers of the detectors.
In these simulations, the initial energies of emitted particles were sampled
from a uniform energy distribution covering the full kinematically relevant
range for each particle species.

The use of a uniform energy distribution ensures that the response matrices
represent purely the detector response and energy-loss effects, without
introducing any model-dependent assumptions on the underlying physical energy
spectra.
The resulting response matrices therefore describe the probability of observing
a given measured energy for a particle emitted with a certain initial energy,
independent of the assumed emission model.

In the unfolding procedure, the detector response functions were weighted
according to the three-dimensional muon stopping-position distribution
inside the target.
The stopping-position distribution was estimated using dedicated simulations
using G4beamline~\cite{Roberts2007},
which model the transport and energy loss of the
incident muon beam in the experimental setup.
The initial beam profile and momentum distribution in the simulation
were tuned to reproduce the measured beam characteristics reported in
Ref.~\cite{Matsuzaki2001}.
In addition, the central momentum of the muon beam was adjusted such that
the simulated stopping ratio reproduces the experimentally measured value
in the present experiment.
The weighted response functions thus account for the realistic spatial
distribution of muon stopping positions within the target.

The unfolding was carried out using the \texttt{RooUnfold} unfolding framework~\cite{Brenner2020},
employing the iterative Bayesian unfolding method~\cite{DAgostini1995} implemented in the
\texttt{RooUnfoldBayes} class.
The number of iterations was set to the default value of four.

Measured energy spectra obtained with different target thicknesses
(27\,{\textmu}m and 200\,{\textmu}m) and with multiple detector telescopes
were included simultaneously in a single unfolding procedure.
Each data set was described by its corresponding response function,
while a common initial energy spectrum was assumed.

Statistical uncertainties of the unfolded spectra were evaluated using a toy Monte Carlo approach.
A set of 10,000 pseudo-data spectra was generated by fluctuating the measured spectra according to Poisson statistics in each energy bin. The unfolding procedure was applied to each pseudo-data set,
and the resulting distribution of unfolded spectra was used to estimate the statistical uncertainties.
Systematic uncertainties associated with the unfolding procedure are discussed in the following subsection.

\subsection{Systematic uncertainties}

\subsubsection{Number of stopped muons}

The uncertainty in the number of stopped muons constitutes a source of systematic uncertainty in the absolute normalization of the unfolded spectra. In the present analysis, the unfolding was performed simultaneously using data sets obtained with different target thicknesses and detector configurations, which is expected to reduce the sensitivity to the uncertainty in the stopped-muon determination.

Nevertheless, a conservative estimate was adopted.
For the 18.0\,MeV/$c$ run with the thin target, propagation of the statistical uncertainties in the fitted target-out and target-in muon yields resulted in a relative uncertainty of approximately 3\% in the number of stopped muons.
For the 21.5\,MeV/$c$ run, a separate relative uncertainty of 3\% was assigned to the pileup correction.
These normalization uncertainties were propagated to the final results.

\subsubsection{Detector placement}

Uncertainties in the relative positioning of the detectors and the target constitute a source of systematic uncertainty in the measured yields. Taking into account possible mechanical tolerances in the experimental setup, relative misalignments of the order of 1\,mm in any direction cannot be excluded.

The effect of such misalignments on the overall yields was evaluated,
and the resulting systematic uncertainty was estimated to be 2.4\%.
This uncertainty was applied as an overall normalization uncertainty in the analysis.

\subsubsection{Unfolding procedure}

Systematic uncertainties associated with the unfolding procedure were evaluated by varying key inputs to the response matrix and by assessing the stability of the unfolding with respect to the number of iterations.

First, the sensitivity to the muon stopping distribution inside the target was investigated.
Starting from the response matrices constructed using the tuned muon beam conditions
that reproduce the experimentally measured stopping ratio,
the unfolding was repeated using response matrices obtained from simulations
in which the incident muon momentum was varied such that the mean stopping position
of muons inside the target was shifted along the beam axis by $\pm$10\% of the target thickness.
The resulting changes in the unfolded spectra were found to be negligible compared to
the corresponding statistical uncertainties and were therefore not included as an
additional systematic contribution.

The thicknesses of the Si targets were measured using a micrometer. 
The observed variation was smaller than the instrumental resolution of 1\,{\textmu}m. 
This level of uncertainty is negligible for the present analysis and is effectively absorbed into the uncertainty associated with the muon stopping distribution inside the target.

Second, the dependence on the number of iterations in the iterative Bayesian unfolding was examined.
The unfolded spectra were found to be well converged, with bin-by-bin differences between
the results obtained with 4 and 20 iterations remaining within 3\% for most energy bins.
These differences were taken as a conservative estimate of the systematic uncertainty
associated with the unfolding procedure.

\subsubsection{Energy calibration}

Systematic uncertainties associated with the energy calibration were evaluated by
assessing the impact of plausible variations in the energy scale on the unfolded
spectra.

For telescope TL, the energy calibration uncertainty was conservatively estimated
by introducing an offset of $\pm 100$\,keV combined with a scaling uncertainty of
$\pm 1\%$.
These variations account for possible residual uncertainties in the offset and gain
determination of the nTD-Si detector.

For telescopes TM and TH, the energy calibration uncertainty
was conservatively estimated by applying a uniform scaling variation of $\pm 3\%$ to
the reconstructed energies.
This choice reflects the dominant uncertainty associated with the energy response of
the CsI(Tl) scintillator and possible nonlinearity effects.

The analysis was repeated after applying these energy-scale variations, and the
resulting changes in the spectra and extracted observables were taken as systematic
uncertainties associated with the energy calibration.

\section{Results}

\begin{figure*}[htb]
  \centering
  \includegraphics[width=\textwidth]{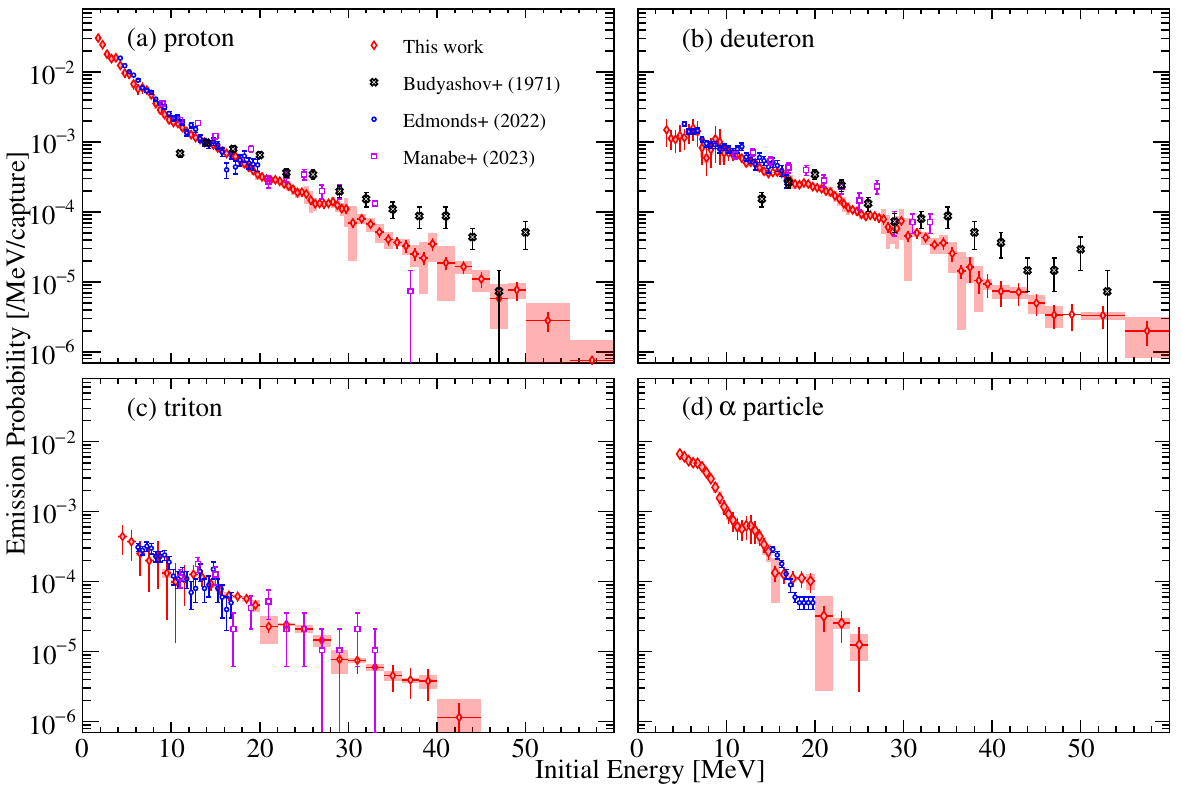}
  \caption{
  Initial energy spectra obtained in this work for (a) protons, (b) deuterons, (c) tritons, and (d) $\alpha$ particles, compared with preceding experimental data~\cite{Budyashov1971, Edmonds2022, Manabe2023}.
  The shaded areas represent systematic uncertainties associated with the unfolding
iterations and the energy calibration, while overall normalization uncertainties
from the muon number determination and detector alignment are not included.
}  \label{fig:spectra_exp}
\end{figure*}

Figure~\ref{fig:spectra_exp} shows the initial energy spectra of emitted protons, deuterons, tritons, and $\alpha$ particles obtained in the present experiment.
For all four particle species, energy spectra were successfully extracted over a wide energy range, extending from the low-energy region up to several tens of MeV, demonstrating the broad kinematic coverage achieved in this measurement.

The error bars shown in Fig.~\ref{fig:spectra_exp} represent the uncertainties propagated through the unfolding procedure. While these uncertainties are dominated by statistical errors of the measured spectra, the unfolding process introduces correlations between neighboring energy bins, and therefore the uncertainties should not be regarded as strictly independent point by point.

As a global feature common to all particle species, the emission probabilities decrease approximately exponentially with increasing emission energy. This overall behavior is qualitatively consistent with the generic characteristics expected for particle emission following \munc.

Coulomb-barrier effects are expected to cause a drop-off in the emission probability at low energies, with a preceding flattening of the spectral slope. In the present measurement, a slight flattening is seen only in the $\alpha$-particle spectrum, but the limited low-energy coverage prevents any quantitative assessment.

At the same time, noticeable differences are observed in the spectral slopes among protons, deuterons, tritons, and $\alpha$ particles. These differences indicate that the energy distributions depend on the type of emitted particle, suggesting that the underlying emission processes are not universal but instead reflect intrinsic differences associated with the emission mechanism and particle properties.

\section{Discussion}

\subsection{Energy spectra}

\subsubsection{Comparison with previous experiments}\label{sec:comp_exp}

Figure~\ref{fig:spectra_exp} compares the initial energy spectra obtained in the present work with all
previously published experimental data~\cite{Budyashov1971, Edmonds2022, Manabe2023}
on charged-particle emission following \munc on Si.

Overall, the present measurement provides charged-particle energy spectra over a
significantly broader energy range than any of the previous experiments.
In particular, continuous coverage is achieved from the low-energy region up to
several tens of MeV for protons, deuterons, and tritons, demonstrating the
comprehensive nature of the present data set.

In comparison with the data reported by Budyashov \textit{et al.}~\cite{Budyashov1971}, the present spectra
exhibit a systematic deviation in the high-energy region, where the Budyashov data
show anomalously large emission probabilities.
This discrepancy may be related to effects such as energy loss or distortion in the
target material; however, detailed information on the data analysis procedures is not
provided in the original publication, and the origin of the discrepancy therefore
remains unclear.

The spectra reported by Manabe \textit{et al.}~\cite{Manabe2023}
exhibit good overall agreement with the present results in terms of spectral
shape over the energy range where the data sets overlap.
The absolute yields reported in Ref.~\cite{Manabe2023} are, however, systematically
slightly higher than those obtained in the present work.
This difference is most likely attributable to the normalization procedure
adopted in Ref.~\cite{Manabe2023}, where the spectra were normalized to the
17\,MeV proton data~\cite{Budyashov1971} reported by Budyashov \textit{et al.}
Despite this difference in normalization, the overall spectral behavior is
consistent between the two measurements.

Among the existing data, the spectra reported by Edmonds \textit{et al.}~\cite{Edmonds2022}
represent the most comprehensive prior measurement in terms of particle species
and energy coverage.
The present results are in good agreement with these data for protons, deuterons,
and tritons over most of the overlapping energy range.
A small systematic difference is observed in the low-energy region of the proton
spectrum, where the yields reported by Edmonds \textit{et al.} are slightly higher
than those obtained in the present work.
This difference may be related to a possible energy offset in the calibration,
for example arising from differences in the treatment of insensitive layers in
the detector telescopes.
In Ref.~\cite{Edmonds2022}, the treatment of the energy calibration uncertainty
focused on the overall energy scale, while a potential offset component was not
discussed explicitly.

A key advancement of the present work is the extension of the $\alpha$-particle
spectrum toward significantly lower energies than previously achieved, providing
new experimental constraints on the low-energy evaporation component.

\subsubsection{Comparison with theoretical model calculations}

To interpret the measured charged-particle energy spectra,
the present results are compared with representative theoretical model calculations.
In the following, we focus on two approaches that have been applied to describe particle
emission following \munc: the microscopic and evaporation model (MEM) by Minato
\textit{et al.}~\cite{Minato2023},
and the particle transport code PHITS~\cite{Sato2024}.

Natural silicon consists predominantly of $^{28}$Si (92.2\%), with smaller
contributions from $^{29}$Si (4.7\%) and $^{30}$Si (3.1\%)~\cite{Berglund2011}.
Isotope-resolved activation measurements on Si~\cite{Mizuno2025b} indicate
that, although individual emission channels vary among $^{28}$Si, $^{29}$Si,
and $^{30}$Si, the reported summed branching ratios of the $xn1p$ and $xn2p$
channels are of the same order of magnitude.
The comparison of the measured
inclusive spectra with the $^{28}$Si calculations discussed below should
therefore be regarded as approximate and is intended to assess the overall
spectral behavior rather than to provide an isotope-resolved quantitative
test.

The MEM calculation~\cite{Minato2023} described particle emission following \munc as a three-step process
that explicitly incorporates nuclear many-body effects.
In the first step, the weak one-body operators associated with muon capture~\cite{OConnell1972}
excite the nucleus and generate particle-hole configurations in the daughter nucleus.
In Ref.~\cite{Minato2023}, this initial excitation was treated using the second Tamm-Dancoff approximation (STDA)~\cite{Minato2016}, which incorporates two-particle-two-hole (2p-2h) configurations through the coupling of one-particle-one-hole (1p-1h) and 2p-2h states.
Although Ref.~\cite{Minato2023} employed both SkO'~\cite{Reinhard1999} and SGII~\cite{VanGiai1981a},
the present calculations were performed using SkO' only.
The dependence of the charged-particle energy spectra on the choice of these
effective interactions was found to be weak.
This extension allows us to examine the sensitivity of the strength distributions to the treatment of nuclear correlations over a wide excitation-energy range.
In addition to the one-body weak interaction, the effect of meson-exchange currents (MEC)~\cite{Lifshitz1988}
is incorporated phenomenologically.
MEC-induced transitions are treated as an additional contribution to the initial excitation,
enhancing the population of high-energy configurations
that are not sufficiently described by one-body operators alone.
This contribution is particularly important for reproducing the high-energy components of proton emission spectra~\cite{Minato2023}.

In the second step, the subsequent de-excitation is modeled within the framework of
the two-component exciton model~\cite{Kalbach1986, Koning2004}.
This model describes the evolution from the initially populated particle-hole configurations
toward statistical equilibrium,
allowing for preequilibrium emission of nucleons and light clusters during the equilibration process.
Once a compound-nuclear configuration is formed, the remaining excitation energy is dissipated through statistical evaporation, treated using a Hauser-Feshbach formalism~\cite{Hauser1952, Iwamoto2016}.

In contrast, PHITS~\cite{Sato2024} is a general-purpose Monte Carlo particle transport code designed to simulate the production, interaction, and transport of a wide variety of particles over a broad energy range.
As part of its comprehensive physics framework, PHITS includes models for muon-induced reactions~\cite{Abe2017}, which enable simulations of processes following muon capture, such as preequilibrium emission and statistical evaporation.
In the muon capture module of PHITS, the excitation energy of the nucleus after \munc is sampled from the phenomenological excitation function proposed by Singer~\cite{Singer1962}, rather than being derived from an explicit microscopic description of nuclear structure.
Particle emission following \munc is then modeled using a combination of the JAERI quantum molecular dynamics (JQMD) model~\cite{Niita1995, Ogawa2015} for the preequilibrium stage and the General Evaporation Model (GEM)~\cite{Furihata2000, Furihata2001} for subsequent statistical decay, with parameters constrained by a broad range of experimental data.
In addition, to enhance the emission of light composite particles, PHITS incorporates a surface coalescence model (SCM)~\cite{Watanabe2007}, which accounts for the formation of composite particles through the coalescence of emitted nucleons near the nuclear surface.
PHITS provides an event-by-event description of particle production and transport,
allowing direct comparison with experimental observables such as energy spectra and
particle yields.

In the present study, calculations were performed not only with the standard PHITS models
but also with an extended treatment that phenomenologically incorporates
the MEC effect~\cite{Lifshitz1988}, in close analogy to the MEM-based approach.
This extension enhances the population of high-excitation-energy configurations
following \munc and allows us to assess the impact of MEC
on emitted-particle spectra within the PHITS framework.

Both models have been shown to reproduce global features of charged-particle emission
following \munc, such as overall yields and general spectral trends~\cite{Minato2023,Manabe2023}.
However, they rely on different assumptions regarding the treatment of preequilibrium
dynamics and the formation of composite particles.
A comparison between the present data and these model calculations therefore provides
valuable insight into the mechanisms governing charged-particle emission, as well as
into the limitations of existing theoretical descriptions.

\begin{figure*}[htb]
  \centering
  \includegraphics[width=\textwidth]{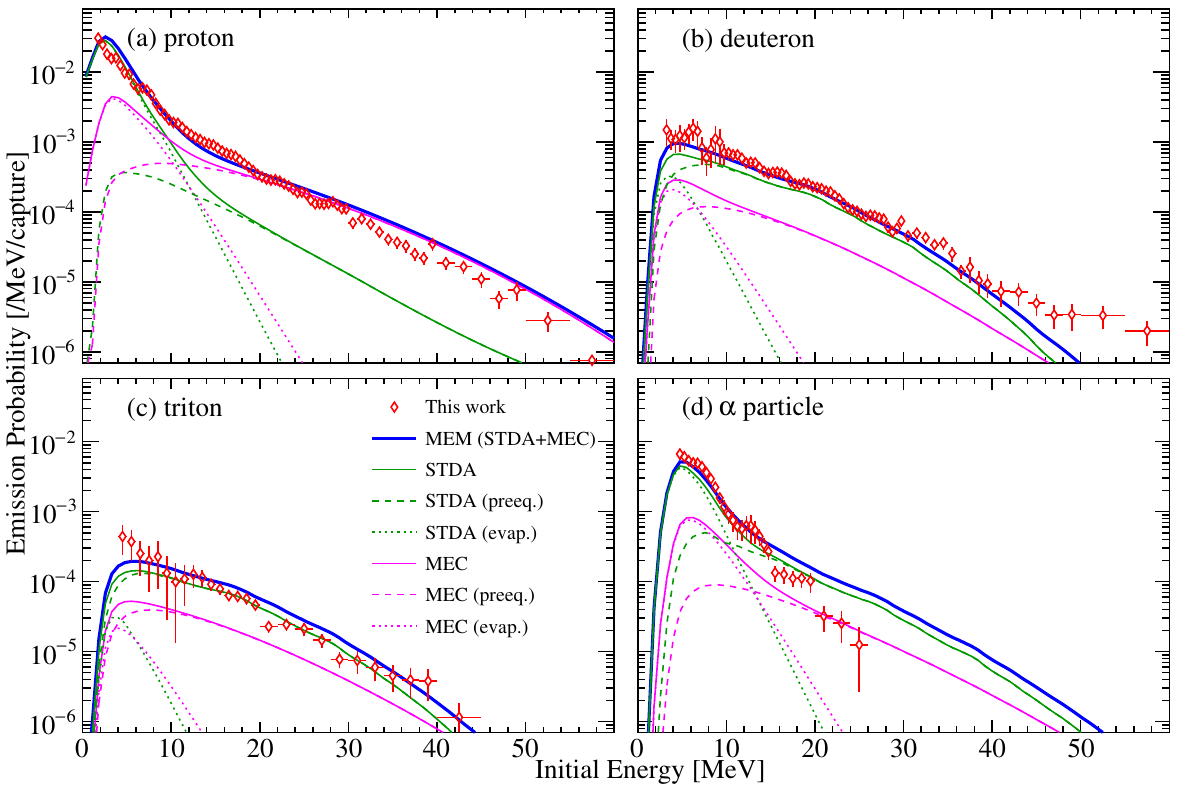}
  \caption{
  Initial energy spectra obtained in this work for (a) protons, (b) deuterons, (c) tritons, and (d) $\alpha$ particles. The experimental data are compared with MEM calculations~\cite{Minato2023} performed using the SkO' effective interaction. In the MEM calculations, the total spectra are decomposed into contributions from the initial excitation described by the STDA and from the MEC component. Each of these contributions is further separated into preequilibrium and evaporation components.
}  \label{fig:spectra_mem}
\end{figure*}

Figure~\ref{fig:spectra_mem} compares the experimentally obtained initial energy spectra
with the MEM calculations.
The calculated spectra are further decomposed into contributions from STDA and MEC,
each of which is separated into preequilibrium and evaporation components.
Overall, the MEM calculation reproduces the global features of the measured spectra
reasonably well for all particle species.
At the same time, a more detailed comparison reveals systematic deviations
that depend on the particle type and the energy region.

For protons, the inclusion of the MEC component is essential to reproduce
the high-energy part of the preequilibrium emission.
A more detailed comparison reveals systematic deviations that depend on the energy region.
The MEM calculation tends to overestimate the yield in the low-energy region,
while underestimating the emission probability in the intermediate energy range
of approximately 10--20\,MeV.
In the high-energy region, the calculated spectrum decreases more slowly than
the experimental one, resulting in a slight overestimation of the yield.
These features suggest that, while the present two-component description consisting of
preequilibrium and compound contributions captures the gross behavior,
a more refined treatment of the transition between different emission mechanisms may be required.

For deuterons and tritons, the overall agreement between the experimental data and
the MEM calculation is reasonably good over a wide energy range,
indicating that the description of composite-particle emission within the
Iwamoto-Harada model~\cite{Iwamoto1982,Sato1983}
captures the essential features of deuteron and triton emission following \munc.
In contrast to the proton case, the contribution from MEC
is found to be less pronounced for these composite particles.
For deuterons, a slight underestimation is observed in the high-energy region.
For tritons, on the other hand, the calculated yield tends to be systematically larger
than the experimental data, although the overall spectral shape is well reproduced.
These discrepancies suggest that further refinement of the preequilibrium component
would be required for a more quantitative description.

In contrast, the comparison for $\alpha$ particles reveals a more pronounced
energy-dependent discrepancy.
The low-energy part of the $\alpha$ spectrum is well reproduced by the MEM
calculation, indicating that the evaporation component, which dominates this energy
region, provides an adequate description of low-energy $\alpha$ emission.
However, at energies above 10\,MeV, the MEM calculation significantly
overestimates the measured $\alpha$ yield.
This discrepancy suggests limitations in the current treatment of composite-particle
formation within the Iwamoto-Harada model~\cite{Iwamoto1982, Sato1983} for $\alpha$ emission, particularly in the
high-energy region where non-evaporative processes are expected to contribute.
The present results therefore point to the need for further refinement of the model,
in particular with respect to the description of composite-particle formation and
preequilibrium $\alpha$ emission.

\begin{figure*}[htb]
  \centering
  \includegraphics[width=\textwidth]{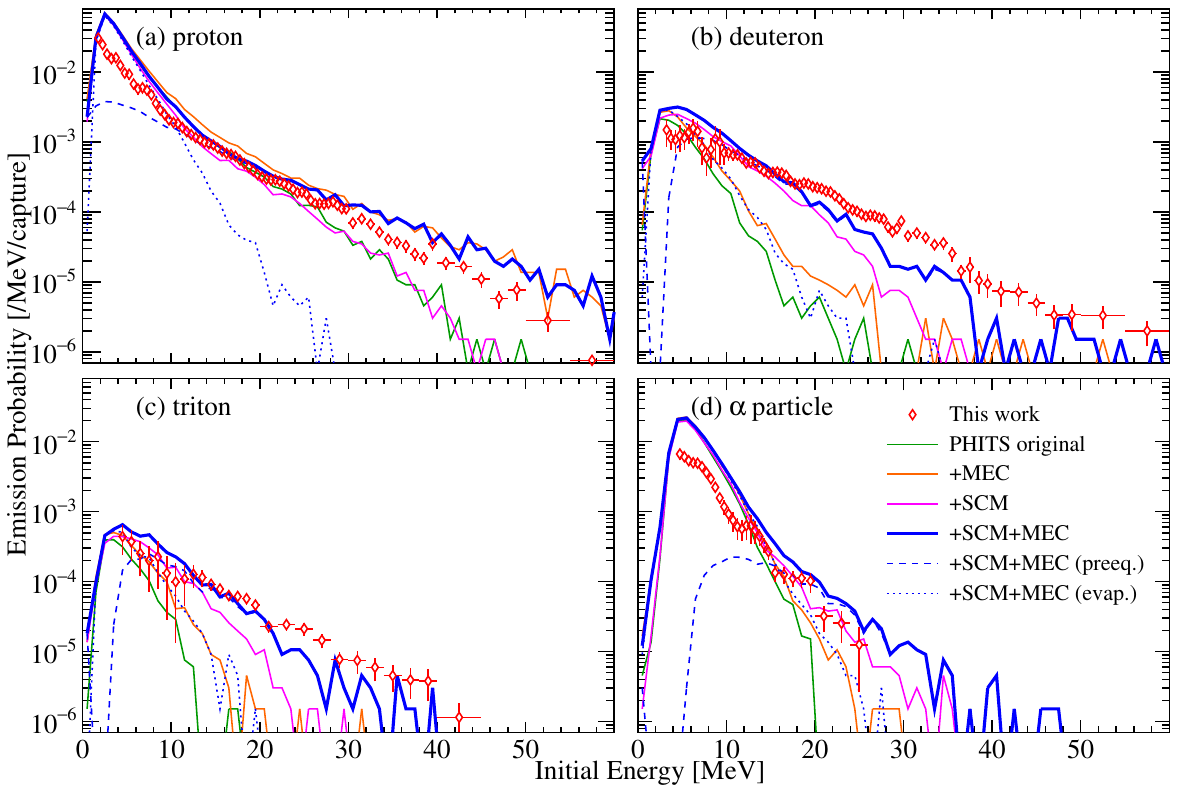}
  \caption{
  Initial energy spectra obtained in this work for (a) protons, (b) deuterons, (c) tritons, and (d) $\alpha$ particles, compared with PHITS calculations using the original model (PHITS), with SCM (+SCM) and with the MEC extension (PHITS+MEC), and with both extensions (+SCM+MEC).
  For the PHITS+SCM+MEC calculations, the total spectra are shown together with
the separate contributions from the preequilibrium (preeq.) and evaporation
(evap.) processes.
}  \label{fig:spectra_phits}
\end{figure*}

Figure~\ref{fig:spectra_phits} shows a comparison of the experimental initial energy spectra with PHITS calculations.
Results obtained with the original PHITS model and with extensions
including the SCM and MEC effects are shown.
For the PHITS calculation including both the SCM and MEC effects (PHITS+SCM+MEC),
the breakdown into preequilibrium and evaporation components is also presented.
Overall, the inclusion of both the SCM and MEC effects improves aspects of
the measured spectral shapes for all charged-particle species, although
noticeable particle-dependent discrepancies in the absolute yields remain.

For protons, the overall spectral trend is reproduced, although less closely than in the MEM calculation.
The PHITS+SCM+MEC calculation tends to overestimate the yield over most of the measured energy range.
However, in the intermediate-energy region of 10--20\,MeV, where the MEM calculation underestimates the yield, the absolute values predicted by PHITS are closer to the experimental data.
The original PHITS model alone shows a significantly poorer description of the high-energy component, indicating that the inclusion of the MEC effect is important for achieving a reasonable reproduction of the proton spectrum.
The overestimation of the proton yield found here is qualitatively consistent with the overestimation of proton-emission probabilities by PHITS reported for Pd isotopes from residual-nucleus production branching ratios~\cite{Niikura2024}.

For deuterons and tritons, the low-energy peak is reproduced to some extent, whereas the high-energy components are significantly underestimated.
Although the inclusion of SCM and the MEC effect substantially improves the overall agreement, particularly below 20\,MeV, noticeable discrepancies persist in the high-energy region.
The agreement achieved by the PHITS calculations is therefore not yet fully comparable to that obtained with the MEM calculations.

For $\alpha$ particles, the agreement between the experimental data and the calculations shows a pronounced energy dependence.
The absolute yield is overestimated over most of the measured energy range.
Nevertheless, the PHITS+SCM+MEC calculation reproduces the overall spectral slope well, including the high-energy behavior above approximately 15\,MeV, more successfully than the MEM calculation.
These results indicate that further refinement in the modeling of $\alpha$-particle emission following \munc is required for a quantitative description of the absolute yield.

Overall, the level of agreement depends on both the particle species and the modeling approach.
MEM reproduces the spectral shapes of protons, deuterons, and tritons more closely and describes the low-energy $\alpha$-particle spectrum well, but overestimates the $\alpha$-particle yield at higher energies.
PHITS tends to overestimate the evaporation component for all four particle species.
Its description of the high-energy behavior is particle dependent: it significantly underestimates the deuteron and triton yields, whereas it reproduces the slope of the $\alpha$-particle spectrum well despite overestimating its absolute yield.

\subsection{Charged-particle yields}

\begin{table*}[tb]
\centering
\caption{Comparison of integrated yields of emitted charged particles (in $10^{-3}$ per capture)
following \munc on silicon.
The present results are compared with previous direct measurements, theoretical calculations,
and residual-nucleus branching ratios deduced from activation and prompt $\gamma$-ray
measurements following \munc in $^{28}$Si, except where otherwise noted.
The theoretical values from Ref.~\cite{Lifshitz1980} correspond to either inclusive or exclusive
emission, as specified.
For the direct measurements, the yields are integrated over the energy ranges indicated in brackets in MeV.
For the residual-nucleus data, the corresponding nuclei and reaction channels are listed.}
\label{tab:comp_yields}
\begin{tabular}{cccccc|ccc}
\hline\hline
& \multicolumn{3}{c}{Direct measurements} & \multicolumn{2}{c|}{Theory} & \multicolumn{3}{c}{Branching-ratio data}\\
emitted & This work & Edmonds & Sobottka-Wills & Lifshitz-Singer & MEM(SkO') & residual & decay & Mizuno \\
particle & & \cite{Edmonds2022} & \cite{Sobottka1968} & \cite{Lifshitz1980} & \cite{Minato2023} & nucleus & channel & \cite{Mizuno2025b}\\\hline
$p$              & 99.6(27)(38) & --- & & 32 & 144 & $^{27}$Mg & $0n1p$ & 28.7(17) \\
                 & \multicolumn{2}{c}{$[1.5 \le E \le 60]$} & & (exclusive) & &          &        & \\
                 & 44.9(15)(17) & 52.5(6)(18) & & & &          &        & \\
                 & \multicolumn{2}{c}{$[4 \le E \le 20]$} & & & &           &        & \\
$d$              & 13.4(8)(5) & --- & & 21 & 11.2 & $^{26}$Mg & $1n1p$ & $>84(8)$\footnotemark[1]\\
                 & \multicolumn{2}{c}{$[3 \le E \le 60]$} & & (inclusive)        &        & \\ %
                 & 8.6(6)(3) & 9.80(22)(41) & & 8.2 & &          &        & \\ %
                 & \multicolumn{2}{c}{$[5 \le E \le 17]$} & & (exclusive) & &          &        & \\ %
$t$              & 2.72(33)(10) & --- & & & 2.68 &$^{25}$Mg & $2n1p$ & $>15(1)$\footnotemark[1]\\
                 & \multicolumn{2}{c}{$[4 \le E \le 60]$} & & & &          &        & \\ %
                 & 1.49(24)(5) & 1.70(8)(10) & & & &          &        & \\ %
                 & \multicolumn{2}{c}{$[6 \le E \le 17]$} & & & &          &        & \\\hline
$Z=1$ sum      & $>115.7(28)(44)$ & & & & 158 & $^{A}$Mg  & $xn1p$ & $>128(8)$ \\\hline
$^{3}$He         & $>0.04$ & --- & & & 0.514 & $^{25}$Na & $1n2p$ & 2.4(17)\\
$\alpha$         & 25.5(18)(10) & --- & & 34 & 28.6 & $^{24}$Na & $2n2p$ & 17.1(13)\\
                 & \multicolumn{2}{c}{$[4.5 \le E \le 26]$}& & (inclusive) & &          &        & \\ %
                 & 0.59(8)(2) & 0.57(3)(10) & & 17 & &          &        & \\
                 & \multicolumn{2}{c}{$[15 \le E \le 20]$} & & (exclusive) & &          &        & \\ %
($\alpha+n$)     & --- & --- & & & &$^{23}$Na & $3n2p$ & $>5(5)$\footnotemark[1]\\
$^{6}$He         & $>0.03$ & --- & & & &$^{22}$Na & $4n2p$ & 1.5(3)\footnotemark[2]\\\hline
$Z=2$ sum      & $>25.6(18)(10)$ & & & & 29.1 & $^{A}$Na  & $xn2p$ & $>26(5)$ \\\hline
$Z>2$            & $>0.013$ & & & & & $^{20}$F & $4n4p$  & 1.16(21) \\\hline
charged-particle & $>141(3)(5)$& 171(30) & 150(20) & 144 & & & $xnyp$ &  $>155(9)$\\
total            & &\multicolumn{2}{c}{$[1.4 \le E \le 26]$} &   & &       & $(y>0)$& \\ %
\hline\hline
\end{tabular}
\footnotetext[1]{These values were obtained for natural silicon and converted to branching ratios for $^{28}$Si using the assumptions described in Ref.~\cite{Measday2007}.}
\footnotetext[2]{This value was obtained for natural silicon in Ref.~\cite{Heisinger2002}; no correction to $^{28}$Si was applied.}
\end{table*}

\subsubsection{Comparison with previous experiments}

A comparison of the integrated yields of emitted charged particles obtained in the
present work with previous measurements by Edmonds~\cite{Edmonds2022}
and Sobottka and Wills~\cite{Sobottka1968} is summarized in Table~\ref{tab:comp_yields}.
Since the present measurement covers a wider energy range than the previous studies,
the integrated yields shown in Table~\ref{tab:comp_yields} are evaluated over the same
energy intervals as those adopted in the corresponding reference data, in order to
enable a direct comparison.

For $Z=1$ charged particles, the integrated yields obtained in the present work are
systematically smaller than those reported by Edmonds~\cite{Edmonds2022}.
As discussed in Sec.~\ref{sec:comp_exp}, this difference may be attributed to systematic effects in the low-energy region of the spectra, where the treatment of energy calibration and energy-loss corrections is particularly sensitive.

For $\alpha$ particles, the integrated yields obtained in the present work are
consistent with those reported by Edmonds~\cite{Edmonds2022} within
the quoted uncertainties.
This agreement is likely due to the fact that the $\alpha$-particle energy range
covered in Ref.~\cite{Edmonds2022} is sufficiently high that the results are largely
insensitive to systematic effects associated with energy calibration in the
low-energy region.

\subsubsection{Comparison with theoretical model calculations}

Table~\ref{tab:comp_yields} also compares the results with theoretical model calculations.
The theoretical values considered here are taken from the work of
Lifshitz and Singer~\cite{Lifshitz1980}, which provides both inclusive and exclusive
predictions for charged-particle emission following \munc in $^{28}$Si,
as well as from the MEM calculations~\cite{Minato2023} discussed in the previous section.

First, the present results are compared with the calculations by Lifshitz and
Singer~\cite{Lifshitz1980}.
In that work, theoretical predictions are provided only for protons, deuterons,
$\alpha$ particles, and the total charged-particle yield.

For proton emission, the exclusive theoretical yield is smaller than the
experimentally determined integrated yield.
However, this difference does not necessarily indicate an inconsistency, since the
exclusive calculation corresponds to proton emission without accompanying neutron
emission, whereas the experimental yield includes contributions from channels in
which protons are emitted in coincidence with one or more neutrons.
For deuteron and $\alpha$ emission, the inclusive theoretical yields appear to be somewhat
larger than the corresponding experimental values.

For the total charged-particle yield, the theoretical value is comparable to the
experimentally determined lower limit.
Given the limited coverage of the charged-particle spectrum in the experiment,
particularly at low energies, the calculation may underestimate the total
charged-particle yield to some extent.
Nevertheless, considering the age of the calculation and the simplified treatment of
particle emission available at the time, the overall level of agreement can be
considered quite good.

The experimental results are next compared with the MEM calculations~\cite{Minato2023}.
For proton emission, the calculated yield significantly exceeds the experimental
result.
As discussed in Sec.~V and illustrated in Fig.~\ref{fig:spectra_mem},
this discrepancy originates primarily from an overestimation of
the low-energy evaporation component in the calculation.

In contrast, the calculated yields for deuterons and tritons show good agreement
with the experimental data within uncertainties.
The $\alpha$-particle yield is also reproduced reasonably well by the MEM calculation,
indicating that the model provides a satisfactory description of composite-particle
emission channels.

The selective overestimation observed only for protons may be related to
the competition with neutron emission in the calculation, particularly in the low-energy region.
Since neutron emission is not directly constrained by the present experiment,
further clarification will require dedicated measurements of neutron energy spectra extending down to low energies.

In this context, it is worth noting that, in the original MEM literature~\cite{Minato2023},
the calculated neutron spectra were compared with the experimental data by Sundelin~\cite{Sundelin1973} and Kozlowski~\cite{Kozlowski1985},
and were found to slightly underestimate the measured yields
in the low-energy region.
Moreover, the available neutron data in those measurements have a lower energy limit of 4.25\,MeV
and therefore do not provide constraints on the low-energy region extending to energies as low as those accessible for protons,
where the neutron yield is expected to increase significantly.
The lack of experimental information in this energy range may thus contribute to
the uncertainty in the balance between proton and neutron emission in the model calculations.

\subsubsection{Comparison to branching ratio measurement}

Branching-ratio measurements of residual nuclei provide complementary insight into the competition between composite-particle emission and multiple-particle emission following \munc. In this subsection, the charged-particle yields obtained in the present work are compared with the branching-ratio measurement reported in Ref.~\cite{Mizuno2025b}.
A quantitative comparison of the integrated charged-particle yields obtained in the present
work with the branching ratios deduced from activation measurements, together with the lower limits obtained using prompt $\gamma$-ray measurements, is summarized in Table~\ref{tab:comp_yields}.
Except for the $4n2p$ value inferred from $^{22}$Na production, the branching ratios quoted here correspond to \munc in $^{28}$Si rather than in $^\mathrm{nat}$Si,
ensuring an unambiguous correspondence between charged-particle emission channels and residual nuclei.
The $4n2p$ value was obtained for natural silicon and was not corrected to $^{28}$Si, as noted in Table~\ref{tab:comp_yields}.

First, the integrated yield of the proton emission spectrum is significantly larger than the branching ratio for the $0n1p$ channel, whereas the integrated yield of the deuteron emission spectrum is substantially smaller than the branching ratio for the
$1n1p$ channel.
This indicates that proton emission is mostly accompanied by the emission of one or more neutrons.
Such behavior suggests that, in the de-excitation following \munc,
the emission of protons and neutrons as separate particles is favored over the emission of bound deuterons.
Consistently, the summed branching ratios of the $xn1p$ channels are found to be in
overall agreement with the sum of the integrated proton, deuteron, and triton yields,
noting that both quantities represent lower limits.

A similar behavior, though less pronounced, is observed for $\alpha$-particle emission. The integrated yield of the $\alpha$-emission spectrum is significantly larger than the branching ratio for the $2n2p$ channel, while the lower limit of the branching ratio for the $3n2p$ channel amounts to a substantial fraction of that for the $2n2p$ channel.
 This suggests that $\alpha$ emission is also likely to be accompanied by neutron emission.

To elucidate how such multiple particles are emitted and whether they exhibit
correlations, coincidence measurements are necessary.
Nevertheless,
charged-particle energy spectra and residual-nucleus branching-ratio measurements provide complementary information. Their combined use offers important constraints for refining theoretical descriptions of \munc.

\subsection{Observation of rare charged-particle emission channels}

In addition to protons, deuterons, tritons, and $\alpha$ particles,
events attributable to rarer charged-particle emission channels were observed in the present
experiment.
In the 200-{\textmu}m target run, a few events consistent with the emission of
$^{3}$He or $^{6}$He were identified based on DPSA.
Owing to the extremely limited statistics and the substantial energy loss in the thick
target, it was not possible to reconstruct their initial energy spectra.

Furthermore, two events corresponding to particles with $Z>2$ were also
observed.
These events may be associated with the emission of heavier clusters, such as $^{8}$Be,
although the present data do not allow a definitive identification or a quantitative
characterization of these channels.

To our knowledge, emissions of $^{3}$He, $^{6}$He, and particles with $Z>2$ following
\munc have not been reported in previous experimental studies.
The present observation therefore extends the range of charged particles identified
after \munc, albeit at a purely qualitative level.

At present, it remains unclear whether these rare channels originate from
evaporation-like processes, preequilibrium mechanisms, or other reaction pathways
activated at high excitation energies.
In this context, it is worth noting that a non-negligible contribution from
$4n4p$ decay branches following \munc at $^{28}$Si, corresponding in terms of particle composition to $2\alpha$ emission,
has been reported in the branching-ratio measurement~\cite{Mizuno2025b}.
Such observations suggest that multi-nucleon emission channels involving
strong correlations among constituent nucleons may play a significant role
in the decay process.

A detailed investigation of the production mechanisms of rare charged particles,
including possible connections to correlated multi-nucleon emission,
would require dedicated measurements with improved statistics and optimized
sensitivity to heavy-cluster emission, which are beyond the scope of the present work.

\section{Conclusions}

In this work, we measured the energy spectra of charged particles emitted
following \munc in silicon at the RIKEN-RAL muon facility.
Compared with existing measurements, a more comprehensive set of spectral
data was obtained over a wide energy range.
In particular, the present experiment provides, for the first time,
experimental information on the low-energy $\alpha$-particle spectrum,
which had not been available in previous studies.

The measured spectra were compared with MEM and PHITS calculations.
MEM reproduces the spectral shapes of protons, deuterons, and tritons more closely and describes the low-energy $\alpha$-particle spectrum well, whereas PHITS reproduces the overall slope of the $\alpha$-particle spectrum well.
Nevertheless, particle-dependent discrepancies in the absolute yields remain, including an overestimation of the evaporation component by PHITS for all four particle species.

The present data provide valuable constraints for theoretical model descriptions
of charged-particle emission after \munc.
Systematic measurements of charged-particle energy spectra for other target
nuclei will further contribute to improving models of preequilibrium and
evaporation processes in \munc reactions.

While the present work is based on single-particle measurements, the results presented here clearly indicate that future coincidence measurements of charged particles, in particular in combination with neutron detection, will be essential for a more complete understanding of multi-particle emission processes following \munc.

\section*{Acknowledgments}

The authors would like to thank the accelerator staff at RAL for providing a high-quality beam.
We also thank Y.~Gomi of Kyoto University for valuable discussions related to the PHITS calculations.
The use of the RIKEN-RAL muon beam was approved by the RIKEN Nishina Center Materials and Life Science PAC (Experiment No.~RB2070004).
This work was partially supported by JSPS KAKENHI Grant Numbers JP19H05664 and JP21H01863.

\appendix

\section{Fitting Function for Peak Timing Distribution}
\label{appendix:fit_function}

The peak timing distribution shown in Sec.~\ref{sec:muon_number} was fitted using a composite function that accounts for two identical beam bunches separated by a fixed time interval \(\Delta t\). The total fitting function is defined as
\begin{align}
f(t) = f_1(t) + f_1(t - \Delta t),
\end{align}
where \(f_1(t)\) represents the contribution from a single beam bunch and is defined by
\begin{align}
f_1(t) =\ 
& A_e\, \mathcal{N}(t; \mu_e, \sigma_e)
+ A_\mu\, \mathcal{N}(t; \mu_\mu, \sigma_\mu) \nonumber \\
&+ A_{\mathrm{decay}}\, \mathcal{D}(t; \mu_\mu, \sigma_\mu, \tau).
\end{align}

The first term represents prompt signals from electrons produced simultaneously with the muon beam, the second term corresponds to prompt signals from muons that are stopped in the plastic scintillator, and the third term describes delayed signals originating from decay electrons emitted by stopped muons, modeled as the convolution of an exponential decay with a Gaussian resolution function.

Here, $\mathcal{N}(t; \mu, \sigma)$ denotes a Gaussian distribution:
\begin{align}
\mathcal{N}(t; \mu, \sigma) =
\frac{1}{\sqrt{2\pi} \sigma} \exp\left( -\frac{(t - \mu)^2}{2\sigma^2} \right),
\end{align}
and $\mathcal{D}(t; \mu, \sigma, \tau)$ is the convolution of a Gaussian with an exponential decay characterized by a decay time constant $\tau$:
\begin{align}
\mathcal{D}(t; \mu, \sigma, \tau) =\ 
&\int_{-\infty}^{t} \mathcal{N}(t'; \mu, \sigma)\, e^{-(t - t')/\tau} \, dt' \\
=\ &\frac{1}{2} \exp\left( -\frac{t - \mu}{\tau} + \frac{\sigma^2}{2\tau^2} \right)\nonumber\\
&\times\operatorname{erfc}\left( \frac{\mu - t + \sigma^2/\tau}{\sqrt{2} \sigma} \right). 
\end{align}

The parameters \(A_\mu\), \(A_e\), and \(A_{\mathrm{decay}}\) denote the amplitudes of the muon, electron, and decay components, respectively. The parameter \(\tau\) is the decay time constant of the exponential, and \(\Delta t\) is the time interval between the two beam bunches. 

In total, the fitting function includes nine free parameters:
\(A_e\), \(\mu_e\), \(\sigma_e\),
\(A_\mu\), \(\mu_\mu\), \(\sigma_\mu\),
\(A_{\mathrm{decay}}\), \(\tau\), and \(\Delta t\).

\bibliography{reference}%

\begin{thebibliography}{62}%
\makeatletter
\providecommand \@ifxundefined [1]{%
 \@ifx{#1\undefined}
}%
\providecommand \@ifnum [1]{%
 \ifnum #1\expandafter \@firstoftwo
 \else \expandafter \@secondoftwo
 \fi
}%
\providecommand \@ifx [1]{%
 \ifx #1\expandafter \@firstoftwo
 \else \expandafter \@secondoftwo
 \fi
}%
\providecommand \natexlab [1]{#1}%
\providecommand \enquote  [1]{``#1''}%
\providecommand \bibnamefont  [1]{#1}%
\providecommand \bibfnamefont [1]{#1}%
\providecommand \citenamefont [1]{#1}%
\providecommand \href@noop [0]{\@secondoftwo}%
\providecommand \href [0]{\begingroup \@sanitize@url \@href}%
\providecommand \@href[1]{\@@startlink{#1}\@@href}%
\providecommand \@@href[1]{\endgroup#1\@@endlink}%
\providecommand \@sanitize@url [0]{\catcode `\\12\catcode `\$12\catcode `\&12\catcode `\#12\catcode `\^12\catcode `\_12\catcode `\%12\relax}%
\providecommand \@@startlink[1]{}%
\providecommand \@@endlink[0]{}%
\providecommand \url  [0]{\begingroup\@sanitize@url \@url }%
\providecommand \@url [1]{\endgroup\@href {#1}{\urlprefix }}%
\providecommand \urlprefix  [0]{URL }%
\providecommand \Eprint [0]{\href }%
\providecommand \doibase [0]{https://doi.org/}%
\providecommand \selectlanguage [0]{\@gobble}%
\providecommand \bibinfo  [0]{\@secondoftwo}%
\providecommand \bibfield  [0]{\@secondoftwo}%
\providecommand \translation [1]{[#1]}%
\providecommand \BibitemOpen [0]{}%
\providecommand \bibitemStop [0]{}%
\providecommand \bibitemNoStop [0]{.\EOS\space}%
\providecommand \EOS [0]{\spacefactor3000\relax}%
\providecommand \BibitemShut  [1]{\csname bibitem#1\endcsname}%
\let\auto@bib@innerbib\@empty
\bibitem [{\citenamefont {Measday}(2001)}]{Measday2001}%
  \BibitemOpen
  \bibfield  {author} {\bibinfo {author} {\bibfnamefont {D.}~\bibnamefont {Measday}},\ }\bibfield  {title} {\bibinfo {title} {The nuclear physics of muon capture},\ }\href {https://doi.org/10.1016/S0370-1573(01)00012-6} {\bibfield  {journal} {\bibinfo  {journal} {Physics Reports}\ }\textbf {\bibinfo {volume} {354}},\ \bibinfo {pages} {243} (\bibinfo {year} {2001})}\BibitemShut {NoStop}%
\bibitem [{\citenamefont {Schr{\"o}der}\ \emph {et~al.}(1974)\citenamefont {Schr{\"o}der}, \citenamefont {Jahnke}, \citenamefont {Lindenberger}, \citenamefont {R{\"o}schert}, \citenamefont {Engfer},\ and\ \citenamefont {Walter}}]{Schroder1974}%
  \BibitemOpen
  \bibfield  {author} {\bibinfo {author} {\bibfnamefont {W.~U.}\ \bibnamefont {Schr{\"o}der}}, \bibinfo {author} {\bibfnamefont {U.}~\bibnamefont {Jahnke}}, \bibinfo {author} {\bibfnamefont {K.~H.}\ \bibnamefont {Lindenberger}}, \bibinfo {author} {\bibfnamefont {G.}~\bibnamefont {R{\"o}schert}}, \bibinfo {author} {\bibfnamefont {R.}~\bibnamefont {Engfer}},\ and\ \bibinfo {author} {\bibfnamefont {H.~K.}\ \bibnamefont {Walter}},\ }\bibfield  {title} {\bibinfo {title} {Spectra of neutrons from {$\mu$} capture in thallium, lead and bismuth},\ }\href {https://doi.org/10.1007/BF01670062} {\bibfield  {journal} {\bibinfo  {journal} {Zeitschrift f\"ur Physik}\ }\textbf {\bibinfo {volume} {268}},\ \bibinfo {pages} {57} (\bibinfo {year} {1974})}\BibitemShut {NoStop}%
\bibitem [{\citenamefont {Plett}\ and\ \citenamefont {Sobottka}(1971)}]{Plett1971}%
  \BibitemOpen
  \bibfield  {author} {\bibinfo {author} {\bibfnamefont {M.~E.}\ \bibnamefont {Plett}}\ and\ \bibinfo {author} {\bibfnamefont {S.~E.}\ \bibnamefont {Sobottka}},\ }\bibfield  {title} {\bibinfo {title} {Effects of the giant resonance on the energy spectra of neutrons emitted following muon capture in {$^{12}$C} and {$^{16}$O}},\ }\href {https://doi.org/10.1103/PhysRevC.3.1003} {\bibfield  {journal} {\bibinfo  {journal} {Physical Review C}\ }\textbf {\bibinfo {volume} {3}},\ \bibinfo {pages} {1003} (\bibinfo {year} {1971})}\BibitemShut {NoStop}%
\bibitem [{\citenamefont {Measday}\ and\ \citenamefont {Stocki}(2006)}]{Measday2006}%
  \BibitemOpen
  \bibfield  {author} {\bibinfo {author} {\bibfnamefont {D.~F.}\ \bibnamefont {Measday}}\ and\ \bibinfo {author} {\bibfnamefont {T.~J.}\ \bibnamefont {Stocki}},\ }\bibfield  {title} {\bibinfo {title} {{$gamma$} rays from muon capture in natural {{Ca}}, {{Fe}}, and {{Ni}}},\ }\href {https://doi.org/10.1103/PhysRevC.73.045501} {\bibfield  {journal} {\bibinfo  {journal} {Physical Review C}\ }\textbf {\bibinfo {volume} {73}},\ \bibinfo {pages} {045501} (\bibinfo {year} {2006})}\BibitemShut {NoStop}%
\bibitem [{\citenamefont {Wyttenbach}\ \emph {et~al.}(1978)\citenamefont {Wyttenbach}, \citenamefont {Baertschi}, \citenamefont {Bajo}, \citenamefont {Hadermann}, \citenamefont {Junker}, \citenamefont {Katcoff}, \citenamefont {Hermes},\ and\ \citenamefont {Pruys}}]{Wyttenbach1978}%
  \BibitemOpen
  \bibfield  {author} {\bibinfo {author} {\bibfnamefont {A.}~\bibnamefont {Wyttenbach}}, \bibinfo {author} {\bibfnamefont {P.}~\bibnamefont {Baertschi}}, \bibinfo {author} {\bibfnamefont {S.}~\bibnamefont {Bajo}}, \bibinfo {author} {\bibfnamefont {J.}~\bibnamefont {Hadermann}}, \bibinfo {author} {\bibfnamefont {K.}~\bibnamefont {Junker}}, \bibinfo {author} {\bibfnamefont {S.}~\bibnamefont {Katcoff}}, \bibinfo {author} {\bibfnamefont {E.~A.}\ \bibnamefont {Hermes}},\ and\ \bibinfo {author} {\bibfnamefont {H.~S.}\ \bibnamefont {Pruys}},\ }\bibfield  {title} {\bibinfo {title} {Probabilities of muon induced nuclear reactions involving charged particle emission},\ }\href {https://doi.org/10.1016/0375-9474(78)90218-X} {\bibfield  {journal} {\bibinfo  {journal} {Nuclear Physics A}\ }\textbf {\bibinfo {volume} {294}},\ \bibinfo {pages} {278} (\bibinfo {year} {1978})}\BibitemShut {NoStop}%
\bibitem [{\citenamefont {Niikura}\ \emph {et~al.}(2024)\citenamefont {Niikura}, \citenamefont {Saito}, \citenamefont {Matsuzaki}, \citenamefont {Ishida},\ and\ \citenamefont {Hillier}}]{Niikura2024}%
  \BibitemOpen
  \bibfield  {author} {\bibinfo {author} {\bibfnamefont {M.}~\bibnamefont {Niikura}}, \bibinfo {author} {\bibfnamefont {T.~Y.}\ \bibnamefont {Saito}}, \bibinfo {author} {\bibfnamefont {T.}~\bibnamefont {Matsuzaki}}, \bibinfo {author} {\bibfnamefont {K.}~\bibnamefont {Ishida}},\ and\ \bibinfo {author} {\bibfnamefont {A.}~\bibnamefont {Hillier}},\ }\bibfield  {title} {\bibinfo {title} {Measurement of the production branching ratios following nuclear muon capture for palladium isotopes using the in-beam activation method},\ }\href {https://doi.org/10.1103/PhysRevC.109.014328} {\bibfield  {journal} {\bibinfo  {journal} {Physical Review C}\ }\textbf {\bibinfo {volume} {109}},\ \bibinfo {pages} {014328} (\bibinfo {year} {2024})}\BibitemShut {NoStop}%
\bibitem [{\citenamefont {Yamaguchi}\ \emph {et~al.}(2025)\citenamefont {Yamaguchi}, \citenamefont {Niikura}, \citenamefont {Mizuno}, \citenamefont {Tampo}, \citenamefont {Harada}, \citenamefont {Kawamura}, \citenamefont {Umegaki}, \citenamefont {Takeshita},\ and\ \citenamefont {Haga}}]{Yamaguchi2025a}%
  \BibitemOpen
  \bibfield  {author} {\bibinfo {author} {\bibfnamefont {Y.}~\bibnamefont {Yamaguchi}}, \bibinfo {author} {\bibfnamefont {M.}~\bibnamefont {Niikura}}, \bibinfo {author} {\bibfnamefont {R.}~\bibnamefont {Mizuno}}, \bibinfo {author} {\bibfnamefont {M.}~\bibnamefont {Tampo}}, \bibinfo {author} {\bibfnamefont {M.}~\bibnamefont {Harada}}, \bibinfo {author} {\bibfnamefont {N.}~\bibnamefont {Kawamura}}, \bibinfo {author} {\bibfnamefont {I.}~\bibnamefont {Umegaki}}, \bibinfo {author} {\bibfnamefont {S.}~\bibnamefont {Takeshita}},\ and\ \bibinfo {author} {\bibfnamefont {K.}~\bibnamefont {Haga}},\ }\bibfield  {title} {\bibinfo {title} {Measurement of radionuclide production probabilities in negative muon nuclear capture and validation of {{Monte Carlo}} simulation model},\ }\href {https://doi.org/10.1016/j.nimb.2025.165801} {\bibfield  {journal} {\bibinfo  {journal} {Nuclear Instruments and Methods in Physics Research Section B: Beam Interactions with Materials and Atoms}\ }\textbf {\bibinfo {volume} {567}},\ \bibinfo {pages} {165801} (\bibinfo {year} {2025})}\BibitemShut {NoStop}%
\bibitem [{\citenamefont {Mizuno}\ \emph {et~al.}(2025)\citenamefont {Mizuno}, \citenamefont {Niikura}, \citenamefont {Saito}, \citenamefont {Matsuzaki}, \citenamefont {Abe}, \citenamefont {Fukuda}, \citenamefont {Hashimoto}, \citenamefont {Hillier}, \citenamefont {Ishida}, \citenamefont {Kawamura}, \citenamefont {Kawase}, \citenamefont {Kawata}, \citenamefont {Kitafuji}, \citenamefont {Minato}, \citenamefont {Oishi}, \citenamefont {Sato}, \citenamefont {Shimomura}, \citenamefont {Strasser}, \citenamefont {Takeshita}, \citenamefont {Tomono},\ and\ \citenamefont {Watanabe}}]{Mizuno2025b}%
  \BibitemOpen
  \bibfield  {author} {\bibinfo {author} {\bibfnamefont {R.}~\bibnamefont {Mizuno}}, \bibinfo {author} {\bibfnamefont {M.}~\bibnamefont {Niikura}}, \bibinfo {author} {\bibfnamefont {T.~Y.}\ \bibnamefont {Saito}}, \bibinfo {author} {\bibfnamefont {T.}~\bibnamefont {Matsuzaki}}, \bibinfo {author} {\bibfnamefont {S.}~\bibnamefont {Abe}}, \bibinfo {author} {\bibfnamefont {H.}~\bibnamefont {Fukuda}}, \bibinfo {author} {\bibfnamefont {M.}~\bibnamefont {Hashimoto}}, \bibinfo {author} {\bibfnamefont {A.~D.}\ \bibnamefont {Hillier}}, \bibinfo {author} {\bibfnamefont {K.}~\bibnamefont {Ishida}}, \bibinfo {author} {\bibfnamefont {N.}~\bibnamefont {Kawamura}}, \bibinfo {author} {\bibfnamefont {S.}~\bibnamefont {Kawase}}, \bibinfo {author} {\bibfnamefont {T.}~\bibnamefont {Kawata}}, \bibinfo {author} {\bibfnamefont {K.}~\bibnamefont {Kitafuji}}, \bibinfo {author} {\bibfnamefont {F.}~\bibnamefont {Minato}}, \bibinfo {author} {\bibfnamefont {M.}~\bibnamefont {Oishi}}, \bibinfo {author} {\bibfnamefont {A.}~\bibnamefont {Sato}}, \bibinfo {author} {\bibfnamefont {K.}~\bibnamefont {Shimomura}}, \bibinfo {author} {\bibfnamefont {P.}~\bibnamefont {Strasser}}, \bibinfo {author} {\bibfnamefont {S.}~\bibnamefont {Takeshita}}, \bibinfo {author} {\bibfnamefont {D.}~\bibnamefont {Tomono}},\ and\ \bibinfo {author} {\bibfnamefont {Y.}~\bibnamefont {Watanabe}},\ }\bibfield  {title} {\bibinfo {title} {Measurement of production branching ratio after muon nuclear capture reaction of {{Al}} and {{Si}} isotopes},\ }\href {https://doi.org/10.1103/kycz-qprw} {\bibfield  {journal} {\bibinfo  {journal} {Physical Review C}\ }\textbf {\bibinfo {volume} {112}},\ \bibinfo {pages} {054305} (\bibinfo {year} {2025})}\BibitemShut {NoStop}%
\bibitem [{\citenamefont {Backenstoss}\ \emph {et~al.}(1971)\citenamefont {Backenstoss}, \citenamefont {Charalambus}, \citenamefont {Daniel}, \citenamefont {Hamilton}, \citenamefont {Lynen}, \citenamefont {Von Der~Malsburg}, \citenamefont {Poelz},\ and\ \citenamefont {Povel}}]{Backenstoss1971}%
  \BibitemOpen
  \bibfield  {author} {\bibinfo {author} {\bibfnamefont {G.}~\bibnamefont {Backenstoss}}, \bibinfo {author} {\bibfnamefont {S.}~\bibnamefont {Charalambus}}, \bibinfo {author} {\bibfnamefont {H.}~\bibnamefont {Daniel}}, \bibinfo {author} {\bibfnamefont {W.}~\bibnamefont {Hamilton}}, \bibinfo {author} {\bibfnamefont {U.}~\bibnamefont {Lynen}}, \bibinfo {author} {\bibfnamefont {{\relax Ch}.}~\bibnamefont {Von Der~Malsburg}}, \bibinfo {author} {\bibfnamefont {G.}~\bibnamefont {Poelz}},\ and\ \bibinfo {author} {\bibfnamefont {H.}~\bibnamefont {Povel}},\ }\bibfield  {title} {\bibinfo {title} {Nuclear {$\gamma$}-rays following muon capture},\ }\href {https://doi.org/10.1016/0375-9474(71)90253-3} {\bibfield  {journal} {\bibinfo  {journal} {Nuclear Physics A}\ }\textbf {\bibinfo {volume} {162}},\ \bibinfo {pages} {541} (\bibinfo {year} {1971})}\BibitemShut {NoStop}%
\bibitem [{\citenamefont {Johnson}\ \emph {et~al.}(1996)\citenamefont {Johnson}, \citenamefont {Gorringe}, \citenamefont {Armstrong}, \citenamefont {Bauer}, \citenamefont {Hasinoff}, \citenamefont {Kovash}, \citenamefont {Measday}, \citenamefont {Moftah}, \citenamefont {Porter},\ and\ \citenamefont {Wright}}]{Johnson1996}%
  \BibitemOpen
  \bibfield  {author} {\bibinfo {author} {\bibfnamefont {B.~L.}\ \bibnamefont {Johnson}}, \bibinfo {author} {\bibfnamefont {T.~P.}\ \bibnamefont {Gorringe}}, \bibinfo {author} {\bibfnamefont {D.~S.}\ \bibnamefont {Armstrong}}, \bibinfo {author} {\bibfnamefont {J.}~\bibnamefont {Bauer}}, \bibinfo {author} {\bibfnamefont {M.~D.}\ \bibnamefont {Hasinoff}}, \bibinfo {author} {\bibfnamefont {M.~A.}\ \bibnamefont {Kovash}}, \bibinfo {author} {\bibfnamefont {D.~F.}\ \bibnamefont {Measday}}, \bibinfo {author} {\bibfnamefont {B.~A.}\ \bibnamefont {Moftah}}, \bibinfo {author} {\bibfnamefont {R.}~\bibnamefont {Porter}},\ and\ \bibinfo {author} {\bibfnamefont {D.~H.}\ \bibnamefont {Wright}},\ }\bibfield  {title} {\bibinfo {title} {Observables in muon capture on {$^{23}$Na} and the effective weak couplings {$\tilde{g}_a$} and {$\tilde{g}_p$}},\ }\href {https://doi.org/10.1103/PhysRevC.54.2714} {\bibfield  {journal} {\bibinfo  {journal} {Physical Review C}\ }\textbf {\bibinfo {volume} {54}},\ \bibinfo {pages} {2714} (\bibinfo {year} {1996})}\BibitemShut {NoStop}%
\bibitem [{\citenamefont {Gorringe}\ \emph {et~al.}(1999)\citenamefont {Gorringe}, \citenamefont {Armstrong}, \citenamefont {Arole}, \citenamefont {Boleman}, \citenamefont {Gete}, \citenamefont {Kuzmin}, \citenamefont {Moftah}, \citenamefont {Sedlar}, \citenamefont {Stocki},\ and\ \citenamefont {Tetereva}}]{Gorringe1999}%
  \BibitemOpen
  \bibfield  {author} {\bibinfo {author} {\bibfnamefont {T.~P.}\ \bibnamefont {Gorringe}}, \bibinfo {author} {\bibfnamefont {D.~S.}\ \bibnamefont {Armstrong}}, \bibinfo {author} {\bibfnamefont {S.}~\bibnamefont {Arole}}, \bibinfo {author} {\bibfnamefont {M.}~\bibnamefont {Boleman}}, \bibinfo {author} {\bibfnamefont {E.}~\bibnamefont {Gete}}, \bibinfo {author} {\bibfnamefont {V.}~\bibnamefont {Kuzmin}}, \bibinfo {author} {\bibfnamefont {B.~A.}\ \bibnamefont {Moftah}}, \bibinfo {author} {\bibfnamefont {R.}~\bibnamefont {Sedlar}}, \bibinfo {author} {\bibfnamefont {T.~J.}\ \bibnamefont {Stocki}},\ and\ \bibinfo {author} {\bibfnamefont {T.}~\bibnamefont {Tetereva}},\ }\bibfield  {title} {\bibinfo {title} {Measurement of partial muon capture rates in {$1s$}-{$0d$} shell nuclei},\ }\href {https://doi.org/10.1103/PhysRevC.60.055501} {\bibfield  {journal} {\bibinfo  {journal} {Physical Review C}\ }\textbf {\bibinfo {volume} {60}},\ \bibinfo {pages} {055501} (\bibinfo {year} {1999})}\BibitemShut {NoStop}%
\bibitem [{\citenamefont {Stocki}\ \emph {et~al.}(2002)\citenamefont {Stocki}, \citenamefont {Measday}, \citenamefont {Gete}, \citenamefont {Saliba}, \citenamefont {Moftah},\ and\ \citenamefont {Gorringe}}]{Stocki2002}%
  \BibitemOpen
  \bibfield  {author} {\bibinfo {author} {\bibfnamefont {T.}~\bibnamefont {Stocki}}, \bibinfo {author} {\bibfnamefont {D.}~\bibnamefont {Measday}}, \bibinfo {author} {\bibfnamefont {E.}~\bibnamefont {Gete}}, \bibinfo {author} {\bibfnamefont {M.}~\bibnamefont {Saliba}}, \bibinfo {author} {\bibfnamefont {B.}~\bibnamefont {Moftah}},\ and\ \bibinfo {author} {\bibfnamefont {T.}~\bibnamefont {Gorringe}},\ }\bibfield  {title} {\bibinfo {title} {Gamma rays from muon capture in {{$^{14}$N}}},\ }\href {https://doi.org/10.1016/S0375-9474(01)01249-0} {\bibfield  {journal} {\bibinfo  {journal} {Nuclear Physics A}\ }\textbf {\bibinfo {volume} {697}},\ \bibinfo {pages} {55} (\bibinfo {year} {2002})}\BibitemShut {NoStop}%
\bibitem [{\citenamefont {Measday}\ \emph {et~al.}(2007{\natexlab{a}})\citenamefont {Measday}, \citenamefont {Stocki},\ and\ \citenamefont {Tam}}]{Measday2007a}%
  \BibitemOpen
  \bibfield  {author} {\bibinfo {author} {\bibfnamefont {D.~F.}\ \bibnamefont {Measday}}, \bibinfo {author} {\bibfnamefont {T.~J.}\ \bibnamefont {Stocki}},\ and\ \bibinfo {author} {\bibfnamefont {H.}~\bibnamefont {Tam}},\ }\bibfield  {title} {\bibinfo {title} {{$\gamma$} rays from muon capture in {{I}}, {{Au}}, and {{Bi}}},\ }\href {https://doi.org/10.1103/PhysRevC.75.045501} {\bibfield  {journal} {\bibinfo  {journal} {Physical Review C}\ }\textbf {\bibinfo {volume} {75}},\ \bibinfo {pages} {045501} (\bibinfo {year} {2007}{\natexlab{a}})}\BibitemShut {NoStop}%
\bibitem [{\citenamefont {Measday}\ \emph {et~al.}(2007{\natexlab{b}})\citenamefont {Measday}, \citenamefont {Stocki}, \citenamefont {Moftah},\ and\ \citenamefont {Tam}}]{Measday2007}%
  \BibitemOpen
  \bibfield  {author} {\bibinfo {author} {\bibfnamefont {D.~F.}\ \bibnamefont {Measday}}, \bibinfo {author} {\bibfnamefont {T.~J.}\ \bibnamefont {Stocki}}, \bibinfo {author} {\bibfnamefont {B.~A.}\ \bibnamefont {Moftah}},\ and\ \bibinfo {author} {\bibfnamefont {H.}~\bibnamefont {Tam}},\ }\bibfield  {title} {\bibinfo {title} {{$\gamma$} rays from muon capture in {{$^{27}$Al}} and natural {{Si}}},\ }\href {https://doi.org/10.1103/PhysRevC.76.035504} {\bibfield  {journal} {\bibinfo  {journal} {Physical Review C}\ }\textbf {\bibinfo {volume} {76}},\ \bibinfo {pages} {035504} (\bibinfo {year} {2007}{\natexlab{b}})}\BibitemShut {NoStop}%
\bibitem [{\citenamefont {Zinatulina}\ \emph {et~al.}(2019)\citenamefont {Zinatulina}, \citenamefont {Brudanin}, \citenamefont {Egorov}, \citenamefont {Petitjean}, \citenamefont {Shirchenko}, \citenamefont {Suhonen},\ and\ \citenamefont {Yutlandov}}]{Zinatulina2019}%
  \BibitemOpen
  \bibfield  {author} {\bibinfo {author} {\bibfnamefont {D.}~\bibnamefont {Zinatulina}}, \bibinfo {author} {\bibfnamefont {V.}~\bibnamefont {Brudanin}}, \bibinfo {author} {\bibfnamefont {V.}~\bibnamefont {Egorov}}, \bibinfo {author} {\bibfnamefont {C.}~\bibnamefont {Petitjean}}, \bibinfo {author} {\bibfnamefont {M.}~\bibnamefont {Shirchenko}}, \bibinfo {author} {\bibfnamefont {J.}~\bibnamefont {Suhonen}},\ and\ \bibinfo {author} {\bibfnamefont {I.}~\bibnamefont {Yutlandov}},\ }\bibfield  {title} {\bibinfo {title} {Ordinary muon capture studies for the matrix elements in {$\beta\beta$} decay},\ }\href {https://doi.org/10.1103/PhysRevC.99.024327} {\bibfield  {journal} {\bibinfo  {journal} {Physical Review C}\ }\textbf {\bibinfo {volume} {99}},\ \bibinfo {pages} {024327} (\bibinfo {year} {2019})}\BibitemShut {NoStop}%
\bibitem [{\citenamefont {Heusser}\ and\ \citenamefont {Kirsten}(1972)}]{Heusser1972}%
  \BibitemOpen
  \bibfield  {author} {\bibinfo {author} {\bibfnamefont {G.}~\bibnamefont {Heusser}}\ and\ \bibinfo {author} {\bibfnamefont {T.}~\bibnamefont {Kirsten}},\ }\bibfield  {title} {\bibinfo {title} {Radioisotope production rates by muon capture},\ }\href {https://doi.org/10.1016/0375-9474(72)91065-2} {\bibfield  {journal} {\bibinfo  {journal} {Nuclear Physics A}\ }\textbf {\bibinfo {volume} {195}},\ \bibinfo {pages} {369} (\bibinfo {year} {1972})}\BibitemShut {NoStop}%
\bibitem [{\citenamefont {Heisinger}\ \emph {et~al.}(2002)\citenamefont {Heisinger}, \citenamefont {Lal}, \citenamefont {Jull}, \citenamefont {Kubik}, \citenamefont {{Ivy-Ochs}}, \citenamefont {Knie},\ and\ \citenamefont {Nolte}}]{Heisinger2002}%
  \BibitemOpen
  \bibfield  {author} {\bibinfo {author} {\bibfnamefont {B.}~\bibnamefont {Heisinger}}, \bibinfo {author} {\bibfnamefont {D.}~\bibnamefont {Lal}}, \bibinfo {author} {\bibfnamefont {A.~J.~T.}\ \bibnamefont {Jull}}, \bibinfo {author} {\bibfnamefont {P.}~\bibnamefont {Kubik}}, \bibinfo {author} {\bibfnamefont {S.}~\bibnamefont {{Ivy-Ochs}}}, \bibinfo {author} {\bibfnamefont {K.}~\bibnamefont {Knie}},\ and\ \bibinfo {author} {\bibfnamefont {E.}~\bibnamefont {Nolte}},\ }\bibfield  {title} {\bibinfo {title} {Production of selected cosmogenic radionuclides by muons: 2. {{Capture}} of negative muons},\ }\href {https://doi.org/10.1016/S0012-821X(02)00641-6} {\bibfield  {journal} {\bibinfo  {journal} {Earth and Planetary Science Letters}\ }\textbf {\bibinfo {volume} {200}},\ \bibinfo {pages} {357} (\bibinfo {year} {2002})}\BibitemShut {NoStop}%
\bibitem [{\citenamefont {Sobottka}\ and\ \citenamefont {Wills}(1968)}]{Sobottka1968}%
  \BibitemOpen
  \bibfield  {author} {\bibinfo {author} {\bibfnamefont {S.~E.}\ \bibnamefont {Sobottka}}\ and\ \bibinfo {author} {\bibfnamefont {E.~L.}\ \bibnamefont {Wills}},\ }\bibfield  {title} {\bibinfo {title} {Energy {{Spectrum}} of {{Charged Particles Emitted Following Muon Capture}} in {{Si${}^{28}$}}},\ }\href {https://doi.org/10.1103/PhysRevLett.20.596} {\bibfield  {journal} {\bibinfo  {journal} {Physical Review Letters}\ }\textbf {\bibinfo {volume} {20}},\ \bibinfo {pages} {596} (\bibinfo {year} {1968})}\BibitemShut {NoStop}%
\bibitem [{\citenamefont {Budyashov}\ \emph {et~al.}(1971)\citenamefont {Budyashov}, \citenamefont {Zinov}, \citenamefont {Konin}, \citenamefont {Mukhin},\ and\ \citenamefont {Chatrchyan}}]{Budyashov1971}%
  \BibitemOpen
  \bibfield  {author} {\bibinfo {author} {\bibfnamefont {Y.~G.}\ \bibnamefont {Budyashov}}, \bibinfo {author} {\bibfnamefont {V.~G.}\ \bibnamefont {Zinov}}, \bibinfo {author} {\bibfnamefont {A.~D.}\ \bibnamefont {Konin}}, \bibinfo {author} {\bibfnamefont {A.~I.}\ \bibnamefont {Mukhin}},\ and\ \bibinfo {author} {\bibfnamefont {A.~M.}\ \bibnamefont {Chatrchyan}},\ }\bibfield  {title} {\bibinfo {title} {Charged particles from the capture of negative muons by the nuclei {\textsuperscript{28}}{Si}, {\textsuperscript{32}}{S}, {\textsuperscript{40}}{Ca}, and {\textsuperscript{64}}{Cu}},\ }\href@noop {} {\bibfield  {journal} {\bibinfo  {journal} {Journal of Experimental and Theoretical Physics}\ }\textbf {\bibinfo {volume} {33}},\ \bibinfo {pages} {11} (\bibinfo {year} {1971})}\BibitemShut {NoStop}%
\bibitem [{\citenamefont {Edmonds}\ \emph {et~al.}(2022)\citenamefont {Edmonds}, \citenamefont {Quirk}, \citenamefont {Wong}, \citenamefont {Alexander}, \citenamefont {Bernstein}, \citenamefont {Daniel}, \citenamefont {Diociaiuti}, \citenamefont {Donghia}, \citenamefont {Gillies}, \citenamefont {Hungerford}, \citenamefont {Kammel}, \citenamefont {Krikler}, \citenamefont {Kuno}, \citenamefont {Lancaster}, \citenamefont {Litchfield}, \citenamefont {Miller}, \citenamefont {Palladino}, \citenamefont {Repond}, \citenamefont {Sato}, \citenamefont {Sarra}, \citenamefont {Soleti}, \citenamefont {Tishchenko}, \citenamefont {Tran}, \citenamefont {Uchida}, \citenamefont {Winter},\ and\ \citenamefont {Wu}}]{Edmonds2022}%
  \BibitemOpen
  \bibfield  {author} {\bibinfo {author} {\bibfnamefont {A.}~\bibnamefont {Edmonds}}, \bibinfo {author} {\bibfnamefont {J.}~\bibnamefont {Quirk}}, \bibinfo {author} {\bibfnamefont {M.~L.}\ \bibnamefont {Wong}}, \bibinfo {author} {\bibfnamefont {D.}~\bibnamefont {Alexander}}, \bibinfo {author} {\bibfnamefont {R.~H.}\ \bibnamefont {Bernstein}}, \bibinfo {author} {\bibfnamefont {A.}~\bibnamefont {Daniel}}, \bibinfo {author} {\bibfnamefont {E.}~\bibnamefont {Diociaiuti}}, \bibinfo {author} {\bibfnamefont {R.}~\bibnamefont {Donghia}}, \bibinfo {author} {\bibfnamefont {E.~L.}\ \bibnamefont {Gillies}}, \bibinfo {author} {\bibfnamefont {E.~V.}\ \bibnamefont {Hungerford}}, \bibinfo {author} {\bibfnamefont {P.}~\bibnamefont {Kammel}}, \bibinfo {author} {\bibfnamefont {B.~E.}\ \bibnamefont {Krikler}}, \bibinfo {author} {\bibfnamefont {Y.}~\bibnamefont {Kuno}}, \bibinfo {author} {\bibfnamefont {M.}~\bibnamefont {Lancaster}}, \bibinfo {author} {\bibfnamefont {R.~P.}\ \bibnamefont {Litchfield}}, \bibinfo {author} {\bibfnamefont {J.~P.}\ \bibnamefont {Miller}}, \bibinfo {author} {\bibfnamefont {A.}~\bibnamefont {Palladino}}, \bibinfo {author} {\bibfnamefont {J.}~\bibnamefont {Repond}}, \bibinfo {author} {\bibfnamefont {A.}~\bibnamefont {Sato}}, \bibinfo {author} {\bibfnamefont {I.}~\bibnamefont {Sarra}}, \bibinfo {author} {\bibfnamefont {S.~R.}\ \bibnamefont {Soleti}}, \bibinfo {author} {\bibfnamefont {V.}~\bibnamefont {Tishchenko}}, \bibinfo {author} {\bibfnamefont {N.~H.}\ \bibnamefont {Tran}}, \bibinfo {author} {\bibfnamefont {Y.}~\bibnamefont {Uchida}}, \bibinfo {author} {\bibfnamefont {P.}~\bibnamefont {Winter}},\ and\ \bibinfo {author} {\bibfnamefont {C.}~\bibnamefont {Wu}},\ }\bibfield  {title} {\bibinfo {title} {Measurement of proton, deuteron, triton, and {$\alpha$} particle emission after nuclear muon capture on {{Al}}, {{Si}}, and {{Ti}} with the {{AlCap}} experiment},\ }\href {https://doi.org/10.1103/PhysRevC.105.035501} {\bibfield  {journal} {\bibinfo  {journal} {Physical Review C}\ }\textbf {\bibinfo {volume} {105}},\ \bibinfo {pages} {035501} (\bibinfo {year} {2022})}\BibitemShut {NoStop}%
\bibitem [{\citenamefont {Manabe}\ \emph {et~al.}(2023)\citenamefont {Manabe}, \citenamefont {Watanabe}, \citenamefont {Niikura}, \citenamefont {Nakano}, \citenamefont {Saito}, \citenamefont {Suzuki}, \citenamefont {Kawashima}, \citenamefont {Tomono}, \citenamefont {Sato},\ and\ \citenamefont {Harano}}]{Manabe2023}%
  \BibitemOpen
  \bibfield  {author} {\bibinfo {author} {\bibfnamefont {S.}~\bibnamefont {Manabe}}, \bibinfo {author} {\bibfnamefont {Y.}~\bibnamefont {Watanabe}}, \bibinfo {author} {\bibfnamefont {M.}~\bibnamefont {Niikura}}, \bibinfo {author} {\bibfnamefont {K.}~\bibnamefont {Nakano}}, \bibinfo {author} {\bibfnamefont {T.~Y.}\ \bibnamefont {Saito}}, \bibinfo {author} {\bibfnamefont {D.}~\bibnamefont {Suzuki}}, \bibinfo {author} {\bibfnamefont {Y.}~\bibnamefont {Kawashima}}, \bibinfo {author} {\bibfnamefont {D.}~\bibnamefont {Tomono}}, \bibinfo {author} {\bibfnamefont {A.}~\bibnamefont {Sato}},\ and\ \bibinfo {author} {\bibfnamefont {H.}~\bibnamefont {Harano}},\ }\bibfield  {title} {\bibinfo {title} {Emissions of {{Hydrogen Isotopes}} from the {{Nuclear Muon Capture Reaction}} in {\textsuperscript{nat}}{{Si}}},\ }\href {https://doi.org/10.1051/epjconf/202328401029} {\bibfield  {journal} {\bibinfo  {journal} {EPJ Web of Conferences}\ }\textbf {\bibinfo {volume} {284}},\ \bibinfo {pages} {01029} (\bibinfo {year} {2023})}\BibitemShut {NoStop}%
\bibitem [{\citenamefont {Sierawski}\ \emph {et~al.}(2014)\citenamefont {Sierawski}, \citenamefont {Bhuva}, \citenamefont {Reed}, \citenamefont {Tam}, \citenamefont {Narasimham}, \citenamefont {Ishida}, \citenamefont {Hillier}, \citenamefont {Trinczek}, \citenamefont {Blackmore}, \citenamefont {Wen},\ and\ \citenamefont {Wong}}]{Sierawski2014}%
  \BibitemOpen
  \bibfield  {author} {\bibinfo {author} {\bibfnamefont {B.~D.}\ \bibnamefont {Sierawski}}, \bibinfo {author} {\bibfnamefont {B.}~\bibnamefont {Bhuva}}, \bibinfo {author} {\bibfnamefont {R.}~\bibnamefont {Reed}}, \bibinfo {author} {\bibfnamefont {N.}~\bibnamefont {Tam}}, \bibinfo {author} {\bibfnamefont {B.}~\bibnamefont {Narasimham}}, \bibinfo {author} {\bibfnamefont {K.}~\bibnamefont {Ishida}}, \bibinfo {author} {\bibfnamefont {A.}~\bibnamefont {Hillier}}, \bibinfo {author} {\bibfnamefont {M.}~\bibnamefont {Trinczek}}, \bibinfo {author} {\bibfnamefont {E.}~\bibnamefont {Blackmore}}, \bibinfo {author} {\bibfnamefont {S.-J.}\ \bibnamefont {Wen}},\ and\ \bibinfo {author} {\bibfnamefont {R.}~\bibnamefont {Wong}},\ }\bibfield  {title} {\bibinfo {title} {Bias dependence of muon-induced single event upsets in 28 nm static random access memories},\ }in\ \href {https://doi.org/10.1109/IRPS.2014.6860585} {\emph {\bibinfo {booktitle} {2014 IEEE International Reliability Physics Symposium}}}\ (\bibinfo {year} {2014})\ pp.\ \bibinfo {pages} {2B.2.1--2B.2.5}\BibitemShut {NoStop}%
\bibitem [{\citenamefont {Trippe}\ \emph {et~al.}(2016)\citenamefont {Trippe}, \citenamefont {Reed}, \citenamefont {Sierawski}, \citenamefont {Weller}, \citenamefont {Austin}, \citenamefont {Massengill}, \citenamefont {Bhuva}, \citenamefont {Warren},\ and\ \citenamefont {Narasimham}}]{Trippe2016}%
  \BibitemOpen
  \bibfield  {author} {\bibinfo {author} {\bibfnamefont {J.~M.}\ \bibnamefont {Trippe}}, \bibinfo {author} {\bibfnamefont {R.~A.}\ \bibnamefont {Reed}}, \bibinfo {author} {\bibfnamefont {B.~D.}\ \bibnamefont {Sierawski}}, \bibinfo {author} {\bibfnamefont {R.~A.}\ \bibnamefont {Weller}}, \bibinfo {author} {\bibfnamefont {R.~A.}\ \bibnamefont {Austin}}, \bibinfo {author} {\bibfnamefont {L.~W.}\ \bibnamefont {Massengill}}, \bibinfo {author} {\bibfnamefont {B.~L.}\ \bibnamefont {Bhuva}}, \bibinfo {author} {\bibfnamefont {K.~M.}\ \bibnamefont {Warren}},\ and\ \bibinfo {author} {\bibfnamefont {B.}~\bibnamefont {Narasimham}},\ }\bibfield  {title} {\bibinfo {title} {Predicting the vulnerability of memories to muon-induced {{SEUs}} with low-energy proton tests informed by {{Monte Carlo}} simulations},\ }in\ \href {https://doi.org/10.1109/IRPS.2016.7574642} {\emph {\bibinfo {booktitle} {2016 {{IEEE International Reliability Physics Symposium}} ({{IRPS}})}}}\ (\bibinfo {year} {2016})\ pp.\ \bibinfo {pages} {SE--6--1--SE--6--6}\BibitemShut {NoStop}%
\bibitem [{\citenamefont {Manabe}\ \emph {et~al.}(2018)\citenamefont {Manabe}, \citenamefont {Watanabe}, \citenamefont {Liao}, \citenamefont {Hashimoto}, \citenamefont {Nakano}, \citenamefont {Sato}, \citenamefont {Kin}, \citenamefont {Abe}, \citenamefont {Hamada}, \citenamefont {Tampo},\ and\ \citenamefont {Miyake}}]{Manabe2018}%
  \BibitemOpen
  \bibfield  {author} {\bibinfo {author} {\bibfnamefont {S.}~\bibnamefont {Manabe}}, \bibinfo {author} {\bibfnamefont {Y.}~\bibnamefont {Watanabe}}, \bibinfo {author} {\bibfnamefont {W.}~\bibnamefont {Liao}}, \bibinfo {author} {\bibfnamefont {M.}~\bibnamefont {Hashimoto}}, \bibinfo {author} {\bibfnamefont {K.}~\bibnamefont {Nakano}}, \bibinfo {author} {\bibfnamefont {H.}~\bibnamefont {Sato}}, \bibinfo {author} {\bibfnamefont {T.}~\bibnamefont {Kin}}, \bibinfo {author} {\bibfnamefont {S.~I.}\ \bibnamefont {Abe}}, \bibinfo {author} {\bibfnamefont {K.}~\bibnamefont {Hamada}}, \bibinfo {author} {\bibfnamefont {M.}~\bibnamefont {Tampo}},\ and\ \bibinfo {author} {\bibfnamefont {Y.}~\bibnamefont {Miyake}},\ }\bibfield  {title} {\bibinfo {title} {Negative and {{Positive Muon-Induced Single Event Upsets}} in 65-nm {{UTBB SOI SRAMs}}},\ }\href {https://doi.org/10.1109/TNS.2018.2839704} {\bibfield  {journal} {\bibinfo  {journal} {IEEE Transactions on Nuclear Science}\ }\textbf {\bibinfo {volume} {65}},\ \bibinfo {pages} {1742} (\bibinfo {year} {2018})}\BibitemShut {NoStop}%
\bibitem [{\citenamefont {Liao}\ \emph {et~al.}(2019)\citenamefont {Liao}, \citenamefont {Hashimoto}, \citenamefont {Manabe}, \citenamefont {Watanabe}, \citenamefont {Abe}, \citenamefont {Nakano}, \citenamefont {Takeshita}, \citenamefont {Tampo}, \citenamefont {Takeshita},\ and\ \citenamefont {Miyake}}]{Liao2019}%
  \BibitemOpen
  \bibfield  {author} {\bibinfo {author} {\bibfnamefont {W.}~\bibnamefont {Liao}}, \bibinfo {author} {\bibfnamefont {M.}~\bibnamefont {Hashimoto}}, \bibinfo {author} {\bibfnamefont {S.}~\bibnamefont {Manabe}}, \bibinfo {author} {\bibfnamefont {Y.}~\bibnamefont {Watanabe}}, \bibinfo {author} {\bibfnamefont {S.~I.}\ \bibnamefont {Abe}}, \bibinfo {author} {\bibfnamefont {K.}~\bibnamefont {Nakano}}, \bibinfo {author} {\bibfnamefont {H.}~\bibnamefont {Takeshita}}, \bibinfo {author} {\bibfnamefont {M.}~\bibnamefont {Tampo}}, \bibinfo {author} {\bibfnamefont {S.}~\bibnamefont {Takeshita}},\ and\ \bibinfo {author} {\bibfnamefont {Y.}~\bibnamefont {Miyake}},\ }\bibfield  {title} {\bibinfo {title} {Negative and {{Positive Muon-Induced SEU Cross Sections}} in 28-nm and 65-nm {{Planar Bulk CMOS SRAMs}}},\ }\href {https://doi.org/10.1109/IRPS.2019.8720568} {\bibfield  {journal} {\bibinfo  {journal} {IEEE International Reliability Physics Symposium Proceedings}\ }\textbf {\bibinfo {volume} {2019-March}},\ \bibinfo {pages} {1} (\bibinfo {year} {2019})}\BibitemShut {NoStop}%
\bibitem [{\citenamefont {Autran}\ and\ \citenamefont {Munteanu}(2024)}]{Autran2024}%
  \BibitemOpen
  \bibfield  {author} {\bibinfo {author} {\bibfnamefont {J.-L.}\ \bibnamefont {Autran}}\ and\ \bibinfo {author} {\bibfnamefont {D.}~\bibnamefont {Munteanu}},\ }\bibfield  {title} {\bibinfo {title} {Interactions of {{Low-Energy Muons}} with {{Silicon}}: {{Numerical Simulation}} of {{Negative Muon Capture}} and {{Prospects}} for {{Soft Errors}}},\ }\href {https://doi.org/10.3390/jne5010007} {\bibfield  {journal} {\bibinfo  {journal} {Journal of Nuclear Engineering}\ }\textbf {\bibinfo {volume} {5}},\ \bibinfo {pages} {91} (\bibinfo {year} {2024})}\BibitemShut {NoStop}%
\bibitem [{\citenamefont {Fabero}\ \emph {et~al.}(2026)\citenamefont {Fabero}, \citenamefont {Franco}, \citenamefont {Mecha}, \citenamefont {Rezaei}, \citenamefont {Hillier}, \citenamefont {Lord}, \citenamefont {Cottrell}, \citenamefont {Colangeli}, \citenamefont {Fonnesu}, \citenamefont {Pagano},\ and\ \citenamefont {Clemente}}]{Fabero2026}%
  \BibitemOpen
  \bibfield  {author} {\bibinfo {author} {\bibfnamefont {J.~C.}\ \bibnamefont {Fabero}}, \bibinfo {author} {\bibfnamefont {F.~J.}\ \bibnamefont {Franco}}, \bibinfo {author} {\bibfnamefont {H.}~\bibnamefont {Mecha}}, \bibinfo {author} {\bibfnamefont {M.}~\bibnamefont {Rezaei}}, \bibinfo {author} {\bibfnamefont {A.~D.}\ \bibnamefont {Hillier}}, \bibinfo {author} {\bibfnamefont {J.~S.}\ \bibnamefont {Lord}}, \bibinfo {author} {\bibfnamefont {S.~P.}\ \bibnamefont {Cottrell}}, \bibinfo {author} {\bibfnamefont {A.}~\bibnamefont {Colangeli}}, \bibinfo {author} {\bibfnamefont {N.}~\bibnamefont {Fonnesu}}, \bibinfo {author} {\bibfnamefont {G.}~\bibnamefont {Pagano}},\ and\ \bibinfo {author} {\bibfnamefont {J.~A.}\ \bibnamefont {Clemente}},\ }\bibfield  {title} {\bibinfo {title} {Static and {{Dynamic Tests}} on a 40-nm {{Commercial SRAM}} under {{Muons}} and {{Neutrons}}},\ }\href {https://doi.org/10.1109/TNS.2026.3652620} {\bibfield  {journal} {\bibinfo  {journal} {IEEE Transactions on Nuclear Science}\ ,\ \bibinfo {pages} {1}} (\bibinfo {year} {2026})}\BibitemShut {NoStop}%
\bibitem [{\citenamefont {Matsuzaki}\ \emph {et~al.}(2001)\citenamefont {Matsuzaki}, \citenamefont {Ishida}, \citenamefont {Nagamine}, \citenamefont {Watanabe}, \citenamefont {Eaton},\ and\ \citenamefont {Williams}}]{Matsuzaki2001}%
  \BibitemOpen
  \bibfield  {author} {\bibinfo {author} {\bibfnamefont {T.}~\bibnamefont {Matsuzaki}}, \bibinfo {author} {\bibfnamefont {K.}~\bibnamefont {Ishida}}, \bibinfo {author} {\bibfnamefont {K.}~\bibnamefont {Nagamine}}, \bibinfo {author} {\bibfnamefont {I.}~\bibnamefont {Watanabe}}, \bibinfo {author} {\bibfnamefont {G.}~\bibnamefont {Eaton}},\ and\ \bibinfo {author} {\bibfnamefont {W.}~\bibnamefont {Williams}},\ }\bibfield  {title} {\bibinfo {title} {The {{RIKEN-RAL}} pulsed {{Muon Facility}}},\ }\href {https://doi.org/10.1016/S0168-9002(01)00694-5} {\bibfield  {journal} {\bibinfo  {journal} {Nuclear Instruments and Methods in Physics Research Section A: Accelerators, Spectrometers, Detectors and Associated Equipment}\ }\textbf {\bibinfo {volume} {465}},\ \bibinfo {pages} {365} (\bibinfo {year} {2001})}\BibitemShut {NoStop}%
\bibitem [{\citenamefont {Hillier}\ \emph {et~al.}(2018)\citenamefont {Hillier}, \citenamefont {Lord}, \citenamefont {Ishida},\ and\ \citenamefont {Rogers}}]{Hillier2018}%
  \BibitemOpen
  \bibfield  {author} {\bibinfo {author} {\bibfnamefont {A.~D.}\ \bibnamefont {Hillier}}, \bibinfo {author} {\bibfnamefont {J.~S.}\ \bibnamefont {Lord}}, \bibinfo {author} {\bibfnamefont {K.}~\bibnamefont {Ishida}},\ and\ \bibinfo {author} {\bibfnamefont {C.}~\bibnamefont {Rogers}},\ }\bibfield  {title} {\bibinfo {title} {Muons at {{ISIS}}},\ }\href {https://doi.org/10.1098/rsta.2018.0064} {\bibfield  {journal} {\bibinfo  {journal} {Philosophical Transactions of the Royal Society A: Mathematical, Physical and Engineering Sciences}\ }\textbf {\bibinfo {volume} {377}},\ \bibinfo {pages} {20180064} (\bibinfo {year} {2018})}\BibitemShut {NoStop}%
\bibitem [{\citenamefont {Roberts}\ and\ \citenamefont {Kaplan}(2007)}]{Roberts2007}%
  \BibitemOpen
  \bibfield  {author} {\bibinfo {author} {\bibfnamefont {T.~J.}\ \bibnamefont {Roberts}}\ and\ \bibinfo {author} {\bibfnamefont {D.~M.}\ \bibnamefont {Kaplan}},\ }\bibfield  {title} {\bibinfo {title} {G4beamline simulation program for matter-dominated beamlines},\ }in\ \href {https://doi.org/10.1109/PAC.2007.4440461} {\emph {\bibinfo {booktitle} {2007 {{IEEE Particle Accelerator Conference}} ({{PAC}})}}}\ (\bibinfo {year} {2007})\ pp.\ \bibinfo {pages} {3468--3470}\BibitemShut {NoStop}%
\bibitem [{\citenamefont {Kawase}\ \emph {et~al.}(2024)\citenamefont {Kawase}, \citenamefont {Murota}, \citenamefont {Fukuda}, \citenamefont {Oishi}, \citenamefont {Kawata}, \citenamefont {Kitafuji}, \citenamefont {Manabe}, \citenamefont {Watanabe}, \citenamefont {Nishibata}, \citenamefont {Go}, \citenamefont {Kai}, \citenamefont {Nagata}, \citenamefont {Muto}, \citenamefont {Ishibashi}, \citenamefont {Niikura}, \citenamefont {Suzuki}, \citenamefont {Matsuzaki}, \citenamefont {Ishida}, \citenamefont {Mizuno},\ and\ \citenamefont {Kitamura}}]{Kawase2024}%
  \BibitemOpen
  \bibfield  {author} {\bibinfo {author} {\bibfnamefont {S.}~\bibnamefont {Kawase}}, \bibinfo {author} {\bibfnamefont {T.}~\bibnamefont {Murota}}, \bibinfo {author} {\bibfnamefont {H.}~\bibnamefont {Fukuda}}, \bibinfo {author} {\bibfnamefont {M.}~\bibnamefont {Oishi}}, \bibinfo {author} {\bibfnamefont {T.}~\bibnamefont {Kawata}}, \bibinfo {author} {\bibfnamefont {K.}~\bibnamefont {Kitafuji}}, \bibinfo {author} {\bibfnamefont {S.}~\bibnamefont {Manabe}}, \bibinfo {author} {\bibfnamefont {Y.}~\bibnamefont {Watanabe}}, \bibinfo {author} {\bibfnamefont {H.}~\bibnamefont {Nishibata}}, \bibinfo {author} {\bibfnamefont {S.}~\bibnamefont {Go}}, \bibinfo {author} {\bibfnamefont {T.}~\bibnamefont {Kai}}, \bibinfo {author} {\bibfnamefont {Y.}~\bibnamefont {Nagata}}, \bibinfo {author} {\bibfnamefont {T.}~\bibnamefont {Muto}}, \bibinfo {author} {\bibfnamefont {Y.}~\bibnamefont {Ishibashi}}, \bibinfo {author} {\bibfnamefont {M.}~\bibnamefont {Niikura}}, \bibinfo {author} {\bibfnamefont {D.}~\bibnamefont {Suzuki}}, \bibinfo {author} {\bibfnamefont {T.}~\bibnamefont {Matsuzaki}}, \bibinfo {author} {\bibfnamefont {K.}~\bibnamefont {Ishida}}, \bibinfo {author} {\bibfnamefont {R.}~\bibnamefont {Mizuno}},\ and\ \bibinfo {author} {\bibfnamefont {N.}~\bibnamefont {Kitamura}},\ }\bibfield  {title} {\bibinfo {title} {Effect of large-angle incidence on particle identification performance for light-charged ( {{Z}} {$\leq$} 2 ) particles by pulse shape analysis with a pad-type {{nTD}} silicon detector},\ }\href {https://doi.org/10.1016/j.nima.2023.168984} {\bibfield  {journal} {\bibinfo  {journal} {Nuclear Instruments and Methods in Physics Research Section A: Accelerators, Spectrometers, Detectors and Associated Equipment}\ }\textbf {\bibinfo {volume} {1059}},\ \bibinfo {pages} {168984} (\bibinfo {year} {2024})}\BibitemShut {NoStop}%
\bibitem [{\citenamefont {{{CAEN S.p.A.}}}(2025)}]{compass}%
  \BibitemOpen
  \bibfield  {author} {\bibinfo {author} {\bibnamefont {{{CAEN S.p.A.}}}},\ }\href@noop {} {\emph {\bibinfo {title} {CoMPASS User Manual Rev. 25}}} (\bibinfo {year} {2025})\BibitemShut {NoStop}%
\bibitem [{\citenamefont {Suzuki}\ \emph {et~al.}(1987)\citenamefont {Suzuki}, \citenamefont {Measday},\ and\ \citenamefont {Roalsvig}}]{Suzuki1987}%
  \BibitemOpen
  \bibfield  {author} {\bibinfo {author} {\bibfnamefont {T.}~\bibnamefont {Suzuki}}, \bibinfo {author} {\bibfnamefont {D.~F.}\ \bibnamefont {Measday}},\ and\ \bibinfo {author} {\bibfnamefont {J.~P.}\ \bibnamefont {Roalsvig}},\ }\bibfield  {title} {\bibinfo {title} {Total nuclear capture rates for negative muons},\ }\href {https://doi.org/10.1103/PhysRevC.35.2212} {\bibfield  {journal} {\bibinfo  {journal} {Physical Review C}\ }\textbf {\bibinfo {volume} {35}},\ \bibinfo {pages} {2212} (\bibinfo {year} {1987})}\BibitemShut {NoStop}%
\bibitem [{\citenamefont {Agostinelli}\ \emph {et~al.}(2003)\citenamefont {Agostinelli}, \citenamefont {Allison}, \citenamefont {Amako}, \citenamefont {Apostolakis}, \citenamefont {Araujo}, \citenamefont {Arce}, \citenamefont {Asai}, \citenamefont {Axen}, \citenamefont {Banerjee}, \citenamefont {Barrand}, \citenamefont {Behner}, \citenamefont {Bellagamba}, \citenamefont {Boudreau}, \citenamefont {Broglia}, \citenamefont {Brunengo}, \citenamefont {Burkhardt}, \citenamefont {Chauvie}, \citenamefont {Chuma}, \citenamefont {Chytracek}, \citenamefont {Cooperman}, \citenamefont {Cosmo}, \citenamefont {Degtyarenko}, \citenamefont {Dell'Acqua}, \citenamefont {Depaola}, \citenamefont {Dietrich}, \citenamefont {Enami}, \citenamefont {Feliciello}, \citenamefont {Ferguson}, \citenamefont {Fesefeldt}, \citenamefont {Folger}, \citenamefont {Foppiano}, \citenamefont {Forti}, \citenamefont {Garelli}, \citenamefont {Giani}, \citenamefont {Giannitrapani}, \citenamefont {Gibin}, \citenamefont {G{\'o}mez~Cadenas}, \citenamefont {Gonz{\'a}lez}, \citenamefont {Gracia~Abril}, \citenamefont {Greeniaus}, \citenamefont {Greiner}, \citenamefont {Grichine}, \citenamefont {Grossheim}, \citenamefont {Guatelli}, \citenamefont {Gumplinger}, \citenamefont {Hamatsu}, \citenamefont {Hashimoto}, \citenamefont {Hasui}, \citenamefont {Heikkinen}, \citenamefont {Howard}, \citenamefont {Ivanchenko}, \citenamefont {Johnson}, \citenamefont {Jones}, \citenamefont {Kallenbach}, \citenamefont {Kanaya}, \citenamefont {Kawabata}, \citenamefont {Kawabata}, \citenamefont {Kawaguti}, \citenamefont {Kelner}, \citenamefont {Kent}, \citenamefont {Kimura}, \citenamefont {Kodama}, \citenamefont {Kokoulin}, \citenamefont {Kossov}, \citenamefont {Kurashige}, \citenamefont {Lamanna}, \citenamefont {Lamp{\'e}n}, \citenamefont {Lara}, \citenamefont {Lefebure}, \citenamefont {Lei}, \citenamefont {Liendl}, \citenamefont {Lockman}, \citenamefont {Longo}, \citenamefont {Magni}, \citenamefont {Maire}, \citenamefont {Medernach}, \citenamefont {Minamimoto}, \citenamefont {Mora De~Freitas}, \citenamefont {Morita}, \citenamefont {Murakami}, \citenamefont {Nagamatu}, \citenamefont {Nartallo}, \citenamefont {Nieminen}, \citenamefont {Nishimura}, \citenamefont {Ohtsubo}, \citenamefont {Okamura}, \citenamefont {O'Neale}, \citenamefont {Oohata}, \citenamefont {Paech}, \citenamefont {Perl}, \citenamefont {Pfeiffer}, \citenamefont {Pia}, \citenamefont {Ranjard}, \citenamefont {Rybin}, \citenamefont {Sadilov}, \citenamefont {Di~Salvo}, \citenamefont {Santin}, \citenamefont {Sasaki}, \citenamefont {Savvas}, \citenamefont {Sawada}, \citenamefont {Scherer}, \citenamefont {Sei}, \citenamefont {Sirotenko}, \citenamefont {Smith}, \citenamefont {Starkov}, \citenamefont {Stoecker}, \citenamefont {Sulkimo}, \citenamefont {Takahata}, \citenamefont {Tanaka}, \citenamefont {Tcherniaev}, \citenamefont {Safai~Tehrani}, \citenamefont {Tropeano}, \citenamefont {Truscott}, \citenamefont {Uno}, \citenamefont {Urban}, \citenamefont {Urban}, \citenamefont {Verderi}, \citenamefont {Walkden}, \citenamefont {Wander}, \citenamefont {Weber}, \citenamefont {Wellisch}, \citenamefont {Wenaus}, \citenamefont {Williams}, \citenamefont {Wright}, \citenamefont {Yamada}, \citenamefont {Yoshida},\ and\ \citenamefont {Zschiesche}}]{Agostinelli2003}%
  \BibitemOpen
  \bibfield  {author} {\bibinfo {author} {\bibfnamefont {S.}~\bibnamefont {Agostinelli}}, \bibinfo {author} {\bibfnamefont {J.}~\bibnamefont {Allison}}, \bibinfo {author} {\bibfnamefont {K.}~\bibnamefont {Amako}}, \bibinfo {author} {\bibfnamefont {J.}~\bibnamefont {Apostolakis}}, \bibinfo {author} {\bibfnamefont {H.}~\bibnamefont {Araujo}}, \bibinfo {author} {\bibfnamefont {P.}~\bibnamefont {Arce}}, \bibinfo {author} {\bibfnamefont {M.}~\bibnamefont {Asai}}, \bibinfo {author} {\bibfnamefont {D.}~\bibnamefont {Axen}}, \bibinfo {author} {\bibfnamefont {S.}~\bibnamefont {Banerjee}}, \bibinfo {author} {\bibfnamefont {G.}~\bibnamefont {Barrand}}, \bibinfo {author} {\bibfnamefont {F.}~\bibnamefont {Behner}}, \bibinfo {author} {\bibfnamefont {L.}~\bibnamefont {Bellagamba}}, \bibinfo {author} {\bibfnamefont {J.}~\bibnamefont {Boudreau}}, \bibinfo {author} {\bibfnamefont {L.}~\bibnamefont {Broglia}}, \bibinfo {author} {\bibfnamefont {A.}~\bibnamefont {Brunengo}}, \bibinfo {author} {\bibfnamefont {H.}~\bibnamefont {Burkhardt}}, \bibinfo {author} {\bibfnamefont {S.}~\bibnamefont {Chauvie}}, \bibinfo {author} {\bibfnamefont {J.}~\bibnamefont {Chuma}}, \bibinfo {author} {\bibfnamefont {R.}~\bibnamefont {Chytracek}}, \bibinfo {author} {\bibfnamefont {G.}~\bibnamefont {Cooperman}}, \bibinfo {author} {\bibfnamefont {G.}~\bibnamefont {Cosmo}}, \bibinfo {author} {\bibfnamefont {P.}~\bibnamefont {Degtyarenko}}, \bibinfo {author} {\bibfnamefont {A.}~\bibnamefont {Dell'Acqua}}, \bibinfo {author} {\bibfnamefont {G.}~\bibnamefont {Depaola}}, \bibinfo {author} {\bibfnamefont {D.}~\bibnamefont {Dietrich}}, \bibinfo {author} {\bibfnamefont {R.}~\bibnamefont {Enami}}, \bibinfo {author} {\bibfnamefont {A.}~\bibnamefont {Feliciello}}, \bibinfo {author} {\bibfnamefont {C.}~\bibnamefont {Ferguson}}, \bibinfo {author} {\bibfnamefont {H.}~\bibnamefont {Fesefeldt}}, \bibinfo {author} {\bibfnamefont {G.}~\bibnamefont {Folger}}, \bibinfo {author} {\bibfnamefont {F.}~\bibnamefont {Foppiano}}, \bibinfo {author} {\bibfnamefont {A.}~\bibnamefont {Forti}}, \bibinfo {author} {\bibfnamefont {S.}~\bibnamefont {Garelli}}, \bibinfo {author} {\bibfnamefont {S.}~\bibnamefont {Giani}}, \bibinfo {author} {\bibfnamefont {R.}~\bibnamefont {Giannitrapani}}, \bibinfo {author} {\bibfnamefont {D.}~\bibnamefont {Gibin}}, \bibinfo {author} {\bibfnamefont {J.}~\bibnamefont {G{\'o}mez~Cadenas}}, \bibinfo {author} {\bibfnamefont {I.}~\bibnamefont {Gonz{\'a}lez}}, \bibinfo {author} {\bibfnamefont {G.}~\bibnamefont {Gracia~Abril}}, \bibinfo {author} {\bibfnamefont {G.}~\bibnamefont {Greeniaus}}, \bibinfo {author} {\bibfnamefont {W.}~\bibnamefont {Greiner}}, \bibinfo {author} {\bibfnamefont {V.}~\bibnamefont {Grichine}}, \bibinfo {author} {\bibfnamefont {A.}~\bibnamefont {Grossheim}}, \bibinfo {author} {\bibfnamefont {S.}~\bibnamefont {Guatelli}}, \bibinfo {author} {\bibfnamefont {P.}~\bibnamefont {Gumplinger}}, \bibinfo {author} {\bibfnamefont {R.}~\bibnamefont {Hamatsu}}, \bibinfo {author} {\bibfnamefont {K.}~\bibnamefont {Hashimoto}}, \bibinfo {author} {\bibfnamefont {H.}~\bibnamefont {Hasui}}, \bibinfo {author} {\bibfnamefont {A.}~\bibnamefont {Heikkinen}}, \bibinfo {author} {\bibfnamefont {A.}~\bibnamefont {Howard}}, \bibinfo {author} {\bibfnamefont {V.}~\bibnamefont {Ivanchenko}}, \bibinfo {author} {\bibfnamefont {A.}~\bibnamefont {Johnson}}, \bibinfo {author} {\bibfnamefont {F.}~\bibnamefont {Jones}}, \bibinfo {author} {\bibfnamefont {J.}~\bibnamefont {Kallenbach}}, \bibinfo {author} {\bibfnamefont {N.}~\bibnamefont {Kanaya}}, \bibinfo {author} {\bibfnamefont {M.}~\bibnamefont {Kawabata}}, \bibinfo {author} {\bibfnamefont {Y.}~\bibnamefont {Kawabata}}, \bibinfo {author} {\bibfnamefont {M.}~\bibnamefont {Kawaguti}}, \bibinfo {author} {\bibfnamefont {S.}~\bibnamefont {Kelner}}, \bibinfo {author} {\bibfnamefont {P.}~\bibnamefont {Kent}}, \bibinfo {author} {\bibfnamefont {A.}~\bibnamefont {Kimura}}, \bibinfo {author} {\bibfnamefont {T.}~\bibnamefont {Kodama}}, \bibinfo {author} {\bibfnamefont {R.}~\bibnamefont {Kokoulin}}, \bibinfo {author} {\bibfnamefont {M.}~\bibnamefont {Kossov}}, \bibinfo {author} {\bibfnamefont {H.}~\bibnamefont {Kurashige}}, \bibinfo {author} {\bibfnamefont {E.}~\bibnamefont {Lamanna}}, \bibinfo {author} {\bibfnamefont {T.}~\bibnamefont {Lamp{\'e}n}}, \bibinfo {author} {\bibfnamefont {V.}~\bibnamefont {Lara}}, \bibinfo {author} {\bibfnamefont {V.}~\bibnamefont {Lefebure}}, \bibinfo {author} {\bibfnamefont {F.}~\bibnamefont {Lei}}, \bibinfo {author} {\bibfnamefont {M.}~\bibnamefont {Liendl}}, \bibinfo {author} {\bibfnamefont {W.}~\bibnamefont {Lockman}}, \bibinfo {author} {\bibfnamefont {F.}~\bibnamefont {Longo}}, \bibinfo {author} {\bibfnamefont {S.}~\bibnamefont {Magni}}, \bibinfo {author} {\bibfnamefont {M.}~\bibnamefont {Maire}}, \bibinfo {author} {\bibfnamefont {E.}~\bibnamefont {Medernach}}, \bibinfo {author} {\bibfnamefont {K.}~\bibnamefont {Minamimoto}}, \bibinfo {author} {\bibfnamefont {P.}~\bibnamefont {Mora De~Freitas}}, \bibinfo {author} {\bibfnamefont {Y.}~\bibnamefont {Morita}}, \bibinfo {author} {\bibfnamefont {K.}~\bibnamefont {Murakami}}, \bibinfo {author} {\bibfnamefont {M.}~\bibnamefont {Nagamatu}}, \bibinfo {author} {\bibfnamefont {R.}~\bibnamefont {Nartallo}}, \bibinfo {author} {\bibfnamefont {P.}~\bibnamefont {Nieminen}}, \bibinfo {author} {\bibfnamefont {T.}~\bibnamefont {Nishimura}}, \bibinfo {author} {\bibfnamefont {K.}~\bibnamefont {Ohtsubo}}, \bibinfo {author} {\bibfnamefont {M.}~\bibnamefont {Okamura}}, \bibinfo {author} {\bibfnamefont {S.}~\bibnamefont {O'Neale}}, \bibinfo {author} {\bibfnamefont {Y.}~\bibnamefont {Oohata}}, \bibinfo {author} {\bibfnamefont {K.}~\bibnamefont {Paech}}, \bibinfo {author} {\bibfnamefont {J.}~\bibnamefont {Perl}}, \bibinfo {author} {\bibfnamefont {A.}~\bibnamefont {Pfeiffer}}, \bibinfo {author} {\bibfnamefont {M.}~\bibnamefont {Pia}}, \bibinfo {author} {\bibfnamefont {F.}~\bibnamefont {Ranjard}}, \bibinfo {author} {\bibfnamefont {A.}~\bibnamefont {Rybin}}, \bibinfo {author} {\bibfnamefont {S.}~\bibnamefont {Sadilov}}, \bibinfo {author} {\bibfnamefont {E.}~\bibnamefont {Di~Salvo}}, \bibinfo {author} {\bibfnamefont {G.}~\bibnamefont {Santin}}, \bibinfo {author} {\bibfnamefont {T.}~\bibnamefont {Sasaki}}, \bibinfo {author} {\bibfnamefont {N.}~\bibnamefont {Savvas}}, \bibinfo {author} {\bibfnamefont {Y.}~\bibnamefont {Sawada}}, \bibinfo {author} {\bibfnamefont {S.}~\bibnamefont {Scherer}}, \bibinfo {author} {\bibfnamefont {S.}~\bibnamefont {Sei}}, \bibinfo {author} {\bibfnamefont {V.}~\bibnamefont {Sirotenko}}, \bibinfo {author} {\bibfnamefont {D.}~\bibnamefont {Smith}}, \bibinfo {author} {\bibfnamefont {N.}~\bibnamefont {Starkov}}, \bibinfo {author} {\bibfnamefont {H.}~\bibnamefont {Stoecker}}, \bibinfo {author} {\bibfnamefont {J.}~\bibnamefont {Sulkimo}}, \bibinfo {author} {\bibfnamefont {M.}~\bibnamefont {Takahata}}, \bibinfo {author} {\bibfnamefont {S.}~\bibnamefont {Tanaka}}, \bibinfo {author} {\bibfnamefont {E.}~\bibnamefont {Tcherniaev}}, \bibinfo {author} {\bibfnamefont {E.}~\bibnamefont {Safai~Tehrani}}, \bibinfo {author} {\bibfnamefont {M.}~\bibnamefont {Tropeano}}, \bibinfo {author} {\bibfnamefont {P.}~\bibnamefont {Truscott}}, \bibinfo {author} {\bibfnamefont {H.}~\bibnamefont {Uno}}, \bibinfo {author} {\bibfnamefont {L.}~\bibnamefont {Urban}}, \bibinfo {author} {\bibfnamefont {P.}~\bibnamefont {Urban}}, \bibinfo {author} {\bibfnamefont {M.}~\bibnamefont {Verderi}}, \bibinfo {author} {\bibfnamefont {A.}~\bibnamefont {Walkden}}, \bibinfo {author} {\bibfnamefont {W.}~\bibnamefont {Wander}}, \bibinfo {author} {\bibfnamefont {H.}~\bibnamefont {Weber}}, \bibinfo {author} {\bibfnamefont {J.}~\bibnamefont {Wellisch}}, \bibinfo {author} {\bibfnamefont {T.}~\bibnamefont {Wenaus}}, \bibinfo {author} {\bibfnamefont {D.}~\bibnamefont {Williams}}, \bibinfo {author} {\bibfnamefont {D.}~\bibnamefont {Wright}}, \bibinfo {author} {\bibfnamefont {T.}~\bibnamefont {Yamada}}, \bibinfo {author} {\bibfnamefont {H.}~\bibnamefont {Yoshida}},\ and\ \bibinfo {author} {\bibfnamefont {D.}~\bibnamefont {Zschiesche}},\ }\bibfield  {title} {\bibinfo {title} {Geant4---a simulation toolkit},\ }\href {https://doi.org/10.1016/S0168-9002(03)01368-8} {\bibfield  {journal} {\bibinfo  {journal} {Nuclear Instruments and Methods in Physics Research Section A: Accelerators, Spectrometers, Detectors and Associated Equipment}\ }\textbf {\bibinfo {volume} {506}},\ \bibinfo {pages} {250} (\bibinfo {year} {2003})}\BibitemShut {NoStop}%
\bibitem [{\citenamefont {Allison}\ \emph {et~al.}(2006)\citenamefont {Allison}, \citenamefont {Amako}, \citenamefont {Apostolakis}, \citenamefont {Araujo}, \citenamefont {Arce~Dubois}, \citenamefont {Asai}, \citenamefont {Barrand}, \citenamefont {Capra}, \citenamefont {Chauvie}, \citenamefont {Chytracek}, \citenamefont {Cirrone}, \citenamefont {Cooperman}, \citenamefont {Cosmo}, \citenamefont {Cuttone}, \citenamefont {Daquino}, \citenamefont {Donszelmann}, \citenamefont {Dressel}, \citenamefont {Folger}, \citenamefont {Foppiano}, \citenamefont {Generowicz}, \citenamefont {Grichine}, \citenamefont {Guatelli}, \citenamefont {Gumplinger}, \citenamefont {Heikkinen}, \citenamefont {Hrivnacova}, \citenamefont {Howard}, \citenamefont {Incerti}, \citenamefont {Ivanchenko}, \citenamefont {Johnson}, \citenamefont {Jones}, \citenamefont {Koi}, \citenamefont {Kokoulin}, \citenamefont {Kossov}, \citenamefont {Kurashige}, \citenamefont {Lara}, \citenamefont {Larsson}, \citenamefont {Lei}, \citenamefont {Link}, \citenamefont {Longo}, \citenamefont {Maire}, \citenamefont {Mantero}, \citenamefont {Mascialino}, \citenamefont {McLaren}, \citenamefont {Mendez~Lorenzo}, \citenamefont {Minamimoto}, \citenamefont {Murakami}, \citenamefont {Nieminen}, \citenamefont {Pandola}, \citenamefont {Parlati}, \citenamefont {Peralta}, \citenamefont {Perl}, \citenamefont {Pfeiffer}, \citenamefont {Pia}, \citenamefont {Ribon}, \citenamefont {Rodrigues}, \citenamefont {Russo}, \citenamefont {Sadilov}, \citenamefont {Santin}, \citenamefont {Sasaki}, \citenamefont {Smith}, \citenamefont {Starkov}, \citenamefont {Tanaka}, \citenamefont {Tcherniaev}, \citenamefont {Tome}, \citenamefont {Trindade}, \citenamefont {Truscott}, \citenamefont {Urban}, \citenamefont {Verderi}, \citenamefont {Walkden}, \citenamefont {Wellisch}, \citenamefont {Williams}, \citenamefont {Wright},\ and\ \citenamefont {Yoshida}}]{Allison2006}%
  \BibitemOpen
  \bibfield  {author} {\bibinfo {author} {\bibfnamefont {J.}~\bibnamefont {Allison}}, \bibinfo {author} {\bibfnamefont {K.}~\bibnamefont {Amako}}, \bibinfo {author} {\bibfnamefont {J.}~\bibnamefont {Apostolakis}}, \bibinfo {author} {\bibfnamefont {H.}~\bibnamefont {Araujo}}, \bibinfo {author} {\bibfnamefont {P.}~\bibnamefont {Arce~Dubois}}, \bibinfo {author} {\bibfnamefont {M.}~\bibnamefont {Asai}}, \bibinfo {author} {\bibfnamefont {G.}~\bibnamefont {Barrand}}, \bibinfo {author} {\bibfnamefont {R.}~\bibnamefont {Capra}}, \bibinfo {author} {\bibfnamefont {S.}~\bibnamefont {Chauvie}}, \bibinfo {author} {\bibfnamefont {R.}~\bibnamefont {Chytracek}}, \bibinfo {author} {\bibfnamefont {G.}~\bibnamefont {Cirrone}}, \bibinfo {author} {\bibfnamefont {G.}~\bibnamefont {Cooperman}}, \bibinfo {author} {\bibfnamefont {G.}~\bibnamefont {Cosmo}}, \bibinfo {author} {\bibfnamefont {G.}~\bibnamefont {Cuttone}}, \bibinfo {author} {\bibfnamefont {G.}~\bibnamefont {Daquino}}, \bibinfo {author} {\bibfnamefont {M.}~\bibnamefont {Donszelmann}}, \bibinfo {author} {\bibfnamefont {M.}~\bibnamefont {Dressel}}, \bibinfo {author} {\bibfnamefont {G.}~\bibnamefont {Folger}}, \bibinfo {author} {\bibfnamefont {F.}~\bibnamefont {Foppiano}}, \bibinfo {author} {\bibfnamefont {J.}~\bibnamefont {Generowicz}}, \bibinfo {author} {\bibfnamefont {V.}~\bibnamefont {Grichine}}, \bibinfo {author} {\bibfnamefont {S.}~\bibnamefont {Guatelli}}, \bibinfo {author} {\bibfnamefont {P.}~\bibnamefont {Gumplinger}}, \bibinfo {author} {\bibfnamefont {A.}~\bibnamefont {Heikkinen}}, \bibinfo {author} {\bibfnamefont {I.}~\bibnamefont {Hrivnacova}}, \bibinfo {author} {\bibfnamefont {A.}~\bibnamefont {Howard}}, \bibinfo {author} {\bibfnamefont {S.}~\bibnamefont {Incerti}}, \bibinfo {author} {\bibfnamefont {V.}~\bibnamefont {Ivanchenko}}, \bibinfo {author} {\bibfnamefont {T.}~\bibnamefont {Johnson}}, \bibinfo {author} {\bibfnamefont {F.}~\bibnamefont {Jones}}, \bibinfo {author} {\bibfnamefont {T.}~\bibnamefont {Koi}}, \bibinfo {author} {\bibfnamefont {R.}~\bibnamefont {Kokoulin}}, \bibinfo {author} {\bibfnamefont {M.}~\bibnamefont {Kossov}}, \bibinfo {author} {\bibfnamefont {H.}~\bibnamefont {Kurashige}}, \bibinfo {author} {\bibfnamefont {V.}~\bibnamefont {Lara}}, \bibinfo {author} {\bibfnamefont {S.}~\bibnamefont {Larsson}}, \bibinfo {author} {\bibfnamefont {F.}~\bibnamefont {Lei}}, \bibinfo {author} {\bibfnamefont {O.}~\bibnamefont {Link}}, \bibinfo {author} {\bibfnamefont {F.}~\bibnamefont {Longo}}, \bibinfo {author} {\bibfnamefont {M.}~\bibnamefont {Maire}}, \bibinfo {author} {\bibfnamefont {A.}~\bibnamefont {Mantero}}, \bibinfo {author} {\bibfnamefont {B.}~\bibnamefont {Mascialino}}, \bibinfo {author} {\bibfnamefont {I.}~\bibnamefont {McLaren}}, \bibinfo {author} {\bibfnamefont {P.}~\bibnamefont {Mendez~Lorenzo}}, \bibinfo {author} {\bibfnamefont {K.}~\bibnamefont {Minamimoto}}, \bibinfo {author} {\bibfnamefont {K.}~\bibnamefont {Murakami}}, \bibinfo {author} {\bibfnamefont {P.}~\bibnamefont {Nieminen}}, \bibinfo {author} {\bibfnamefont {L.}~\bibnamefont {Pandola}}, \bibinfo {author} {\bibfnamefont {S.}~\bibnamefont {Parlati}}, \bibinfo {author} {\bibfnamefont {L.}~\bibnamefont {Peralta}}, \bibinfo {author} {\bibfnamefont {J.}~\bibnamefont {Perl}}, \bibinfo {author} {\bibfnamefont {A.}~\bibnamefont {Pfeiffer}}, \bibinfo {author} {\bibfnamefont {M.}~\bibnamefont {Pia}}, \bibinfo {author} {\bibfnamefont {A.}~\bibnamefont {Ribon}}, \bibinfo {author} {\bibfnamefont {P.}~\bibnamefont {Rodrigues}}, \bibinfo {author} {\bibfnamefont {G.}~\bibnamefont {Russo}}, \bibinfo {author} {\bibfnamefont {S.}~\bibnamefont {Sadilov}}, \bibinfo {author} {\bibfnamefont {G.}~\bibnamefont {Santin}}, \bibinfo {author} {\bibfnamefont {T.}~\bibnamefont {Sasaki}}, \bibinfo {author} {\bibfnamefont {D.}~\bibnamefont {Smith}}, \bibinfo {author} {\bibfnamefont {N.}~\bibnamefont {Starkov}}, \bibinfo {author} {\bibfnamefont {S.}~\bibnamefont {Tanaka}}, \bibinfo {author} {\bibfnamefont {E.}~\bibnamefont {Tcherniaev}}, \bibinfo {author} {\bibfnamefont {B.}~\bibnamefont {Tome}}, \bibinfo {author} {\bibfnamefont {A.}~\bibnamefont {Trindade}}, \bibinfo {author} {\bibfnamefont {P.}~\bibnamefont {Truscott}}, \bibinfo {author} {\bibfnamefont {L.}~\bibnamefont {Urban}}, \bibinfo {author} {\bibfnamefont {M.}~\bibnamefont {Verderi}}, \bibinfo {author} {\bibfnamefont {A.}~\bibnamefont {Walkden}}, \bibinfo {author} {\bibfnamefont {J.}~\bibnamefont {Wellisch}}, \bibinfo {author} {\bibfnamefont {D.}~\bibnamefont {Williams}}, \bibinfo {author} {\bibfnamefont {D.}~\bibnamefont {Wright}},\ and\ \bibinfo {author} {\bibfnamefont {H.}~\bibnamefont {Yoshida}},\ }\bibfield  {title} {\bibinfo {title} {Geant4 developments and applications},\ }\href {https://doi.org/10.1109/TNS.2006.869826} {\bibfield  {journal} {\bibinfo  {journal} {IEEE Transactions on Nuclear Science}\ }\textbf {\bibinfo {volume} {53}},\ \bibinfo {pages} {270} (\bibinfo {year} {2006})}\BibitemShut {NoStop}%
\bibitem [{\citenamefont {Allison}\ \emph {et~al.}(2016)\citenamefont {Allison}, \citenamefont {Amako}, \citenamefont {Apostolakis}, \citenamefont {Arce}, \citenamefont {Asai}, \citenamefont {Aso}, \citenamefont {Bagli}, \citenamefont {Bagulya}, \citenamefont {Banerjee}, \citenamefont {Barrand}, \citenamefont {Beck}, \citenamefont {Bogdanov}, \citenamefont {Brandt}, \citenamefont {Brown}, \citenamefont {Burkhardt}, \citenamefont {Canal}, \citenamefont {{Cano-Ott}}, \citenamefont {Chauvie}, \citenamefont {Cho}, \citenamefont {Cirrone}, \citenamefont {Cooperman}, \citenamefont {{Cort{\'e}s-Giraldo}}, \citenamefont {Cosmo}, \citenamefont {Cuttone}, \citenamefont {Depaola}, \citenamefont {Desorgher}, \citenamefont {Dong}, \citenamefont {Dotti}, \citenamefont {Elvira}, \citenamefont {Folger}, \citenamefont {Francis}, \citenamefont {Galoyan}, \citenamefont {Garnier}, \citenamefont {Gayer}, \citenamefont {Genser}, \citenamefont {Grichine}, \citenamefont {Guatelli}, \citenamefont {Gu{\`e}ye}, \citenamefont {Gumplinger}, \citenamefont {Howard}, \citenamefont {H{\v r}ivn{\'a}{\v c}ov{\'a}}, \citenamefont {Hwang}, \citenamefont {Incerti}, \citenamefont {Ivanchenko}, \citenamefont {Ivanchenko}, \citenamefont {Jones}, \citenamefont {Jun}, \citenamefont {Kaitaniemi}, \citenamefont {Karakatsanis}, \citenamefont {Karamitros}, \citenamefont {Kelsey}, \citenamefont {Kimura}, \citenamefont {Koi}, \citenamefont {Kurashige}, \citenamefont {Lechner}, \citenamefont {Lee}, \citenamefont {Longo}, \citenamefont {Maire}, \citenamefont {Mancusi}, \citenamefont {Mantero}, \citenamefont {Mendoza}, \citenamefont {Morgan}, \citenamefont {Murakami}, \citenamefont {Nikitina}, \citenamefont {Pandola}, \citenamefont {Paprocki}, \citenamefont {Perl}, \citenamefont {Petrovi{\'c}}, \citenamefont {Pia}, \citenamefont {Pokorski}, \citenamefont {Quesada}, \citenamefont {Raine}, \citenamefont {Reis}, \citenamefont {Ribon}, \citenamefont {Risti{\'c}~Fira}, \citenamefont {Romano}, \citenamefont {Russo}, \citenamefont {Santin}, \citenamefont {Sasaki}, \citenamefont {Sawkey}, \citenamefont {Shin}, \citenamefont {Strakovsky}, \citenamefont {Taborda}, \citenamefont {Tanaka}, \citenamefont {Tom{\'e}}, \citenamefont {Toshito}, \citenamefont {Tran}, \citenamefont {Truscott}, \citenamefont {Urban}, \citenamefont {Uzhinsky}, \citenamefont {Verbeke}, \citenamefont {Verderi}, \citenamefont {Wendt}, \citenamefont {Wenzel}, \citenamefont {Wright}, \citenamefont {Wright}, \citenamefont {Yamashita}, \citenamefont {Yarba},\ and\ \citenamefont {Yoshida}}]{Allison2016}%
  \BibitemOpen
  \bibfield  {author} {\bibinfo {author} {\bibfnamefont {J.}~\bibnamefont {Allison}}, \bibinfo {author} {\bibfnamefont {K.}~\bibnamefont {Amako}}, \bibinfo {author} {\bibfnamefont {J.}~\bibnamefont {Apostolakis}}, \bibinfo {author} {\bibfnamefont {P.}~\bibnamefont {Arce}}, \bibinfo {author} {\bibfnamefont {M.}~\bibnamefont {Asai}}, \bibinfo {author} {\bibfnamefont {T.}~\bibnamefont {Aso}}, \bibinfo {author} {\bibfnamefont {E.}~\bibnamefont {Bagli}}, \bibinfo {author} {\bibfnamefont {A.}~\bibnamefont {Bagulya}}, \bibinfo {author} {\bibfnamefont {S.}~\bibnamefont {Banerjee}}, \bibinfo {author} {\bibfnamefont {G.}~\bibnamefont {Barrand}}, \bibinfo {author} {\bibfnamefont {B.}~\bibnamefont {Beck}}, \bibinfo {author} {\bibfnamefont {A.}~\bibnamefont {Bogdanov}}, \bibinfo {author} {\bibfnamefont {D.}~\bibnamefont {Brandt}}, \bibinfo {author} {\bibfnamefont {J.}~\bibnamefont {Brown}}, \bibinfo {author} {\bibfnamefont {H.}~\bibnamefont {Burkhardt}}, \bibinfo {author} {\bibfnamefont {{\relax Ph}.}~\bibnamefont {Canal}}, \bibinfo {author} {\bibfnamefont {D.}~\bibnamefont {{Cano-Ott}}}, \bibinfo {author} {\bibfnamefont {S.}~\bibnamefont {Chauvie}}, \bibinfo {author} {\bibfnamefont {K.}~\bibnamefont {Cho}}, \bibinfo {author} {\bibfnamefont {G.}~\bibnamefont {Cirrone}}, \bibinfo {author} {\bibfnamefont {G.}~\bibnamefont {Cooperman}}, \bibinfo {author} {\bibfnamefont {M.}~\bibnamefont {{Cort{\'e}s-Giraldo}}}, \bibinfo {author} {\bibfnamefont {G.}~\bibnamefont {Cosmo}}, \bibinfo {author} {\bibfnamefont {G.}~\bibnamefont {Cuttone}}, \bibinfo {author} {\bibfnamefont {G.}~\bibnamefont {Depaola}}, \bibinfo {author} {\bibfnamefont {L.}~\bibnamefont {Desorgher}}, \bibinfo {author} {\bibfnamefont {X.}~\bibnamefont {Dong}}, \bibinfo {author} {\bibfnamefont {A.}~\bibnamefont {Dotti}}, \bibinfo {author} {\bibfnamefont {V.}~\bibnamefont {Elvira}}, \bibinfo {author} {\bibfnamefont {G.}~\bibnamefont {Folger}}, \bibinfo {author} {\bibfnamefont {Z.}~\bibnamefont {Francis}}, \bibinfo {author} {\bibfnamefont {A.}~\bibnamefont {Galoyan}}, \bibinfo {author} {\bibfnamefont {L.}~\bibnamefont {Garnier}}, \bibinfo {author} {\bibfnamefont {M.}~\bibnamefont {Gayer}}, \bibinfo {author} {\bibfnamefont {K.}~\bibnamefont {Genser}}, \bibinfo {author} {\bibfnamefont {V.}~\bibnamefont {Grichine}}, \bibinfo {author} {\bibfnamefont {S.}~\bibnamefont {Guatelli}}, \bibinfo {author} {\bibfnamefont {P.}~\bibnamefont {Gu{\`e}ye}}, \bibinfo {author} {\bibfnamefont {P.}~\bibnamefont {Gumplinger}}, \bibinfo {author} {\bibfnamefont {A.}~\bibnamefont {Howard}}, \bibinfo {author} {\bibfnamefont {I.}~\bibnamefont {H{\v r}ivn{\'a}{\v c}ov{\'a}}}, \bibinfo {author} {\bibfnamefont {S.}~\bibnamefont {Hwang}}, \bibinfo {author} {\bibfnamefont {S.}~\bibnamefont {Incerti}}, \bibinfo {author} {\bibfnamefont {A.}~\bibnamefont {Ivanchenko}}, \bibinfo {author} {\bibfnamefont {V.}~\bibnamefont {Ivanchenko}}, \bibinfo {author} {\bibfnamefont {F.}~\bibnamefont {Jones}}, \bibinfo {author} {\bibfnamefont {S.}~\bibnamefont {Jun}}, \bibinfo {author} {\bibfnamefont {P.}~\bibnamefont {Kaitaniemi}}, \bibinfo {author} {\bibfnamefont {N.}~\bibnamefont {Karakatsanis}}, \bibinfo {author} {\bibfnamefont {M.}~\bibnamefont {Karamitros}}, \bibinfo {author} {\bibfnamefont {M.}~\bibnamefont {Kelsey}}, \bibinfo {author} {\bibfnamefont {A.}~\bibnamefont {Kimura}}, \bibinfo {author} {\bibfnamefont {T.}~\bibnamefont {Koi}}, \bibinfo {author} {\bibfnamefont {H.}~\bibnamefont {Kurashige}}, \bibinfo {author} {\bibfnamefont {A.}~\bibnamefont {Lechner}}, \bibinfo {author} {\bibfnamefont {S.}~\bibnamefont {Lee}}, \bibinfo {author} {\bibfnamefont {F.}~\bibnamefont {Longo}}, \bibinfo {author} {\bibfnamefont {M.}~\bibnamefont {Maire}}, \bibinfo {author} {\bibfnamefont {D.}~\bibnamefont {Mancusi}}, \bibinfo {author} {\bibfnamefont {A.}~\bibnamefont {Mantero}}, \bibinfo {author} {\bibfnamefont {E.}~\bibnamefont {Mendoza}}, \bibinfo {author} {\bibfnamefont {B.}~\bibnamefont {Morgan}}, \bibinfo {author} {\bibfnamefont {K.}~\bibnamefont {Murakami}}, \bibinfo {author} {\bibfnamefont {T.}~\bibnamefont {Nikitina}}, \bibinfo {author} {\bibfnamefont {L.}~\bibnamefont {Pandola}}, \bibinfo {author} {\bibfnamefont {P.}~\bibnamefont {Paprocki}}, \bibinfo {author} {\bibfnamefont {J.}~\bibnamefont {Perl}}, \bibinfo {author} {\bibfnamefont {I.}~\bibnamefont {Petrovi{\'c}}}, \bibinfo {author} {\bibfnamefont {M.}~\bibnamefont {Pia}}, \bibinfo {author} {\bibfnamefont {W.}~\bibnamefont {Pokorski}}, \bibinfo {author} {\bibfnamefont {J.}~\bibnamefont {Quesada}}, \bibinfo {author} {\bibfnamefont {M.}~\bibnamefont {Raine}}, \bibinfo {author} {\bibfnamefont {M.}~\bibnamefont {Reis}}, \bibinfo {author} {\bibfnamefont {A.}~\bibnamefont {Ribon}}, \bibinfo {author} {\bibfnamefont {A.}~\bibnamefont {Risti{\'c}~Fira}}, \bibinfo {author} {\bibfnamefont {F.}~\bibnamefont {Romano}}, \bibinfo {author} {\bibfnamefont {G.}~\bibnamefont {Russo}}, \bibinfo {author} {\bibfnamefont {G.}~\bibnamefont {Santin}}, \bibinfo {author} {\bibfnamefont {T.}~\bibnamefont {Sasaki}}, \bibinfo {author} {\bibfnamefont {D.}~\bibnamefont {Sawkey}}, \bibinfo {author} {\bibfnamefont {J.}~\bibnamefont {Shin}}, \bibinfo {author} {\bibfnamefont {I.}~\bibnamefont {Strakovsky}}, \bibinfo {author} {\bibfnamefont {A.}~\bibnamefont {Taborda}}, \bibinfo {author} {\bibfnamefont {S.}~\bibnamefont {Tanaka}}, \bibinfo {author} {\bibfnamefont {B.}~\bibnamefont {Tom{\'e}}}, \bibinfo {author} {\bibfnamefont {T.}~\bibnamefont {Toshito}}, \bibinfo {author} {\bibfnamefont {H.}~\bibnamefont {Tran}}, \bibinfo {author} {\bibfnamefont {P.}~\bibnamefont {Truscott}}, \bibinfo {author} {\bibfnamefont {L.}~\bibnamefont {Urban}}, \bibinfo {author} {\bibfnamefont {V.}~\bibnamefont {Uzhinsky}}, \bibinfo {author} {\bibfnamefont {J.}~\bibnamefont {Verbeke}}, \bibinfo {author} {\bibfnamefont {M.}~\bibnamefont {Verderi}}, \bibinfo {author} {\bibfnamefont {B.}~\bibnamefont {Wendt}}, \bibinfo {author} {\bibfnamefont {H.}~\bibnamefont {Wenzel}}, \bibinfo {author} {\bibfnamefont {D.}~\bibnamefont {Wright}}, \bibinfo {author} {\bibfnamefont {D.}~\bibnamefont {Wright}}, \bibinfo {author} {\bibfnamefont {T.}~\bibnamefont {Yamashita}}, \bibinfo {author} {\bibfnamefont {J.}~\bibnamefont {Yarba}},\ and\ \bibinfo {author} {\bibfnamefont {H.}~\bibnamefont {Yoshida}},\ }\bibfield  {title} {\bibinfo {title} {Recent developments in {{Geant4}}},\ }\href {https://doi.org/10.1016/j.nima.2016.06.125} {\bibfield  {journal} {\bibinfo  {journal} {Nuclear Instruments and Methods in Physics Research Section A: Accelerators, Spectrometers, Detectors and Associated Equipment}\ }\textbf {\bibinfo {volume} {835}},\ \bibinfo {pages} {186} (\bibinfo {year} {2016})}\BibitemShut {NoStop}%
\bibitem [{\citenamefont {Brenner}\ \emph {et~al.}(2020)\citenamefont {Brenner}, \citenamefont {Balasubramanian}, \citenamefont {Burgard}, \citenamefont {Verkerke}, \citenamefont {Cowan}, \citenamefont {Verschuuren},\ and\ \citenamefont {Croft}}]{Brenner2020}%
  \BibitemOpen
  \bibfield  {author} {\bibinfo {author} {\bibfnamefont {L.}~\bibnamefont {Brenner}}, \bibinfo {author} {\bibfnamefont {R.}~\bibnamefont {Balasubramanian}}, \bibinfo {author} {\bibfnamefont {C.}~\bibnamefont {Burgard}}, \bibinfo {author} {\bibfnamefont {W.}~\bibnamefont {Verkerke}}, \bibinfo {author} {\bibfnamefont {G.}~\bibnamefont {Cowan}}, \bibinfo {author} {\bibfnamefont {P.}~\bibnamefont {Verschuuren}},\ and\ \bibinfo {author} {\bibfnamefont {V.}~\bibnamefont {Croft}},\ }\bibfield  {title} {\bibinfo {title} {Comparison of unfolding methods using {{RooFitUnfold}}},\ }\href {https://doi.org/10.1142/S0217751X20501456} {\bibfield  {journal} {\bibinfo  {journal} {International Journal of Modern Physics A}\ }\textbf {\bibinfo {volume} {35}},\ \bibinfo {pages} {2050145} (\bibinfo {year} {2020})}\BibitemShut {NoStop}%
\bibitem [{\citenamefont {D'Agostini}(1995)}]{DAgostini1995}%
  \BibitemOpen
  \bibfield  {author} {\bibinfo {author} {\bibfnamefont {G.}~\bibnamefont {D'Agostini}},\ }\bibfield  {title} {\bibinfo {title} {A multidimensional unfolding method based on {{Bayes}}' theorem},\ }\href {https://doi.org/10.1016/0168-9002(95)00274-X} {\bibfield  {journal} {\bibinfo  {journal} {Nuclear Instruments and Methods in Physics Research Section A: Accelerators, Spectrometers, Detectors and Associated Equipment}\ }\textbf {\bibinfo {volume} {362}},\ \bibinfo {pages} {487} (\bibinfo {year} {1995})}\BibitemShut {NoStop}%
\bibitem [{\citenamefont {Minato}\ \emph {et~al.}(2023)\citenamefont {Minato}, \citenamefont {Naito},\ and\ \citenamefont {Iwamoto}}]{Minato2023}%
  \BibitemOpen
  \bibfield  {author} {\bibinfo {author} {\bibfnamefont {F.}~\bibnamefont {Minato}}, \bibinfo {author} {\bibfnamefont {T.}~\bibnamefont {Naito}},\ and\ \bibinfo {author} {\bibfnamefont {O.}~\bibnamefont {Iwamoto}},\ }\bibfield  {title} {\bibinfo {title} {Nuclear many-body effects on particle emission following muon capture on {$^{28}$}{{Si}} and {$^{40}$}{{Ca}}},\ }\href {https://doi.org/10.1103/PhysRevC.107.054314} {\bibfield  {journal} {\bibinfo  {journal} {Physical Review C}\ }\textbf {\bibinfo {volume} {107}},\ \bibinfo {pages} {054314} (\bibinfo {year} {2023})}\BibitemShut {NoStop}%
\bibitem [{\citenamefont {Sato}\ \emph {et~al.}(2024)\citenamefont {Sato}, \citenamefont {Iwamoto}, \citenamefont {Hashimoto}, \citenamefont {Ogawa}, \citenamefont {Furuta}, \citenamefont {Abe}, \citenamefont {Kai}, \citenamefont {Matsuya}, \citenamefont {Matsuda}, \citenamefont {Hirata}, \citenamefont {Sekikawa}, \citenamefont {Yao}, \citenamefont {Tsai}, \citenamefont {Ratliff}, \citenamefont {Iwase}, \citenamefont {Sakaki}, \citenamefont {Sugihara}, \citenamefont {Shigyo}, \citenamefont {Sihver},\ and\ \citenamefont {Niita}}]{Sato2024}%
  \BibitemOpen
  \bibfield  {author} {\bibinfo {author} {\bibfnamefont {T.}~\bibnamefont {Sato}}, \bibinfo {author} {\bibfnamefont {Y.}~\bibnamefont {Iwamoto}}, \bibinfo {author} {\bibfnamefont {S.}~\bibnamefont {Hashimoto}}, \bibinfo {author} {\bibfnamefont {T.}~\bibnamefont {Ogawa}}, \bibinfo {author} {\bibfnamefont {T.}~\bibnamefont {Furuta}}, \bibinfo {author} {\bibfnamefont {S.}~\bibnamefont {Abe}}, \bibinfo {author} {\bibfnamefont {T.}~\bibnamefont {Kai}}, \bibinfo {author} {\bibfnamefont {Y.}~\bibnamefont {Matsuya}}, \bibinfo {author} {\bibfnamefont {N.}~\bibnamefont {Matsuda}}, \bibinfo {author} {\bibfnamefont {Y.}~\bibnamefont {Hirata}}, \bibinfo {author} {\bibfnamefont {T.}~\bibnamefont {Sekikawa}}, \bibinfo {author} {\bibfnamefont {L.}~\bibnamefont {Yao}}, \bibinfo {author} {\bibfnamefont {P.-E.}\ \bibnamefont {Tsai}}, \bibinfo {author} {\bibfnamefont {H.~N.}\ \bibnamefont {Ratliff}}, \bibinfo {author} {\bibfnamefont {H.}~\bibnamefont {Iwase}}, \bibinfo {author} {\bibfnamefont {Y.}~\bibnamefont {Sakaki}}, \bibinfo {author} {\bibfnamefont {K.}~\bibnamefont {Sugihara}}, \bibinfo {author} {\bibfnamefont {N.}~\bibnamefont {Shigyo}}, \bibinfo {author} {\bibfnamefont {L.}~\bibnamefont {Sihver}},\ and\ \bibinfo {author} {\bibfnamefont {K.}~\bibnamefont {Niita}},\ }\bibfield  {title} {\bibinfo {title} {Recent improvements of the particle and heavy ion transport code system -- {{PHITS}} version 3.33},\ }\href {https://doi.org/10.1080/00223131.2023.2275736} {\bibfield  {journal} {\bibinfo  {journal} {Journal of Nuclear Science and Technology}\ }\textbf {\bibinfo {volume} {61}},\ \bibinfo {pages} {127} (\bibinfo {year} {2024})}\BibitemShut {NoStop}%
\bibitem [{\citenamefont {Berglund}\ and\ \citenamefont {Wieser}(2011)}]{Berglund2011}%
  \BibitemOpen
  \bibfield  {author} {\bibinfo {author} {\bibfnamefont {M.}~\bibnamefont {Berglund}}\ and\ \bibinfo {author} {\bibfnamefont {M.~E.}\ \bibnamefont {Wieser}},\ }\bibfield  {title} {\bibinfo {title} {Isotopic compositions of the elements 2009 ({{IUPAC Technical Report}})},\ }\href {https://doi.org/10.1351/PAC-REP-10-06-02} {\bibfield  {journal} {\bibinfo  {journal} {Pure and Applied Chemistry}\ }\textbf {\bibinfo {volume} {83}},\ \bibinfo {pages} {397} (\bibinfo {year} {2011})}\BibitemShut {NoStop}%
\bibitem [{\citenamefont {O'Connell}\ \emph {et~al.}(1972)\citenamefont {O'Connell}, \citenamefont {Donnelly},\ and\ \citenamefont {Walecka}}]{OConnell1972}%
  \BibitemOpen
  \bibfield  {author} {\bibinfo {author} {\bibfnamefont {J.~S.}\ \bibnamefont {O'Connell}}, \bibinfo {author} {\bibfnamefont {T.~W.}\ \bibnamefont {Donnelly}},\ and\ \bibinfo {author} {\bibfnamefont {J.~D.}\ \bibnamefont {Walecka}},\ }\bibfield  {title} {\bibinfo {title} {Semileptonic {{Weak Interactions}} with {{C$^{12}$}}},\ }\href {https://doi.org/10.1103/PhysRevC.6.719} {\bibfield  {journal} {\bibinfo  {journal} {Physical Review C}\ }\textbf {\bibinfo {volume} {6}},\ \bibinfo {pages} {719} (\bibinfo {year} {1972})}\BibitemShut {NoStop}%
\bibitem [{\citenamefont {Minato}(2016)}]{Minato2016}%
  \BibitemOpen
  \bibfield  {author} {\bibinfo {author} {\bibfnamefont {F.}~\bibnamefont {Minato}},\ }\bibfield  {title} {\bibinfo {title} {Estimation of a 2p2h effect on {{Gamow-Teller}} transitions within the second {{Tamm-Dancoff}} approximation},\ }\href {https://doi.org/10.1103/PhysRevC.93.044319} {\bibfield  {journal} {\bibinfo  {journal} {Physical Review C}\ }\textbf {\bibinfo {volume} {93}},\ \bibinfo {pages} {044319} (\bibinfo {year} {2016})}\BibitemShut {NoStop}%
\bibitem [{\citenamefont {Reinhard}\ \emph {et~al.}(1999)\citenamefont {Reinhard}, \citenamefont {Dean}, \citenamefont {Nazarewicz}, \citenamefont {Dobaczewski}, \citenamefont {Maruhn},\ and\ \citenamefont {Strayer}}]{Reinhard1999}%
  \BibitemOpen
  \bibfield  {author} {\bibinfo {author} {\bibfnamefont {P.-G.}\ \bibnamefont {Reinhard}}, \bibinfo {author} {\bibfnamefont {D.~J.}\ \bibnamefont {Dean}}, \bibinfo {author} {\bibfnamefont {W.}~\bibnamefont {Nazarewicz}}, \bibinfo {author} {\bibfnamefont {J.}~\bibnamefont {Dobaczewski}}, \bibinfo {author} {\bibfnamefont {J.~A.}\ \bibnamefont {Maruhn}},\ and\ \bibinfo {author} {\bibfnamefont {M.~R.}\ \bibnamefont {Strayer}},\ }\bibfield  {title} {\bibinfo {title} {Shape coexistence and the effective nucleon-nucleon interaction},\ }\href {https://doi.org/10.1103/PhysRevC.60.014316} {\bibfield  {journal} {\bibinfo  {journal} {Physical Review C}\ }\textbf {\bibinfo {volume} {60}},\ \bibinfo {pages} {014316} (\bibinfo {year} {1999})}\BibitemShut {NoStop}%
\bibitem [{\citenamefont {Van~Giai}\ and\ \citenamefont {Sagawa}(1981)}]{VanGiai1981a}%
  \BibitemOpen
  \bibfield  {author} {\bibinfo {author} {\bibfnamefont {N.}~\bibnamefont {Van~Giai}}\ and\ \bibinfo {author} {\bibfnamefont {H.}~\bibnamefont {Sagawa}},\ }\bibfield  {title} {\bibinfo {title} {Spin-isospin and pairing properties of modified {{Skyrme}} interactions},\ }\href {https://doi.org/10.1016/0370-2693(81)90646-8} {\bibfield  {journal} {\bibinfo  {journal} {Physics Letters B}\ }\textbf {\bibinfo {volume} {106}},\ \bibinfo {pages} {379} (\bibinfo {year} {1981})}\BibitemShut {NoStop}%
\bibitem [{\citenamefont {Lifshitz}\ and\ \citenamefont {Singer}(1988)}]{Lifshitz1988}%
  \BibitemOpen
  \bibfield  {author} {\bibinfo {author} {\bibfnamefont {M.}~\bibnamefont {Lifshitz}}\ and\ \bibinfo {author} {\bibfnamefont {P.}~\bibnamefont {Singer}},\ }\bibfield  {title} {\bibinfo {title} {Meson-exchange currents and energetic particle emission from {$\mu^-$} capture},\ }\href {https://doi.org/10.1016/0375-9474(88)90330-2} {\bibfield  {journal} {\bibinfo  {journal} {Nuclear Physics A}\ }\textbf {\bibinfo {volume} {476}},\ \bibinfo {pages} {684} (\bibinfo {year} {1988})}\BibitemShut {NoStop}%
\bibitem [{\citenamefont {Kalbach}(1986)}]{Kalbach1986}%
  \BibitemOpen
  \bibfield  {author} {\bibinfo {author} {\bibfnamefont {C.}~\bibnamefont {Kalbach}},\ }\bibfield  {title} {\bibinfo {title} {Two-component exciton model: {{Basic}} formalism away from shell closures},\ }\href {https://doi.org/10.1103/PhysRevC.33.818} {\bibfield  {journal} {\bibinfo  {journal} {Physical Review C}\ }\textbf {\bibinfo {volume} {33}},\ \bibinfo {pages} {818} (\bibinfo {year} {1986})}\BibitemShut {NoStop}%
\bibitem [{\citenamefont {Koning}\ and\ \citenamefont {Duijvestijn}(2004)}]{Koning2004}%
  \BibitemOpen
  \bibfield  {author} {\bibinfo {author} {\bibfnamefont {A.}~\bibnamefont {Koning}}\ and\ \bibinfo {author} {\bibfnamefont {M.}~\bibnamefont {Duijvestijn}},\ }\bibfield  {title} {\bibinfo {title} {A global pre-equilibrium analysis from 7 to 200 {{MeV}} based on the optical model potential},\ }\href {https://doi.org/10.1016/j.nuclphysa.2004.08.013} {\bibfield  {journal} {\bibinfo  {journal} {Nuclear Physics A}\ }\textbf {\bibinfo {volume} {744}},\ \bibinfo {pages} {15} (\bibinfo {year} {2004})}\BibitemShut {NoStop}%
\bibitem [{\citenamefont {Hauser}\ and\ \citenamefont {Feshbach}(1952)}]{Hauser1952}%
  \BibitemOpen
  \bibfield  {author} {\bibinfo {author} {\bibfnamefont {W.}~\bibnamefont {Hauser}}\ and\ \bibinfo {author} {\bibfnamefont {H.}~\bibnamefont {Feshbach}},\ }\bibfield  {title} {\bibinfo {title} {The {{Inelastic Scattering}} of {{Neutrons}}},\ }\href {https://doi.org/10.1103/PhysRev.87.366} {\bibfield  {journal} {\bibinfo  {journal} {Physical Review}\ }\textbf {\bibinfo {volume} {87}},\ \bibinfo {pages} {366} (\bibinfo {year} {1952})}\BibitemShut {NoStop}%
\bibitem [{\citenamefont {Iwamoto}\ \emph {et~al.}(2016)\citenamefont {Iwamoto}, \citenamefont {Iwamoto}, \citenamefont {Kunieda}, \citenamefont {Minato},\ and\ \citenamefont {Shibata}}]{Iwamoto2016}%
  \BibitemOpen
  \bibfield  {author} {\bibinfo {author} {\bibfnamefont {O.}~\bibnamefont {Iwamoto}}, \bibinfo {author} {\bibfnamefont {N.}~\bibnamefont {Iwamoto}}, \bibinfo {author} {\bibfnamefont {S.}~\bibnamefont {Kunieda}}, \bibinfo {author} {\bibfnamefont {F.}~\bibnamefont {Minato}},\ and\ \bibinfo {author} {\bibfnamefont {K.}~\bibnamefont {Shibata}},\ }\bibfield  {title} {\bibinfo {title} {The {{CCONE Code System}} and its {{Application}} to {{Nuclear Data Evaluation}} for {{Fission}} and {{Other Reactions}}},\ }\href {https://doi.org/10.1016/j.nds.2015.12.004} {\bibfield  {journal} {\bibinfo  {journal} {Nuclear Data Sheets}\ }\textbf {\bibinfo {volume} {131}},\ \bibinfo {pages} {259} (\bibinfo {year} {2016})}\BibitemShut {NoStop}%
\bibitem [{\citenamefont {Abe}\ and\ \citenamefont {Sato}(2017)}]{Abe2017}%
  \BibitemOpen
  \bibfield  {author} {\bibinfo {author} {\bibfnamefont {S.}~\bibnamefont {Abe}}\ and\ \bibinfo {author} {\bibfnamefont {T.}~\bibnamefont {Sato}},\ }\bibfield  {title} {\bibinfo {title} {Implementation of muon interaction models in {{PHITS}}},\ }\href {https://doi.org/10.1080/00223131.2016.1210043} {\bibfield  {journal} {\bibinfo  {journal} {Journal of Nuclear Science and Technology}\ }\textbf {\bibinfo {volume} {54}},\ \bibinfo {pages} {101} (\bibinfo {year} {2017})}\BibitemShut {NoStop}%
\bibitem [{\citenamefont {Singer}(1962)}]{Singer1962}%
  \BibitemOpen
  \bibfield  {author} {\bibinfo {author} {\bibfnamefont {P.}~\bibnamefont {Singer}},\ }\bibfield  {title} {\bibinfo {title} {Neutron emission following muon capture in heavy nuclei},\ }\href {https://doi.org/10.1007/BF02732735} {\bibfield  {journal} {\bibinfo  {journal} {Il Nuovo Cimento (1955-1965)}\ }\textbf {\bibinfo {volume} {23}},\ \bibinfo {pages} {669} (\bibinfo {year} {1962})}\BibitemShut {NoStop}%
\bibitem [{\citenamefont {Niita}\ \emph {et~al.}(1995)\citenamefont {Niita}, \citenamefont {Chiba}, \citenamefont {Maruyama}, \citenamefont {Maruyama}, \citenamefont {Takada}, \citenamefont {Fukahori}, \citenamefont {Nakahara},\ and\ \citenamefont {Iwamoto}}]{Niita1995}%
  \BibitemOpen
  \bibfield  {author} {\bibinfo {author} {\bibfnamefont {K.}~\bibnamefont {Niita}}, \bibinfo {author} {\bibfnamefont {S.}~\bibnamefont {Chiba}}, \bibinfo {author} {\bibfnamefont {T.}~\bibnamefont {Maruyama}}, \bibinfo {author} {\bibfnamefont {T.}~\bibnamefont {Maruyama}}, \bibinfo {author} {\bibfnamefont {H.}~\bibnamefont {Takada}}, \bibinfo {author} {\bibfnamefont {T.}~\bibnamefont {Fukahori}}, \bibinfo {author} {\bibfnamefont {Y.}~\bibnamefont {Nakahara}},\ and\ \bibinfo {author} {\bibfnamefont {A.}~\bibnamefont {Iwamoto}},\ }\bibfield  {title} {\bibinfo {title} {Analysis of the {{$(N,xN^\prime)$}} reactions by quantum molecular dynamics plus statistical decay model},\ }\href {https://doi.org/10.1103/PhysRevC.52.2620} {\bibfield  {journal} {\bibinfo  {journal} {Physical Review C}\ }\textbf {\bibinfo {volume} {52}},\ \bibinfo {pages} {2620} (\bibinfo {year} {1995})}\BibitemShut {NoStop}%
\bibitem [{\citenamefont {Ogawa}\ \emph {et~al.}(2015)\citenamefont {Ogawa}, \citenamefont {Sato}, \citenamefont {Hashimoto}, \citenamefont {Satoh}, \citenamefont {Tsuda},\ and\ \citenamefont {Niita}}]{Ogawa2015}%
  \BibitemOpen
  \bibfield  {author} {\bibinfo {author} {\bibfnamefont {T.}~\bibnamefont {Ogawa}}, \bibinfo {author} {\bibfnamefont {T.}~\bibnamefont {Sato}}, \bibinfo {author} {\bibfnamefont {S.}~\bibnamefont {Hashimoto}}, \bibinfo {author} {\bibfnamefont {D.}~\bibnamefont {Satoh}}, \bibinfo {author} {\bibfnamefont {S.}~\bibnamefont {Tsuda}},\ and\ \bibinfo {author} {\bibfnamefont {K.}~\bibnamefont {Niita}},\ }\bibfield  {title} {\bibinfo {title} {Energy-dependent fragmentation cross sections of relativistic {{$^{12}$C}}},\ }\href {https://doi.org/10.1103/PhysRevC.92.024614} {\bibfield  {journal} {\bibinfo  {journal} {Physical Review C}\ }\textbf {\bibinfo {volume} {92}},\ \bibinfo {pages} {024614} (\bibinfo {year} {2015})}\BibitemShut {NoStop}%
\bibitem [{\citenamefont {Furihata}(2000)}]{Furihata2000}%
  \BibitemOpen
  \bibfield  {author} {\bibinfo {author} {\bibfnamefont {S.}~\bibnamefont {Furihata}},\ }\bibfield  {title} {\bibinfo {title} {Statistical analysis of light fragment production from medium energy proton-induced reactions},\ }\href {https://doi.org/10.1016/S0168-583X(00)00332-3} {\bibfield  {journal} {\bibinfo  {journal} {Nuclear Instruments and Methods in Physics Research Section B: Beam Interactions with Materials and Atoms}\ }\textbf {\bibinfo {volume} {171}},\ \bibinfo {pages} {251} (\bibinfo {year} {2000})}\BibitemShut {NoStop}%
\bibitem [{\citenamefont {Furihata}(2001)}]{Furihata2001}%
  \BibitemOpen
  \bibfield  {author} {\bibinfo {author} {\bibfnamefont {S.}~\bibnamefont {Furihata}},\ }\bibfield  {title} {\bibinfo {title} {The {{GEM Code}} - the {{Generalized Evaporation Model}} and the {{Fission Model}}},\ }in\ \href {https://doi.org/10.1007/978-3-642-18211-2_168} {\emph {\bibinfo {booktitle} {Advanced {{Monte Carlo}} for {{Radiation Physics}}, {{Particle Transport Simulation}} and {{Applications}}}}},\ \bibinfo {editor} {edited by\ \bibinfo {editor} {\bibfnamefont {A.}~\bibnamefont {Kling}}, \bibinfo {editor} {\bibfnamefont {F.~J.~C.}\ \bibnamefont {Bar{\"a}o}}, \bibinfo {editor} {\bibfnamefont {M.}~\bibnamefont {Nakagawa}}, \bibinfo {editor} {\bibfnamefont {L.}~\bibnamefont {T{\'a}vora}},\ and\ \bibinfo {editor} {\bibfnamefont {P.}~\bibnamefont {Vaz}}}\ (\bibinfo  {publisher} {Springer Berlin Heidelberg},\ \bibinfo {address} {Berlin, Heidelberg},\ \bibinfo {year} {2001})\ pp.\ \bibinfo {pages} {1045--1050}\BibitemShut {NoStop}%
\bibitem [{\citenamefont {Watanabe}\ and\ \citenamefont {Kadrev}(2007)}]{Watanabe2007}%
  \BibitemOpen
  \bibfield  {author} {\bibinfo {author} {\bibfnamefont {Y.}~\bibnamefont {Watanabe}}\ and\ \bibinfo {author} {\bibfnamefont {D.~N.}\ \bibnamefont {Kadrev}},\ }\bibfield  {title} {\bibinfo {title} {Extension of quantum molecular dynamics for production of light complex particles in nucleon-induced reactions},\ }in\ \href {https://doi.org/10.1051/ndata:07196} {\emph {\bibinfo {booktitle} {{{ND2007}}}}}\ (\bibinfo  {publisher} {EDP Sciences},\ \bibinfo {address} {Nice, France},\ \bibinfo {year} {2007})\ pp.\ \bibinfo {pages} {1121--1124}\BibitemShut {NoStop}%
\bibitem [{\citenamefont {Iwamoto}\ and\ \citenamefont {Harada}(1982)}]{Iwamoto1982}%
  \BibitemOpen
  \bibfield  {author} {\bibinfo {author} {\bibfnamefont {A.}~\bibnamefont {Iwamoto}}\ and\ \bibinfo {author} {\bibfnamefont {K.}~\bibnamefont {Harada}},\ }\bibfield  {title} {\bibinfo {title} {Mechanism of cluster emission in nucleon-induced preequilibrium reactions},\ }\href {https://doi.org/10.1103/PhysRevC.26.1821} {\bibfield  {journal} {\bibinfo  {journal} {Physical Review C}\ }\textbf {\bibinfo {volume} {26}},\ \bibinfo {pages} {1821} (\bibinfo {year} {1982})}\BibitemShut {NoStop}%
\bibitem [{\citenamefont {Sato}\ \emph {et~al.}(1983)\citenamefont {Sato}, \citenamefont {Iwamoto},\ and\ \citenamefont {Harada}}]{Sato1983}%
  \BibitemOpen
  \bibfield  {author} {\bibinfo {author} {\bibfnamefont {K.}~\bibnamefont {Sato}}, \bibinfo {author} {\bibfnamefont {A.}~\bibnamefont {Iwamoto}},\ and\ \bibinfo {author} {\bibfnamefont {K.}~\bibnamefont {Harada}},\ }\bibfield  {title} {\bibinfo {title} {Pre-equilibrium emission of light composite particles in the framework of the exciton model},\ }\href {https://doi.org/10.1103/PhysRevC.28.1527} {\bibfield  {journal} {\bibinfo  {journal} {Physical Review C}\ }\textbf {\bibinfo {volume} {28}},\ \bibinfo {pages} {1527} (\bibinfo {year} {1983})}\BibitemShut {NoStop}%
\bibitem [{\citenamefont {Lifshitz}\ and\ \citenamefont {Singer}(1980)}]{Lifshitz1980}%
  \BibitemOpen
  \bibfield  {author} {\bibinfo {author} {\bibfnamefont {M.}~\bibnamefont {Lifshitz}}\ and\ \bibinfo {author} {\bibfnamefont {P.}~\bibnamefont {Singer}},\ }\bibfield  {title} {\bibinfo {title} {Nuclear excitation function and particle emission from complex nuclei following muon capture},\ }\href {https://doi.org/10.1103/PhysRevC.22.2135} {\bibfield  {journal} {\bibinfo  {journal} {Physical Review C}\ }\textbf {\bibinfo {volume} {22}},\ \bibinfo {pages} {2135} (\bibinfo {year} {1980})}\BibitemShut {NoStop}%
\bibitem [{\citenamefont {Sundelin}\ and\ \citenamefont {Edelstein}(1973)}]{Sundelin1973}%
  \BibitemOpen
  \bibfield  {author} {\bibinfo {author} {\bibfnamefont {R.~M.}\ \bibnamefont {Sundelin}}\ and\ \bibinfo {author} {\bibfnamefont {R.~M.}\ \bibnamefont {Edelstein}},\ }\bibfield  {title} {\bibinfo {title} {Neutron {{Asymmetries}} and {{Energy Spectra}} from {{Muon Capture}} in {{Si}}, {{S}}, and {{Ca}}},\ }\href {https://doi.org/10.1103/PhysRevC.7.1037} {\bibfield  {journal} {\bibinfo  {journal} {Physical Review C}\ }\textbf {\bibinfo {volume} {7}},\ \bibinfo {pages} {1037} (\bibinfo {year} {1973})}\BibitemShut {NoStop}%
\bibitem [{\citenamefont {Kozlowski}\ \emph {et~al.}(1985)\citenamefont {Kozlowski}, \citenamefont {Bertl}, \citenamefont {Povel}, \citenamefont {Sennhauser}, \citenamefont {Walter}, \citenamefont {Zglinski}, \citenamefont {Engfer}, \citenamefont {Grab}, \citenamefont {Hermes}, \citenamefont {Isaak}, \citenamefont {Van Der~Schaaf}, \citenamefont {Van Der~Pluym},\ and\ \citenamefont {Hesselink}}]{Kozlowski1985}%
  \BibitemOpen
  \bibfield  {author} {\bibinfo {author} {\bibfnamefont {T.}~\bibnamefont {Kozlowski}}, \bibinfo {author} {\bibfnamefont {W.}~\bibnamefont {Bertl}}, \bibinfo {author} {\bibfnamefont {H.~P.}\ \bibnamefont {Povel}}, \bibinfo {author} {\bibfnamefont {U.}~\bibnamefont {Sennhauser}}, \bibinfo {author} {\bibfnamefont {H.~K.}\ \bibnamefont {Walter}}, \bibinfo {author} {\bibfnamefont {A.}~\bibnamefont {Zglinski}}, \bibinfo {author} {\bibfnamefont {R.}~\bibnamefont {Engfer}}, \bibinfo {author} {\bibfnamefont {{\relax CH}.}~\bibnamefont {Grab}}, \bibinfo {author} {\bibfnamefont {E.~A.}\ \bibnamefont {Hermes}}, \bibinfo {author} {\bibfnamefont {H.~P.}\ \bibnamefont {Isaak}}, \bibinfo {author} {\bibfnamefont {A.}~\bibnamefont {Van Der~Schaaf}}, \bibinfo {author} {\bibfnamefont {J.}~\bibnamefont {Van Der~Pluym}},\ and\ \bibinfo {author} {\bibfnamefont {W.~H.~A.}\ \bibnamefont {Hesselink}},\ }\bibfield  {title} {\bibinfo {title} {Energy spectra and asymmetries of neutrons emitted after muon capture},\ }\href {https://doi.org/10.1016/0375-9474(85)90556-1} {\bibfield  {journal} {\bibinfo  {journal} {Nuclear Physics A}\ }\textbf {\bibinfo {volume} {436}},\ \bibinfo {pages} {717} (\bibinfo {year} {1985})}\BibitemShut {NoStop}%
\end{thebibliography}%

\end{document}